\documentclass{elsart}

\usepackage{natbib}
\usepackage{lineno}
\usepackage{units}
\usepackage{graphicx}
\usepackage{rotating,makecell}
\usepackage{icarus}
\usepackage{float}
\usepackage{color}
\usepackage{amssymb}

\usepackage{amsmath}
\usepackage{xcolor}
\usepackage{fmtcount}

\bibpunct{(}{)}{;}{a}{,}{,}

\usepackage{graphicx}

\newcommand{\BibTeX}{ \textrm{B\kern-.05em\textsc{i\kern-.025em b}\kern-.08em
    T\kern-.1667em\lower.7ex\hbox{E}\kern-.125emX} }

\begin{document}

\newcommand{\cfbox}[2]{%
    \colorlet{currentcolor}{.}%
    {\color{#1}%
    \fbox{\color{currentcolor}#2}}%
}

\linenumbers

\begin{frontmatter}



\title{Seasonal melting and the formation
 of sedimentary
rocks on Mars, 
with predictions for the Gale Crater mound }
\author[itay1]{Edwin S. Kite}, 
\author[itay2]{Itay Halevy},
\author[melinda1]{Melinda A. Kahre},
\author[ssi]{Michael J. Wolff}, and
\author[ucb1,ucb2]{Michael Manga}
\address[itay1]{Division of Geological and Planetary Sciences,
California Institute of Technology, Pasadena, California 91125, USA}
\address[itay2]{Center for Planetary Sciences, Weizmann
Institute of Science, P.O. Box 26, Rehovot 76100, Israel}
\address[melinda1]{NASA Ames Research
Center, Mountain View, California 94035, USA}
\address[ssi]{Space Science Institute, 4750 Walnut Street, Suite 205, Boulder, Colorado, USA}
\address[ucb1]{Department of Earth and Planetary Science, University of California Berkeley,
Berkeley, California 94720, USA}
\address[ucb2]{Center for Integrative Planetary Science, University of California Berkeley,
Berkeley, California 94720, USA}



%
%
%
%
%


\end{frontmatter}



\begin{flushleft}
\vspace{1cm}
Number of pages: \pageref{lastpage} \\
Number of tables: 1 \\
Number of figures: 19 \\
\end{flushleft}


\begin{pagetwo}{Seasonal melting and sedimentary rocks on Mars}

Edwin S. Kite \\
Caltech, MC 150-21\\
Geological and Planetary Sciences\\
1200 E California Boulevard\\
Pasadena, CA 91125, USA. \\
\\
Email: edwin.kite@gmail.com\\
Phone: (510) 717-5205 \\
\end{pagetwo}

\begin{abstract}
A model for the formation and distribution of sedimentary rocks on Mars is proposed. The rate--limiting step is supply of liquid water from seasonal melting of snow or ice. The model is run for a \emph{O}(10$^2$) mbar pure CO$_2$ atmosphere, dusty snow, and solar luminosity reduced by 23\%. For these conditions snow only melts near the equator, and only when obliquity $\gtrsim$40$^\circ$, eccentricity $\gtrsim$0.12, and perihelion occurs near equinox. These requirements for melting are satisfied by 0.01--20\% of the probability distribution of Mars' past spin-orbit parameters. Total melt production is sufficient to account for aqueous alteration of the sedimentary rocks. The pattern of seasonal snowmelt is integrated over all spin-orbit parameters and compared to the observed distribution of sedimentary rocks. The global distribution of snowmelt has maxima in Valles Marineris,  Meridiani Planum and Gale Crater. These correspond to maxima in the sedimentary-rock distribution. Higher pressures and especially higher temperatures lead to melting over a broader range of spin-orbit parameters. The pattern of sedimentary rocks on Mars is most consistent with a Mars paleoclimate that only rarely produced enough meltwater to precipitate aqueous cements and indurate sediment. 
The results suggest intermittency of snowmelt and long  globally-dry intervals, unfavorable for past life on Mars. This model makes testable predictions for the Mars Science Laboratory rover at Gale Crater. Gale Crater is predicted to be a hemispheric maximum for snowmelt on Mars.
\end{abstract}


\section{Introduction\label{introduction}}

\noindent Climate models struggle to maintain annual mean temperatures $\bar{T} \gtrsim$ 273K on Early Mars \citep{haberle1998} (Forget et al., submitted manuscript, 2012). Seasonal melting can occur for annual maximum temperatures $T_{max}$ $\gtrsim$ 273K, which is much easier to achieve. Therefore, seasonal melting of snow and ice is a candidate water source for surface runoff and aqueous mineralization on Mars. Surface temperatures $\sim$300K occur at low latitudes on today's Mars. However, seasonal melting of surface-covering, flat-lying snowpack does not occur because of (1) evaporative cooling and (2) cold-trapping of snow and ice near the poles or at depth.
Reduced solar luminosity for Early Mars makes melting more difficult \citep{squyres1994}.

\citet{toon1980} modelled the control of Milankovitch cycles on Mars ice temperatures. Melting is favored when snow is darkened by dust, and when evaporative cooling is reduced by increased pressure. 
\citet{jakosky1985} suggested that equatorial snowpacks would form at high obliquity. They pointed out that melt could contribute to the observed low-latitude erosion. \citet{clow1987} modeled snowmelt due to the solid-state greenhouse effect. He tracked meltwater migration to the base of the snowpack.  Several authors have modeled melting on steep slopes as a candidate water source for young midlatitude gullies \citep{costard2002,christensen2003}. \citet{hecht2002} considered the energy balance for water at the melting point in gully alcoves. \citet{williams2008snowpack,williams2009} modeled melting of relatively clean snow overlain by a thin, dark lag deposit. They found melt rates $\sim$1 kg/m$^2$/hr on steep slopes, but argue that this is sufficient to form gullies through either fluvial or debris--flow incision. \citet{morgan2010} used a 1D atmospheric model to examine water ice melting and CO$_2$ frost accumulation. No author has studied Early Mars cold-traps, although Schorghofer's model \citep{schorghofer2005,schorghofer2007b,schorghofer2010} has been used to track cold-traps for subsurface ice over the last 5 Ma \citep{schorghofer2007a}.

\begin{figure}[ht]
\includegraphics[width=1.0\textwidth, clip=true, trim = 5mm 20mm 20mm 60mm]{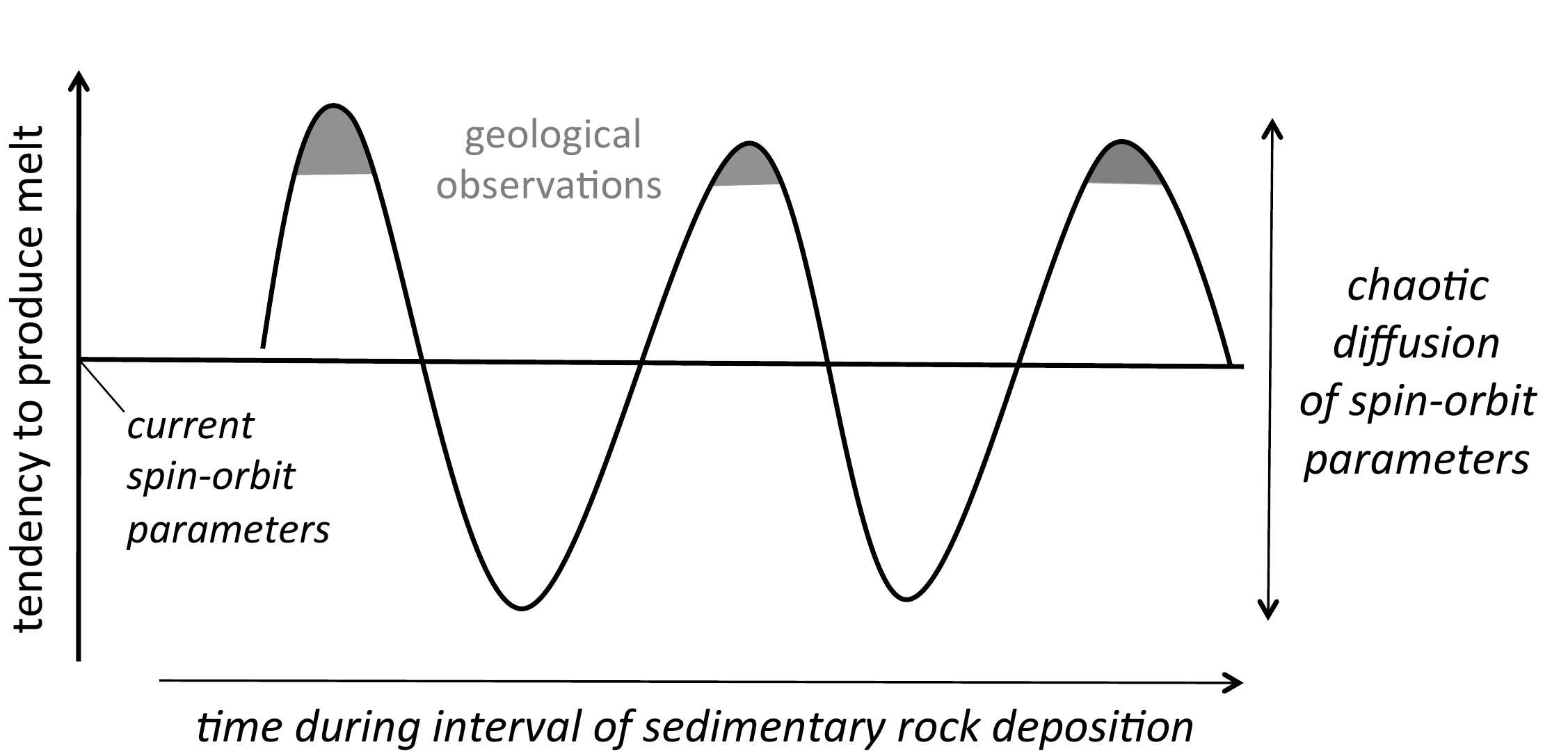}
\caption{Motivation for this paper. Mars underwent tens to thousands of spin-orbit oscillations during the interval of sedimentary-rock deposition. Three are shown schematically in this sketch. The geologic record of metastable surface liquid water is a wet-pass filter of Mars climate history. Mars orbital parameters vary over a wide range, resulting in a correspondingly wide range in tendency to melt. Therefore, evidence that the sedimentary rocks formed in a small fraction of Mars' history \citep{lewis2010} suggests that negligible melting occurred under mean orbital forcing. If Mars sedimentary rock record only records orbital conditions that permitted surface liquid water, modeling average orbital conditions is neither sufficient nor appropriate. Instead, it is necessary to calculate snowmelt  for the full range of orbital elements that Mars likely sampled over the time interval of sedimentary rock deposition. These predictions can then be compared to observations. Because of the evidence for orbital pacing of sedimentary rock accumulation \citep{lewis2008}, transient warming events are not shown, but may have been critical for generating geomorphically effective runoff - see \S8.4. \label{wetpassfilter}}
\end{figure}

The first purpose of this paper is to extend the global snowmelt models by integrating a new thermal model over all spin-orbit parameters, while accounting for cold-traps. Chaotic diffusion in the solar system makes it almost certain that Mars' obliquity ($\phi$) has ranged twenty times more widely than Earth's obliquity over billion-year periods, and that Mars' eccentricity has had a long-term variance twice that of the Earth \citep{touma1993,laskar1993,laskar2004,laskar2008}. These wide swings cause large variations in insolation and propensity to melt (Figure \ref{wetpassfilter}).


The second purpose of this paper is to understand the water source for sedimentary rock formation on Mars \citep{malin2000a}.  We focus on the hypothesis that supply of water from seasonal melting was the limiting step in the formation of sedimentary rocks on Early Mars \citep[e.g.,][]{wendt2011}. Existing evidence for snowmelt-limited sedimentary rock formation is discussed in \S\ref{hypothesis}. 

If surface liquid water availability was necessary for sedimentary rock formation, then the spatial distributions of liquid water availability and sedimentary rock detections should correspond. \S\ref{data} analyzes the global sedimentary rock distribution. 
In the only previous global model of sedimentary rock formation on Mars, \citet{andrewshanna2007} tracked groundwater flow in a global aquifer that is recharged by a broad low-latitude belt of precipitation. Groundwater upwelling is focussed in low-lying areas, generally consistent with the observed distribution of sedimentary rocks \citep{andrewshanna2010,andrewshanna2011}. This model assumes $\bar{T}>$273K, in order to avoid the development of an impermeable cryosphere. Especially in light of the Faint Young Sun predicted by standard solar models, temperatures this high may be unsustainable for the long periods of time required to form the sedimentary rocks \citep{haberle1998,tian2010} (Forget et al., submitted manuscript, 2012). The model described (\S\ref{model}) and analyzed (\S\ref{results}) in this paper assumes that liquid water is supplied from locally-derived snowmelt, rather than a deep global aquifer. Groundwater flow is confined to shallow, local aquifers perched above the cryosphere. Annually-- and planet--averaged temperatures remain similar to today's, which reduces the required change in climate forcing from the present state. If Mars climate once sustained $\bar{T}>$273K, then it must have passed through climate conditions amenable to snowmelt en route to the modern desert \citep{mckay1991duration}. The converse is not true.

Model predictions for different paleoclimate parameters are compared to global data in \S\ref{distribution}. \S\ref{formation} makes testable predictions for Gale Crater,  the target for the Mars Science Laboratory (MSL) mission \citep{milliken2010}. 

The discussion (\S\ref{discussion}) includes comparison of the snowmelt model to  alternative hypotheses such as global groundwater \citep{andrewshanna2007,andrewshanna2010,andrewshanna2011} and cryogenic weathering within ice sheets \citep{catling2006,niles2009}. The conclusions of this study are stated in \S\ref{conclusions}. 

This scope of this paper is forward modeling of snowmelt production as a function of (unknown) Early Mars climate parameters. Beyond a qualitative discussion in \S\ref{formation} and \S\ref{discussion}, there is no attempt to physically model the processes running from snowmelt production to sedimentary rock formation. A computationally inexpensive 1D model allows us to sweep over a large parameter space. The trade-off is that 1D models cannot track the effect of topographically-forced planetary waves on the atmospheric transport of water vapor, which controls snow precipitation \citep{vincendon2010,forget2006,colaprete2005}. Any 1D snow location prescription is therefore an idealization.  


\section{Snowmelt hypothesis\label{hypothesis}}

\noindent  Liquid water is required to explain sedimentary rock texture and bulk geochemistry along the Mars Exploration Rover \emph{Opportunity} traverse, and there is strong evidence for extending this conclusion to other light-toned, sulfate-bearing sedimentary rocks on Mars \citep{bibring2007,mclennan2008,roach2010diagenetic,murchie2009b,weitz2008ophir}. The hypothesis in this paper is that the water source for sedimentary rocks on Early Mars was seasonal melting, and that liquid water was infrequently available so that melt availability was the limiting factor in forming sedimentary rocks. ``Sedimentary rocks'' is used to mean units comprised of chemical precipitates or siliciclastic material cemented by chemical precipitates, usually sulfates. These are recognized from orbit as light-toned layered sedimentary deposits \citep{malin2010} that characteristically show diagnostic sulfate features in the near-infrared. This definition excludes layered phyllosilicates, which usually predate sulfates \citep{bibring2006, ehlmann2011}.



\subsection{What is the evidence that sediment lithification on Mars requires liquid water?} 

\noindent Erosion to form cliffs and boulders \citep{malin2000a}, ejection of meter-size boulders from small, fresh craters \citep{golombek2010}, resistance to crushing by rover wheels, and microscopic texture \citep{okubo2007} show that most light-toned sedimentary deposits (hereafter ``sedimentary rocks") are indurated or lithified. Lithification involves compaction and cementation. Water is required to form aqueous cements, and for fluvial sediment transport. At the \emph{Opportunity} landing site, evaporitic sandstones (60\% chemical precipitates by weight on an anhydrous basis) record groundwater recharge and aqueous cementation, surface runoff, and shallow lithification \citep{mclennan2005,mclennan2008}. Aqueous minerals are present in sedimentary rocks throughout Meridiani and the Valles Marineris \citep{bibring2007}. \citet{murchie2009b} argue for water-limited lithification of the Valles Marineris sedimentary rocks. Subsurface pressure-solution recrystallisation can occlude porosity and lithify weak evaporites at $\sim$30 bars, without aqueous cementation \citep{warren2006}. But in this case water is probably still needed to form evaporites at 60\% by weight. Some layered sedimentary deposits on Mars might not require liquid water to form, but these deposits are usually younger or at higher latitudes than the sulfate-bearing layered sedimentary rocks \citep{hynek2003ash,bridges2010,fenton2010}. 

Outcrop thermal inertia (TI) is almost always low at 100m scale, so these rocks were either never strongly cemented or have been weakened after exhumation \citep{edwards2009}.

\subsection{When did sulfate-bearing sedimentary rocks form?}

\noindent Sulfate-bearing sedimentary rocks occur relatively late in the stratigraphic sequence of evidence for stable surface liquid water on Mars \citep{murchie2009a,fassett2008a,masse2012,mangold2010,thollot2012}.
Formation of sedimentary rocks peaked in the Hesperian \citep{carr2010}, well after the peak of phyllosilicate formation on Mars \citep{ehlmann2011,fassett2011,salvatore2010,bibring2006}. Phyllosilicates are sometimes interbedded with older sulfate-bearing sedimentary rocks \citep{ehlmann2011}, but those phyllosilicates may be reworked \citep{barnhart2011}. Sedimentary rocks also postdate almost all of the large-scale, regionally integrated highland valley networks of the Late Noachian/Early Hesperian \citep{carr2010,fassett2011}, and are spatially separated from these ``classic'' valley networks \citep{hynek2010}. Therefore, the observed sedimentary rocks cannot be the terminal deposits of the classic valley networks. The climate that created the classic valley networks could have been different from the climate that formed the sedimentary rocks \citep{andrewshanna2011}. Sedimentary rocks do contain some channels, often preserved in inverted relief \citep{edgett2005,burr2009,burr2010}. Finally, many sedimentary rocks postdate the large impacts of the Late Heavy Bombardment, and many have quasi-periodic bedding suggesting orbitally-paced deposition \citep{lewis2008,lewis2010}. These observations are inconsistent with brief bursts of rapid sedimentary rock formation during impact-induced greenhouse events. 

\subsection{What existing data supports the snowmelt hypothesis?}
\noindent Liquid water was in short supply even at the time of sedimentary rock formation at the \emph{Opportunity} landing site. Mineralogy indicates low water/rock ratios during alteration, and that the cumulative duration of water-rock interaction was $\le$ 100 ka \citep{berger2009,hurowitz2007,elwoodmadden2009}.
 Weathering at Meridiani Planum was either isochemical or at low water/rock ratio or both \citep{ming2008}, consistent with a rare trickle of snowmelt.  Low specific grind energy of sandstones indicates weak aqueous cementation \citep{herkenhoff2008}. The present day extent of sedimentary rock outcrops is small, and the persistence of opal, jarosite and olivine in rocks (and olivine in soils) indicates minimal water-rock interaction since those minerals crystallized \citep{tosca2009,olsen2007,yen2005}.  Away from the sedimentary rocks themselves, aqueous mineralization was minor or absent elsewhere on the planet at the time when most sulfate-bearing sedimentary rocks formed \citep{murchie2009a,salvatore2010,mustard2009,fassett2011,hausrath2008}. Globally, soils formed ``with little aqueous alteration under conditions similar to those of the current Martian climate'' \citep{bandfield2011}, and elemental profiles indicate top-down mobilization of soluble elements \citep{amundson2008, arvidson2010}. 
 
 Dividing the total thickness of sedimentary rock deposits by the thickness of quasi-periodic layers and then multiplying by the obliquity periods thought to pace accumulation suggests that the sedimentary rocks formed in 1-10 Ma \citep{lewis2010}.  
This is a small fraction of Mars' history. Geomorphic evidence that the Mars surface environment has only marginally supported surface liquid water since the Noachian includes mean erosion rates $\sim$1 atom/year \citep{golombek2006}, together with a sharp post-Noachian decline in valley network formation and crater infilling \citep{fassett2008a,forsbergtaylor2004}. These data argue for a short-lived and downward-infiltrating post-Noachian water supply, suggestive of transient liquid water that is generated only during brief melt events. 

\subsection{How could brief pulses of snowmelt form kilometer-thick accumulations of sedimentary rock?}

\noindent Antarctica's McMurdo Dry Valleys are a terrestrial analog for seasonal-melt-limited fluvial erosion and sedimentary rock formation at $\bar{T}<$ 273K \citep{doran2010,lee2003,marchant2007}. Weathering and mineralization is confined to lakes, hyporheic zones, and a shallow active layer. However, seasonal river discharges reach 20 m$^3$ s$^{-1}$ \citep{mcknight2011}, fluvial valleys incise $>$ 3m deep into granite \citep{shaw1980}, and annually-averaged weathering intensity within the hyporheic zone is greater than in temperate latitudes \citep{nezat2001}. Ions are concentrated within ice-covered lakes by sublimation. Outcrops of gypsum, carbonate evaporites, and algal limestone sediments show that sediments have accumulated at the base of melt-fed perennial lakes for 300,000 years \citep{mckay1985,hendy2000}. Dry Valley Drilling Project cores show lithification in older horizons \citep{mckelvey1981}.

Order-of-magnitude energy and mass balance shows that brief, rare pulses of snowmelt provide enough water to form the kilometers of sedimentary rock observed on Mars. 
For solar luminosity reduced by 23\%, peak noontime insolation at Mars at perihelion on a moderate-eccentricity ($e$=0.15) orbit  is $\approx$ 630 W/m$^2$. If the snowpack is dusty then its albedo will be that of Mars dust, 0.28 \citep{putzig2005}. During melting, radiative losses are $\sigma T_{melt}^4$ $\approx$ 320W/m$^2$, and for a 200 mbar atmosphere a reasonable value for wind-speed dependent sublimation losses into dry air is $\sim$60 W/m$^2$. Conductive losses will be roughly one-half the diurnal temperature range divided by the diurnal skin depth, giving 60 W/m$^2$ for the snowpack material properties in \citet{carr2003a} and a 100K diurnal cycle of surface temperature. Greenhouse forcing from a 200 mbar CO$_2$ atmosphere equilibrated with a 230K daily mean surface temperature is $\sim$ 60 W/m$^2$(from detailed radiative transfer calculations: Appendix \ref{details}). Neglecting all other gains and losses, the net energy available for melting is therefore 630(1-0.28) - 320 - 60 - 60  + 60 $\sim$ 100 W/m$^2$, equivalent to approximately 1 kg/m$^2$/hr snowmelt. 
The total water required to form the 800m-thick Meridiani sediments depends on the water/rock mass ratio (W/R) during alteration. W/R is given as $\lesssim$1 by \citet{berger2009} and $\lesssim$300 by \citet{hurowitz2007}. This corresponds to a time-integrated melt column of either $\lesssim$0.3 km or  $\lesssim$100 km, respectively, for a bulk Meridiani sandstone density of 2.3, and ignoring the contribution of water bound in hydrated minerals to the solid mass of the deposit. Assuming melting occurs for 10\% of each sol, the melt season lasts for 10\% of the year, and  10\% of melt is available for alteration, snowmelt production for only 30 Kyr (for W/R = 1) or 10 Myr (for W/R = 300) provides enough snowmelt to reach the upper limit on W/R for the entire Meridiani sandstone. The period in question lasted $O$(10$^9$) years, so climate and orbital conditions favorable for surface liquid water at Meridiani are only needed for $<$1\% of the time. This is consistent with the wet-pass filter sketched in Figure \ref{wetpassfilter}.
 

\section{Distribution of sedimentary rocks on Mars\label{data}}

\begin{figure*}[h]
\begin{center}$
\begin{array}{cc}
\includegraphics[width=0.5\textwidth, clip=true, trim = 10mm 0mm 10mm 10mm]{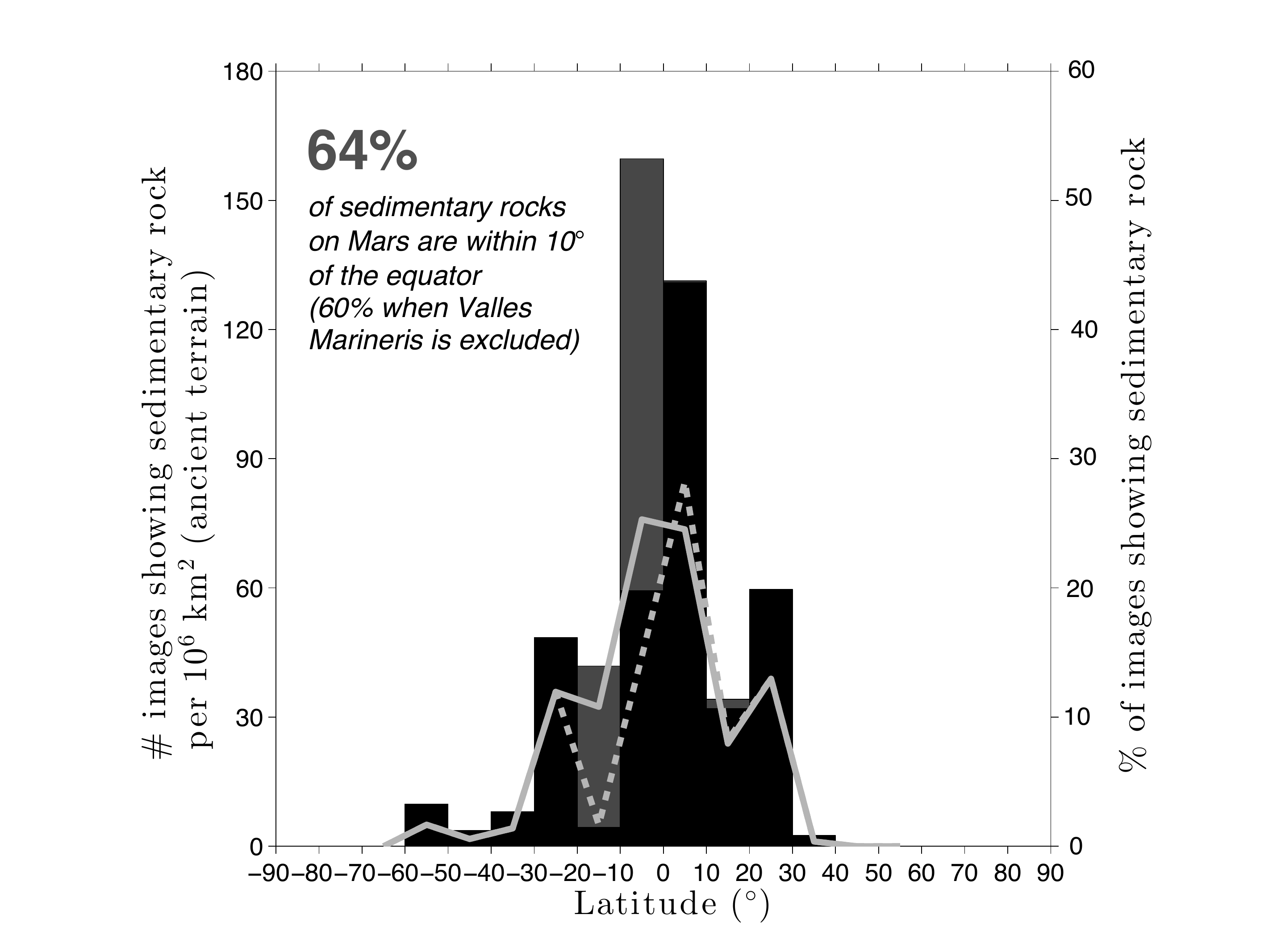} &
\includegraphics[width=0.5\textwidth, clip=true, trim = 10mm 0mm 10mm 10mm]{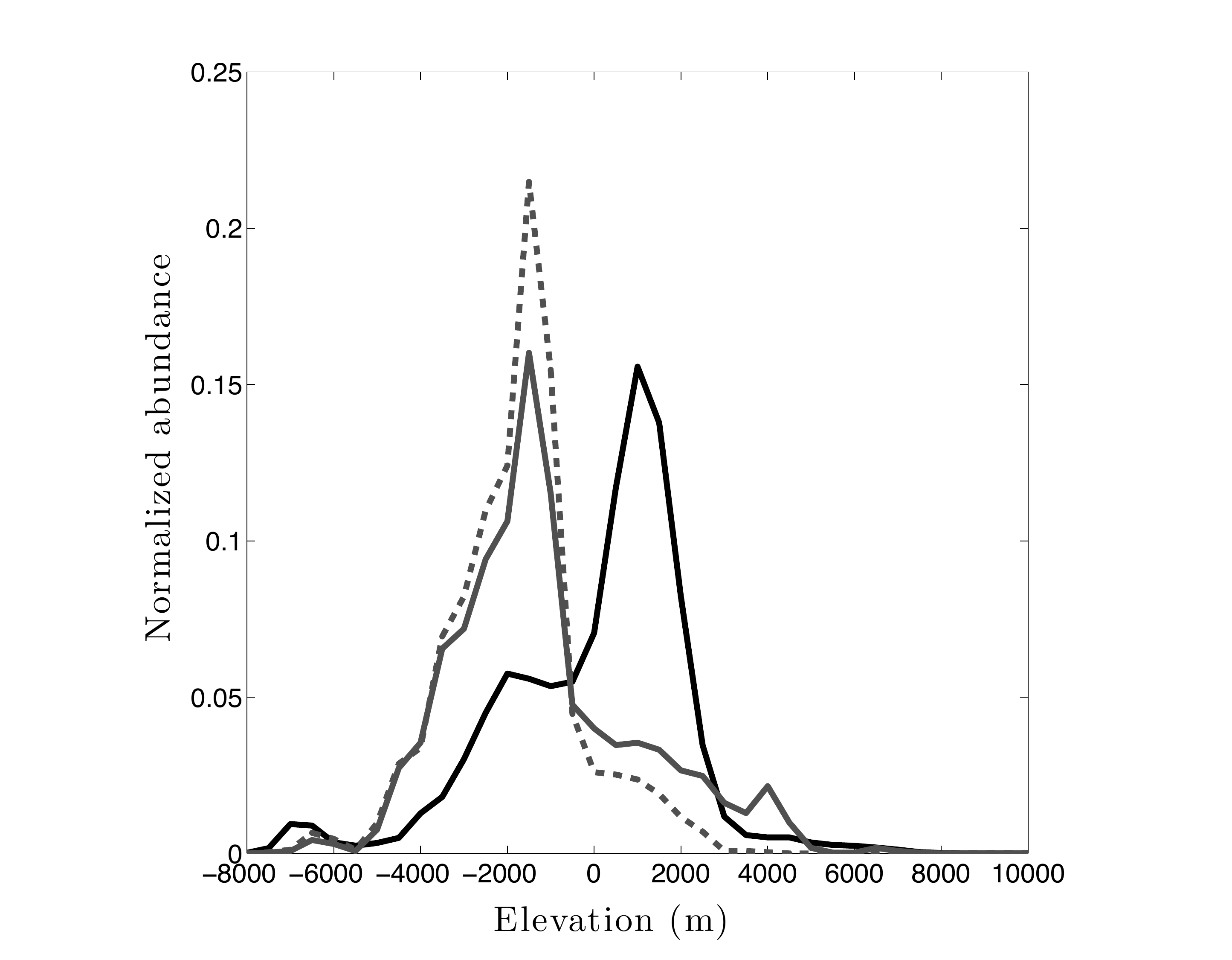} \\
\end{array}$
\end{center}
\includegraphics[width=1.0\textwidth, clip=true, trim = 20mm 80mm 10mm 70mm]{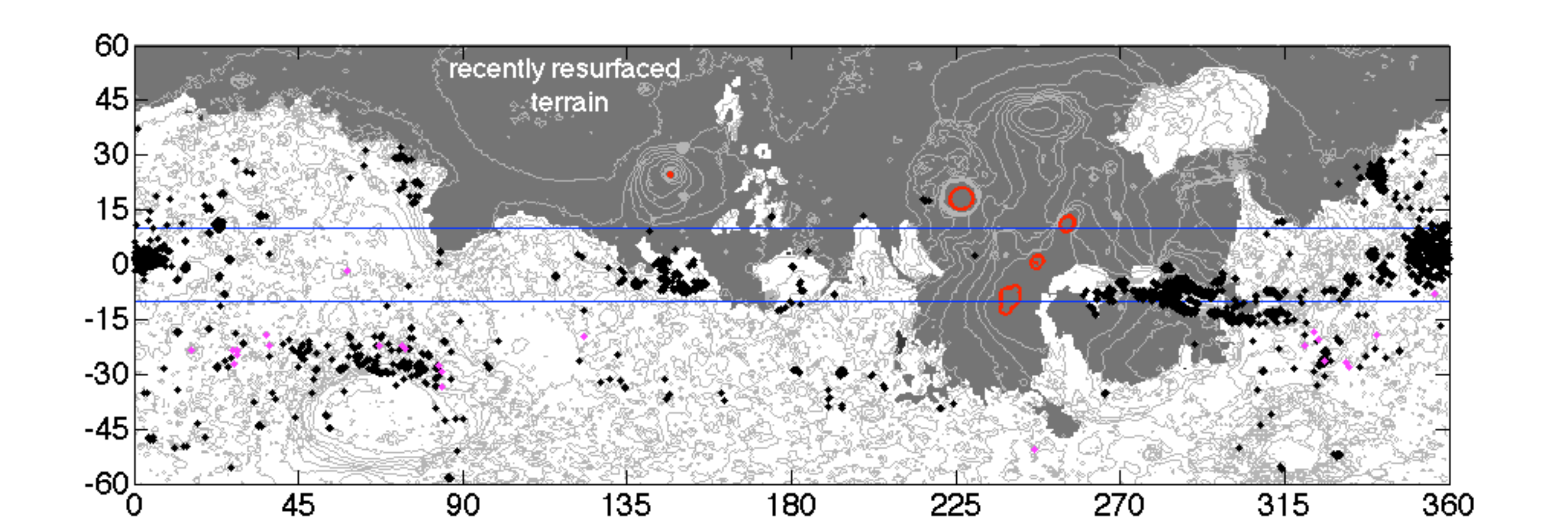}

\caption{\textbf{(a)} Latitudinal dependence of sedimentary rocks, masking out recently resurfaced terrain (Appendix \ref{masking}). Latitude bin size is 10$^\circ$. Histogram corresponds to number of images (left axis). Dark gray bars are the contribution from the Valles Marineris region, and black bars represent the rest of the planet. Lines correspond to the percentage of images showing sedimentary rock (right axis). Dashed line is the percentage of images showing sedimentary rocks once the Valles Marineris region is excluded. \textbf{(b)} Elevation dependence of sedimentary rocks, masking out recently resurfaced terrain. Elevation bin size is 500m. Gray line is normalized histogram of terrain with sedimentary rocks, and black line is histogram of all ancient terrain. Dotted gray line is the normalized histogram of terrain with sedimentary rocks, after masking out Valles Marineris. Median sedimentary rock elevation is $\sim$2km lower than median ancient terrain.  
(c) Distribution of sedimentary rocks on Mars (black dots, from \citet{malin2010}). Alluvial fans are also shown (purple dots, from \citet{kraal2008}). Blue horizontal lines highlight the $\pm$10$^\circ$ latitude band. Dark gray shading corresponds to recently resurfaced terrain. Light gray contours show topography, with the +10km contour shown in red.
\label{figuredata} \label{figure:sedrocksmap}}
\end{figure*}

\noindent The Mars Orbiter Camera Narrow Angle (MOC NA) team documented $\sim$4,000 ``layered rock outcrops of probable or likely sedimentary origin'' \citep{malin2000a,malin2001,malin2010}. Details of our analysis of these data are give in Appendix \ref{masking}. The resulting distribution of sedimentary rocks on Mars (Figure \ref{figuredata}) suggests that surface water availability was narrowly concentrated near the equator and at low elevations.  64\% of sedimentary rocks are within 10$^\circ$ of the equator, 60\% when the Valles Marineris region is excluded (Figure \ref{figuredata}a). Blanketing by young mantling deposits may contribute to the paucity of sedimentary rocks poleward of 35$^\circ$, but cannot explain the rarity of sedimentary rocks at 10-35$^\circ$ latitude relative to the equatorial belt. The $\pm$10$^\circ$ band is not unusual in thermal inertia, dust cover index, albedo, or surface age distribution, so a dependence of sedimentary rock on these parameters could not explain the latitudinal distribution. 

On average, sedimentary rocks are lower than ancient terrain by 2km (Figure \ref{figuredata}b). On Earth sedimentary rocks are low--lying because of sediment transport by regional-integrated channel networks. Evidence for regionally-integrated channel networks on Mars mostly predates the sedimentary rock era \citep{carr2010,fassett2011}. The low-elevation bias is independent of the equatorial concentration. Therefore, the low-elevation bias is reflective of a planetwide, non-fluvial process that occurs preferentially at low elevations.

Sedimentary rock abundance does not decline monotonically away from the equator. Abundance away from the equator is much less than in the equatorial sedimentary-rock belt, but ``wings'' of increased sedimentary rock abundance are found nearly symmetric about the equator at 25-30S and 20-30N.  The sedimentary rocks in the southern wing are regionally associated with clusters of large alluvial fans (Figure \ref{figuredata}d). 


%
%

\begin{figure*}[ht]
\includegraphics[width=1.0\textwidth, clip=true, trim = 10mm 5mm 0mm 10mm]{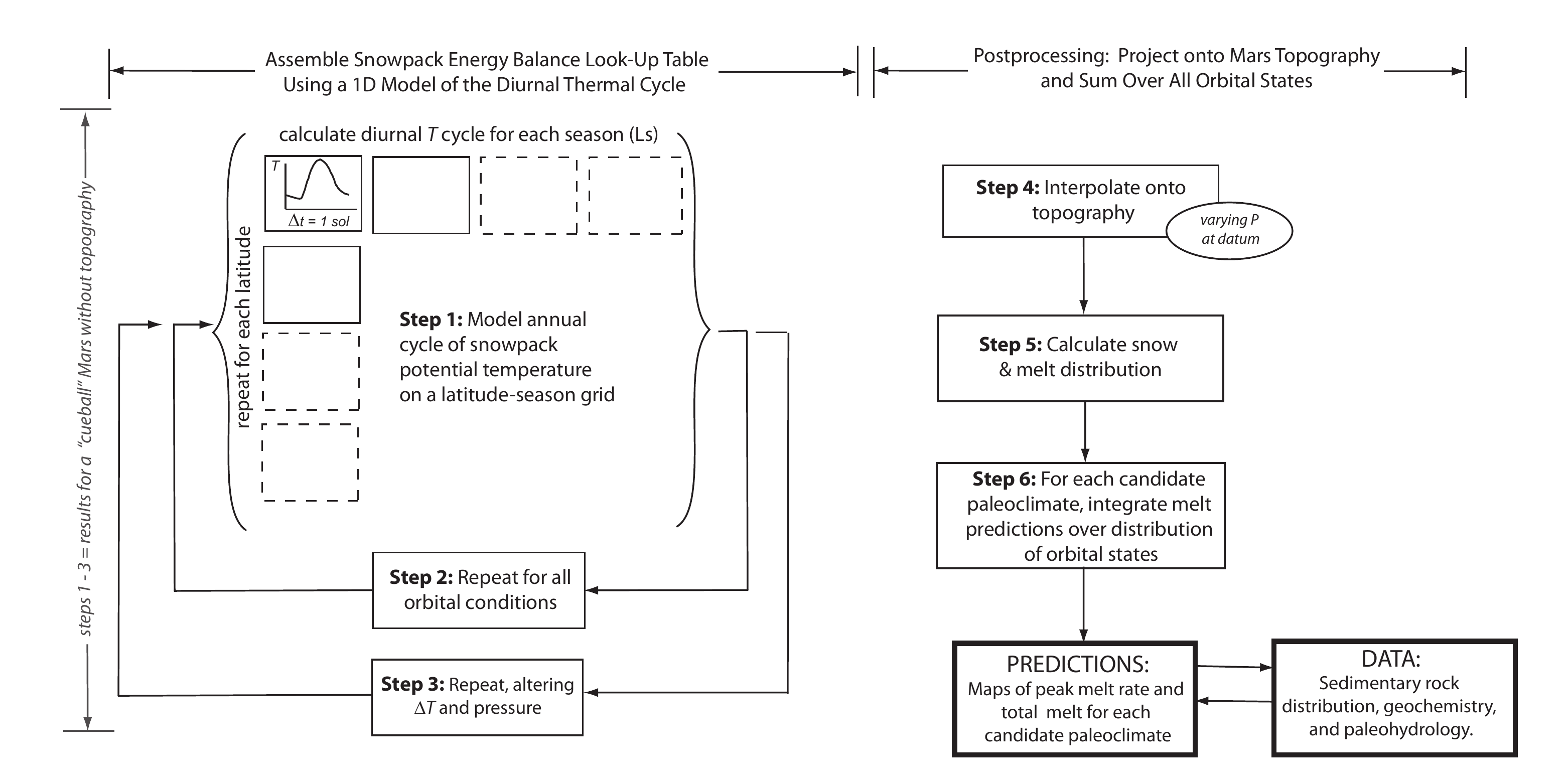}
\vspace{-0.3in}
\caption{Workflow of the Early Mars seasonal melting model. See text for details.}
\label{fig:intro2}
\end{figure*}

\vspace{-0.4in}

\section{Model\label{model}}

\noindent This section describes the seasonal melt model. \S4.1 describes the model framework and assumptions. \S4.2 describes the 1D snowpack thermal model. \S4.3 describes the potential-well approximation for warm-season snow locations, and \S4.4 explains how results from 1D models are combined to produce predictions. 

\subsection{Overview of model framework}
\noindent Controls on Mars snowmelt include:--

\vspace{0.1in}

\begin{itemize}
\item  \emph{Spin-orbit properties} \textbf{O} = \{$\phi$, $e$, $L_p$, $L_s$, latitude\}  control the distribution of sunlight at a given location. These include obliquity $\phi$, eccentricity $e$, solar longitude of perihelion $L_p$, solar longitude $L_s$, and latitude. Milankovitch parameters \textbf{O$^\prime$} = \{$\phi$, $e$, $L_p$\} oscillate or circulate on 10$^{4-6}$ yr frequencies, and $\phi$ shows chaotic shifts at $\sim$250 Myr intervals \citep{head2011}. We iterate over all the spin-orbit properties that have probably been encountered by Mars over the last 3.5 Ga (Steps 1--2 in Figure \ref{fig:intro2}). 
\item \emph{Climate parameters} \textbf{C} =  \{$P$, $\Delta T$, $f_{snow}$\} include atmospheric pressure $P$ (assumed to be mostly CO$_2$),  freezing-point depression/non-CO$_2$ greenhouse forcing $\Delta T$, and snow coverage fraction $f_{snow}$. These are iterated over a large range (Step 3 in Figure \ref{fig:intro2}).
They are assumed to vary slowly relative to changes in spin-orbit properties.
\item  \emph{Surface material properties, insolation}. These are held fixed for the ensemble of model runs. Results are sensitive to snowpack TI and albedo. Parameter choices and sensitivity tests are discussed in Appendix \ref{material}. 

\end{itemize}





\noindent The predictions of the snowmelt hypothesis (\S\ref{hypothesis}.6) are evaluated as follows (Figure \ref{fig:intro2}). First, for a given trial set of past climate parameters \textbf{C} and orbital parameters \textbf{O}, snow temperature for all seasons and latitudes is calculated using a 1D surface energy balance model (Step 1 in Figure \ref{fig:intro2}). Using the 1D model output for a range of $P$ (Step 3), the potential annual-average snow sublimation rates are mapped onto topography (Step 4). Warm-season snow is assigned to locations most favorable for interanually persistent snow (Figure \ref{figure:cueball}). Snowmelt occurs when temperatures at these locations exceed freezing (Step 5). These results provide latitude-season diagrams that are appropriate for a flat Mars. Results for this fictitious ``cueball'' Mars are analyzed in \S\ref{results}.3. To map the results onto real Mars topography, a sequence of latitude-season output for different $P$ but the same \textbf{O$^\prime$} is stacked in elevation (results given in \S\ref{results}.4). The output at this stage consists of maps of snow stability for the given \textbf{C} and \textbf{O$^\prime$}, along with time series of snow temperature and melt rates. 
Next, the framework loops over all possible Early Mars \textbf{O} (Step 6), convolving the outputs with the \textbf{O$^\prime$} probability distribution function \citep{laskar2004}. The output is now a map of predicted snowmelt on Mars integrated over geologic time, and this can be compared to observed sedimentary rock abundance and thickness data (\S\ref{results}.5). 
These maps are computed for many plausible \textbf{C}. Assuming the snowmelt model is correct, the \textbf{C} that produces the best correlation between model predictions and data is the best-fit Early Mars climate (\S\ref{distribution}.1). Interpolating in the melt rate output gives a predicted time series at Gale Crater (\S\ref{formation}).
%

\subsection{Thermal model}

\begin{figure}[h]
\includegraphics[width=1.0\textwidth, clip=true, trim = 25mm 145mm 85mm 55mm]{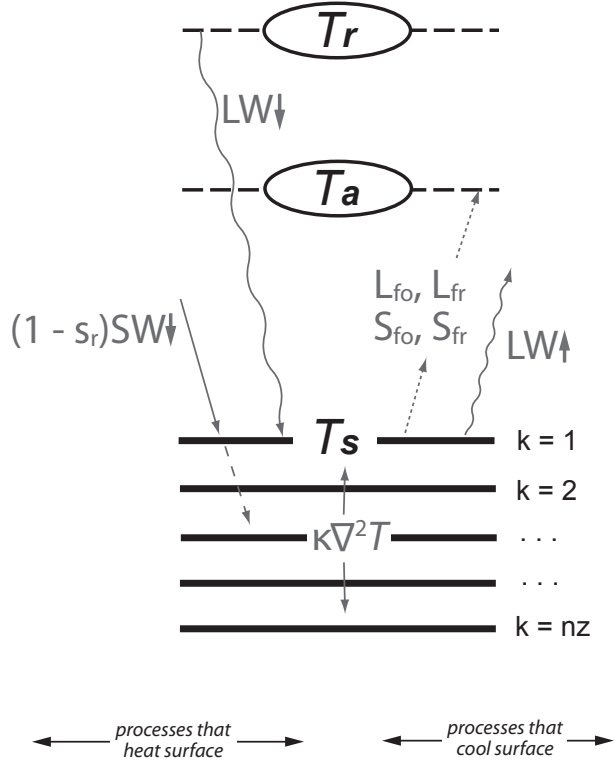}
\caption{Vertical discretization and energy flow in the 1D model. Solid horizontal lines correspond to solid surface layers numbered K = \{1... nz\}, dashed horizontal lines correspond to atmospheric layers. $T_r$ is the effective atmospheric radiative temperature, $T_a$ is the atmospheric surface layer temperature, and $T_s$ is the ground surface temperature. The diagonal arrows correspond to energy fluxes: $LW\!\downarrow$ for greenhouse effect, $(1 - s_r) SW\!\downarrow$ for insolation attenuated by Rayleigh scattering, $LW\!\uparrow$ for backradiation, and \{$L_{fo}, L_{fr}, S_{fo}$ and $S_{fr}$\} for the turbulent fluxes. Some insolation penetrates into the snowpack (dashed continuation of insolation arrow). $\kappa \nabla^2 T$ corresponds to conductive diffusion.\label{SnowmeltPaperFigure1}}
\label{fig:intro1}
\end{figure}

\noindent Surface liquid water is always unstable to evaporation on a desert planet \citep{richardson2008, soto2008}. However, transient liquid water can occur metastably if temperatures exceed the freezing point, and if $P$ exceeds the triple point (in order to prevent internal boiling) \citep{hecht2002}. 

These conditions are modeled using a 1D thermal model (Figure \ref{SnowmeltPaperFigure1}). When temperature exceeds (273.15K - $\Delta T$), melting occurs and buffers the temperature at the melting point. Within the snowpack, material properties are assumed uniform with depth and heat flow is by conduction and solar absorption only. When melt is not present, energy balance in the subsurface layer adjacent to the surface is given for unit surface area by (Figure \ref{fig:intro1}) 




\begin{align}
 \rho C_s \Delta z \frac{\partial T_1}{\partial t} &=& \frac{k}{\Delta z} \frac{\partial^2 T}{\partial z^2}  - \underbrace{\epsilon \sigma T^4 + LW\!\!\downarrow + (1 - s_r) Q(1) SW\!\!\downarrow}_{radiative\,\, terms}  \nonumber \\
 & & - \underbrace{S_{fr}  - L_{fr}}_{free \,\,convection} - \underbrace{S_{fo} - L_{fo}}_{forced\,\,convection}
\end{align}

\noindent and energy balance at depth $z$ within the snowpack is given by (Figure \ref{fig:intro1}) 

\begin{equation}
\begin{centering}
 \rho C_s \Delta z \frac{\partial T_K}{\partial t} = \frac{k}{\Delta z} \frac{\partial^2 T}{\partial z^2}  + \underbrace{(1 - s_r) Q(z) SW\!\!\downarrow}_{solid-state\,\, greenhouse} 
 \end{centering}
 \end{equation}


\noindent Here, $\rho$ is snow density, $C_s$ is snow specific heat capacity, $\Delta z$ is the thickness of the subsurface layer whose upper boundary is the surface,  $T$ is the temperature at subsurface level $K$ = \{1, 2, ... , n$_z$\}(Figure \ref{fig:intro1}), $k$ is snow thermal conductivity, $\epsilon$ is the longwave emissivity of ice, $LW\!\!\downarrow$ is downwelling longwave radiation, $s_r$ is the Rayleigh-scattering correction factor, $Q_{\{1,2,..., \mathrm{n_z}\}}$ is the fraction of sunlight absorbed at level $z$ (Appendix \ref{solidstate}),
$SW\!\!\downarrow$ is solar flux per unit surface area, $S_{fr}$ corresponds to free sensible heat losses driven by atmosphere-surface temperature differences, $L_{fr}$ corresponds to evaporative cooling by free convection when the atmosphere has relative humidity $<$1, $S_{fo}$ corresponds to forced sensible heat losses caused by cool breezes over warm ground, and $L_{fo}$ corresponds to additional evaporative cooling when the wind is nonzero. Snow albedo, $\alpha$, is 1 - $\displaystyle\int Q_z \mathrm\,\,{d}z$. All results presented in this paper are for the 3.5 Gya solar luminosity reported by \citet{bahcall2001}, 23\% below present.

The 1D model draws on previous work by \citet{toon1980,clow1987,richardson2005,williams2008snowpack,dundas2010}; and \citet{liston2005}. A simpler version is discussed in \citet{kite2011a,kite2011b}.  
Representative output is shown in Figure \ref{FigureThermal}. 
Details of the flux parameterizations, melt handling, and run conditions are given in Appendix \ref{details}. 

\begin{figure}[h]
\includegraphics[width=1.0\textwidth, clip=true,trim=22mm 3mm 25mm 5mm]{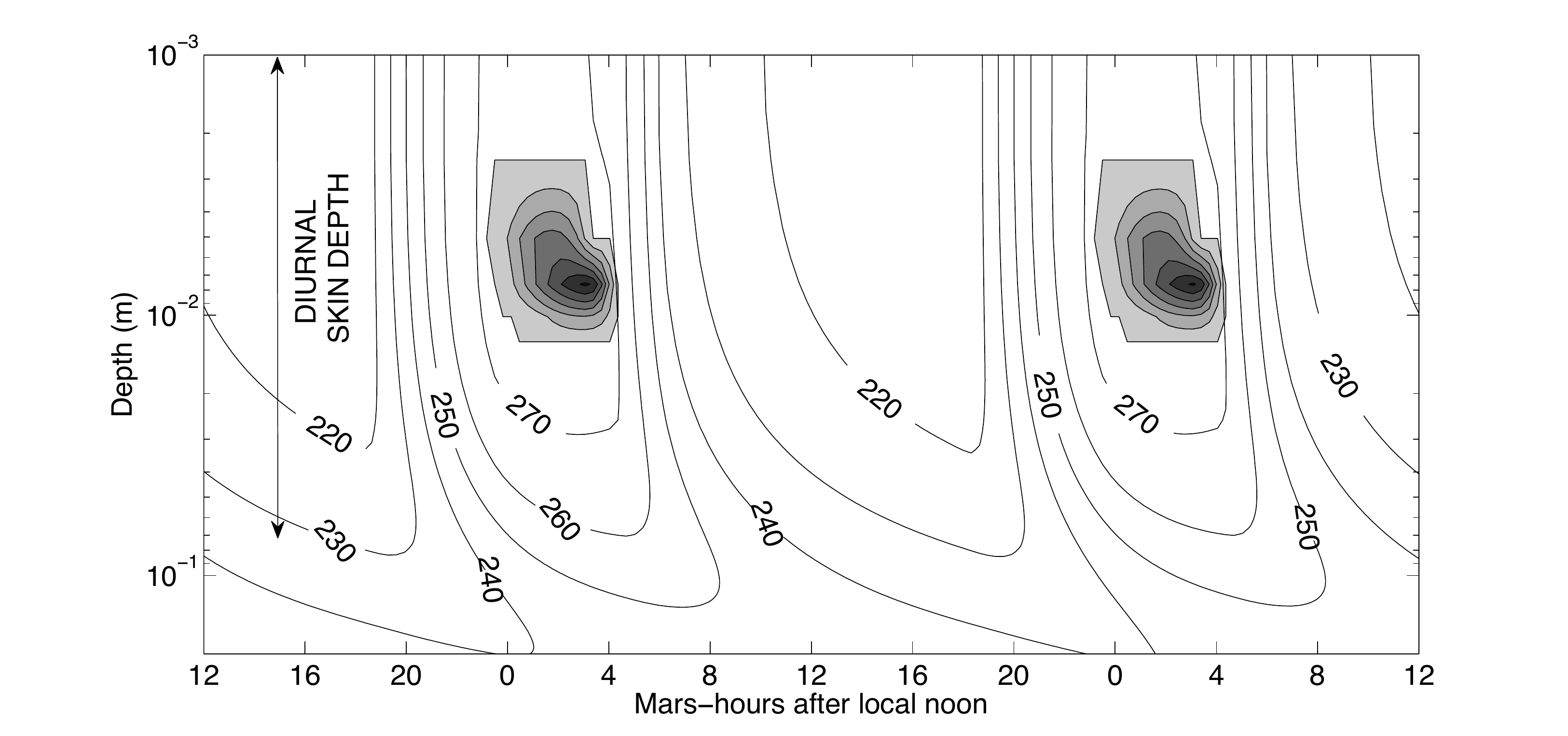}
\caption{Daily cycle of temperature and melting in the upper 15cm of snowpack. Grayscale corresponds to melt fraction at a constant temperature of 273.15K, with contours of 0.1 (edge of gray region corresponds to zero melt). Temperature contours are shown wherever the snow is not melting, at intervals of 10K. The blocky outline of the ``melt = 0'' contour corresponds to the 1200s resolution at which model output is stored for subsequent processing: the underlying timestep is 12s. $L_s$ = 0$^\circ$, $L_p$ = 0$^\circ$, $\phi$ = 50$^\circ$.  \label{FigureThermal} }
\end{figure}

\subsection{Snow location prescription: the potential-well approximation}

\noindent For $\sim$100 mbar CO$_2$, the model predicts melt only during the warmest season, and usually within a diurnal skin depth of the surface. Experiments, observation and theory agree that warm-season snow and ice within this depth range on Mars is in cold-trap equilibrium with orbital forcing \citep{mellon1995,hudson2008,hudson2009,schorghofer2005,boynton2002}. Because we are interested in above-freezing $T$ only where snow is present, the quantity of interest is the annual-maximum $T$ experienced by the cold traps, whose location depends on orbital conditions and topography. For most orbital conditions, this $T$ is below freezing, so the greatest interest is in the orbital conditions that maximize the cold-traps' annual maximum $T$. 

\begin{figure*}[h]
\begin{center}$
\begin{array}{cc}
\includegraphics[width=3.0in, clip=true, trim = 0mm 0mm 0mm 0mm]{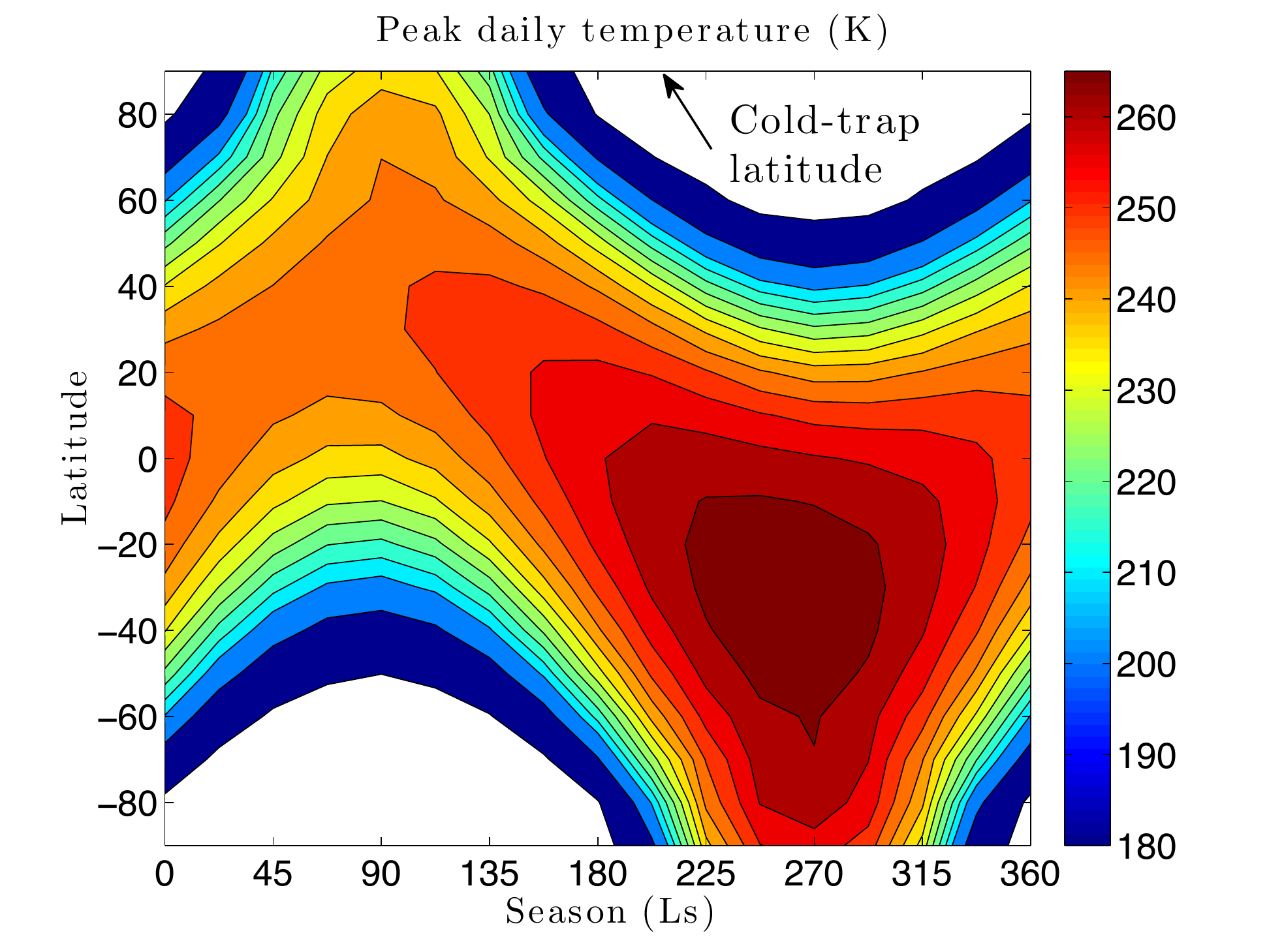} &
\includegraphics[width=3.0in, clip=true, trim = 0mm 0mm 0mm 0mm]{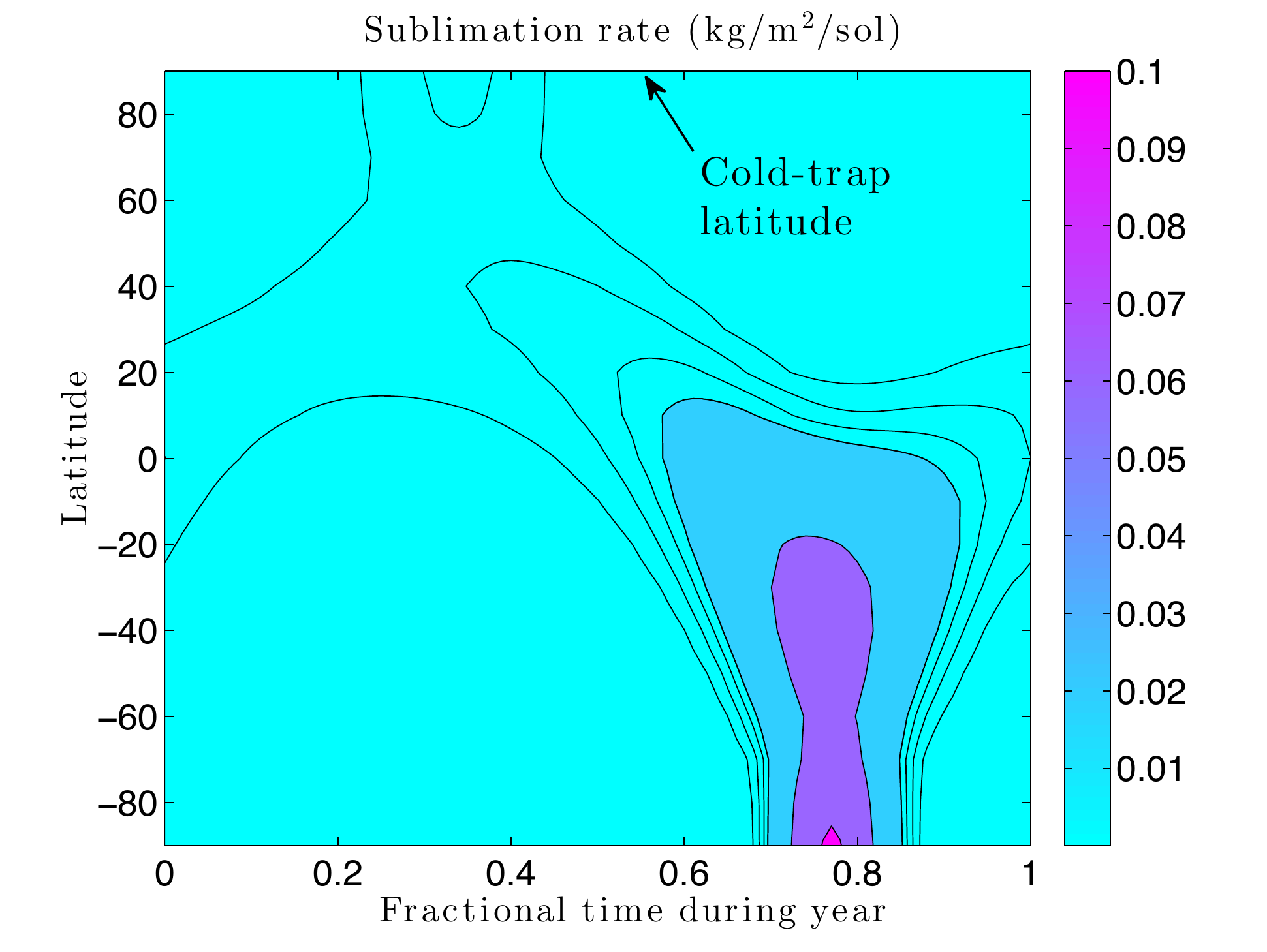} \\ \includegraphics[width=3.0in, clip=true, trim = 0mm 0mm 0mm 0mm]{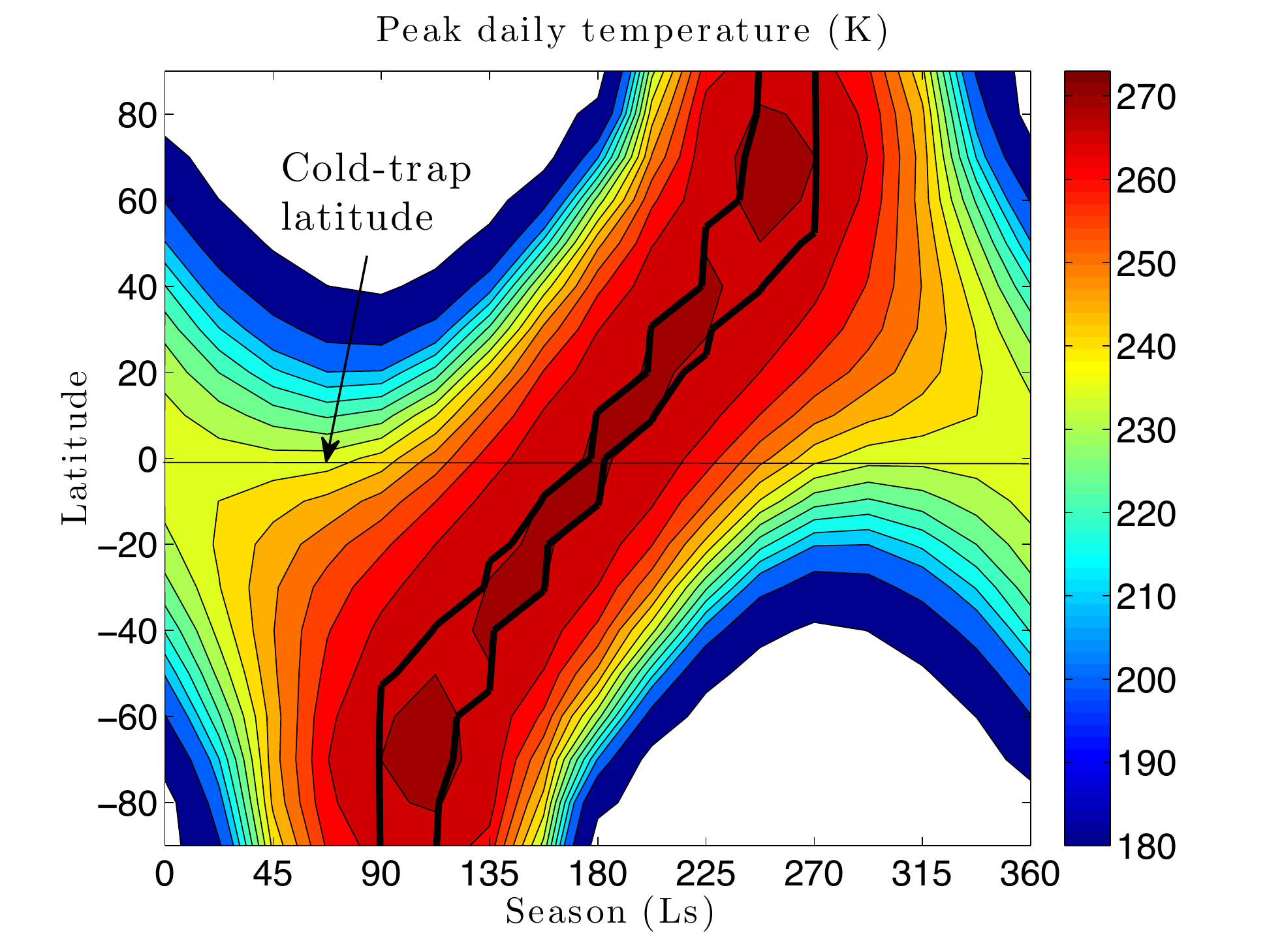} &
\includegraphics[width=3.0in, clip=true, trim = 0mm 0mm 0mm 0mm]{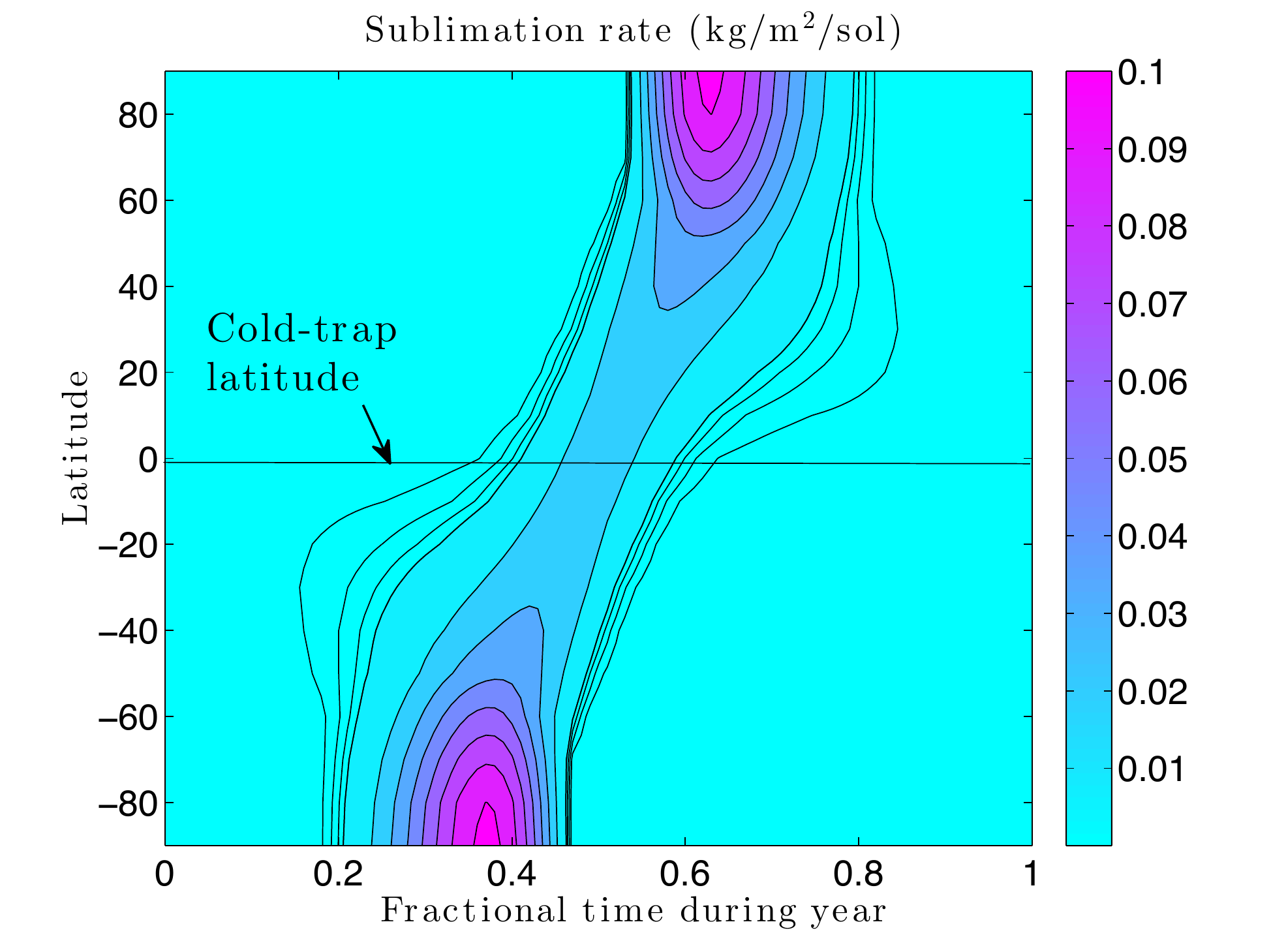}
\end{array}$
\end{center}
\caption{Seasonal cycle of diurnal--peak temperature and diurnal--mean free sublimation rate for 3.5 Gya insolation, assuming flat topography. 145 mbar pure CO$_2$ atmosphere. \textbf{(a,b)} present-day orbital forcing ($\phi$ = 25.2$^\circ$, $e$ = 0.093, $L_p$ = 251$^\circ$); \textbf{(c,d)} optimal conditions for melting -- high-$\phi$, moderate $e$, and $L_p$ aligned with equinox. Contours of daily maximum surface temperature are drawn at 180K, 200K, and 210K and then at intervals of 5K up to a maximum of 270K, only reached in \textbf{(c)}. White shading corresponds to CO$_2$ condensation at the surface. Sublimation--rate contours are drawn intervals of 0.025 kg/m$^2$/sol from 0 to 0.1 kg/m$^2$/sol and then at intervals of 0.2 kg/m$^2$/sol. At low $e$ and low $\phi$ (similar to today, \textbf{(a,b)}), ice is stable at the poles, where temperatures never exceed
freezing. In \textbf{(c,d)}, ice is most stable at the equator, and annual peak temperature exceeds freezing everywhere at some point during the year. The thick black line in \textbf{(c)} corresponds to subsurface melting at some point during the day. (No melting is predicted for modern orbital conditions.) The blockiness of this line corresponds to the underlying seasonal resolution (22.5$^\circ$ in $L_s$). Solid-state greenhouse effect warms the subsurface relative to the surface by up to several K. \label{figure:seasonal}}
\end{figure*}

\begin{figure*}
\begin{center}$
\vspace{0.3in}
\begin{array}{cc}
\includegraphics[width=0.55\textwidth, clip=true, trim = 13mm 3mm 6mm 6mm]{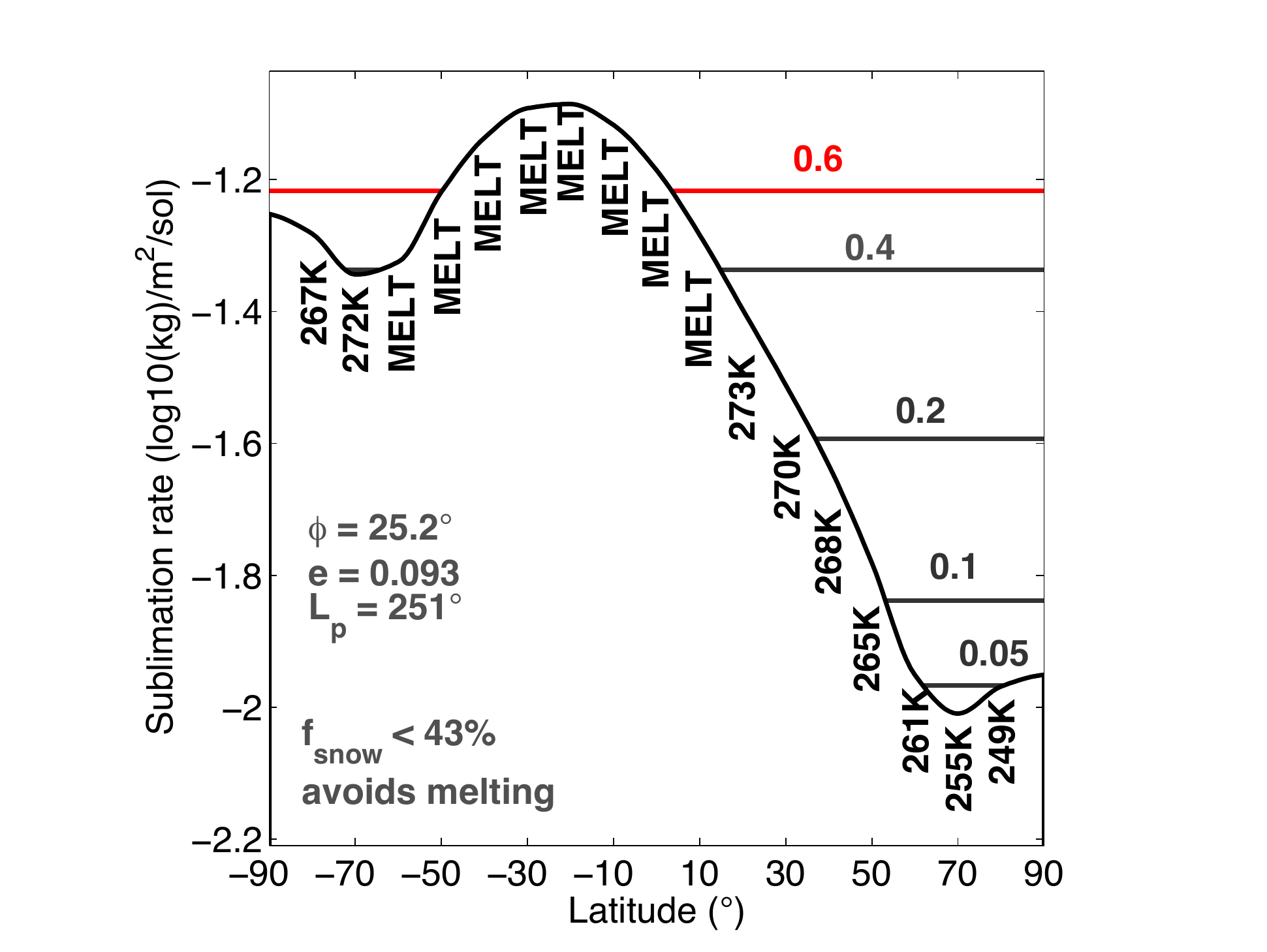} &
\includegraphics[width=0.55\textwidth, clip=true, trim = 13mm 3mm 6mm 6mm]{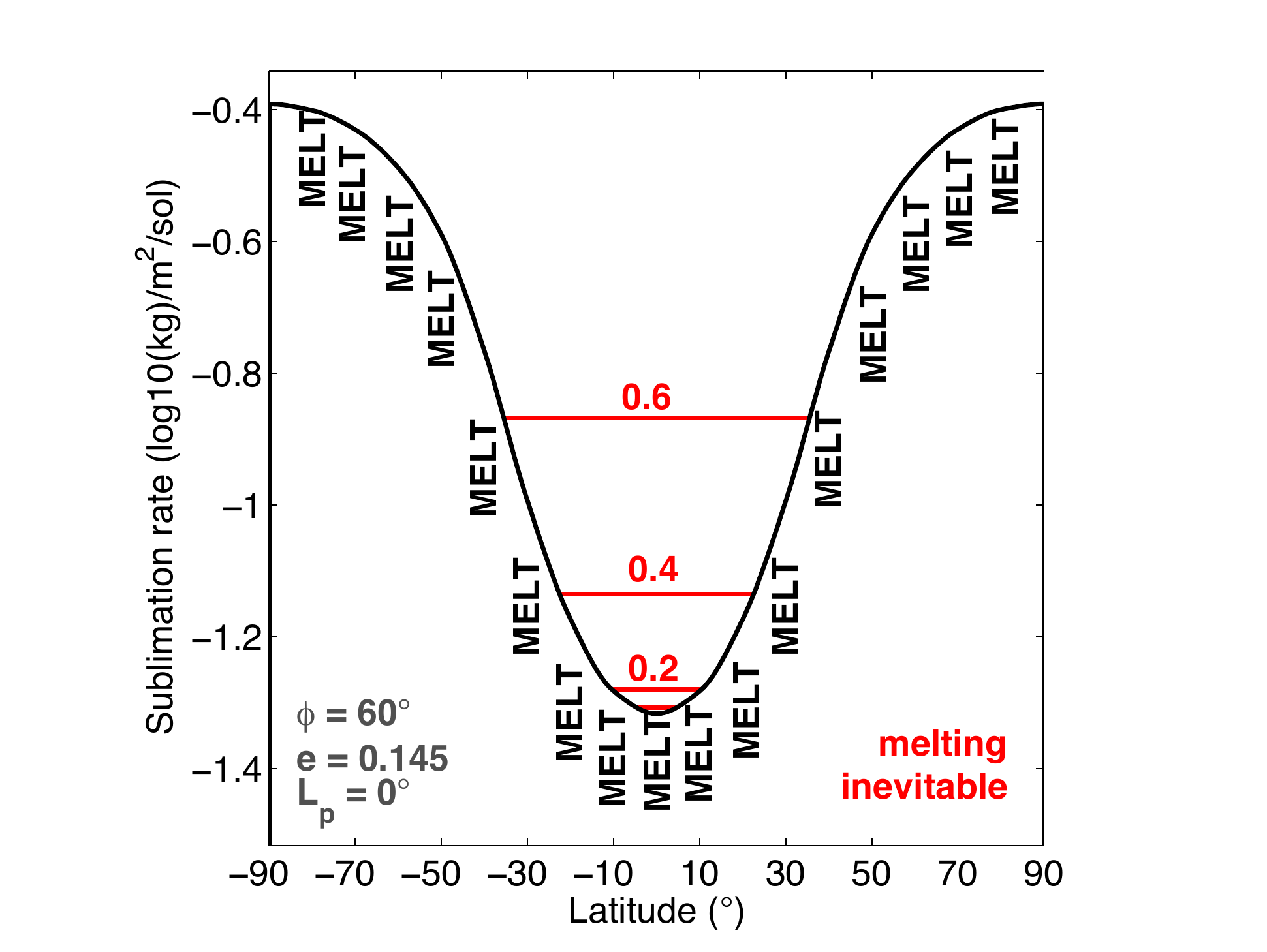} \\
\includegraphics[width=0.55\textwidth, clip=true, trim = 13mm 3mm 6mm 6mm]{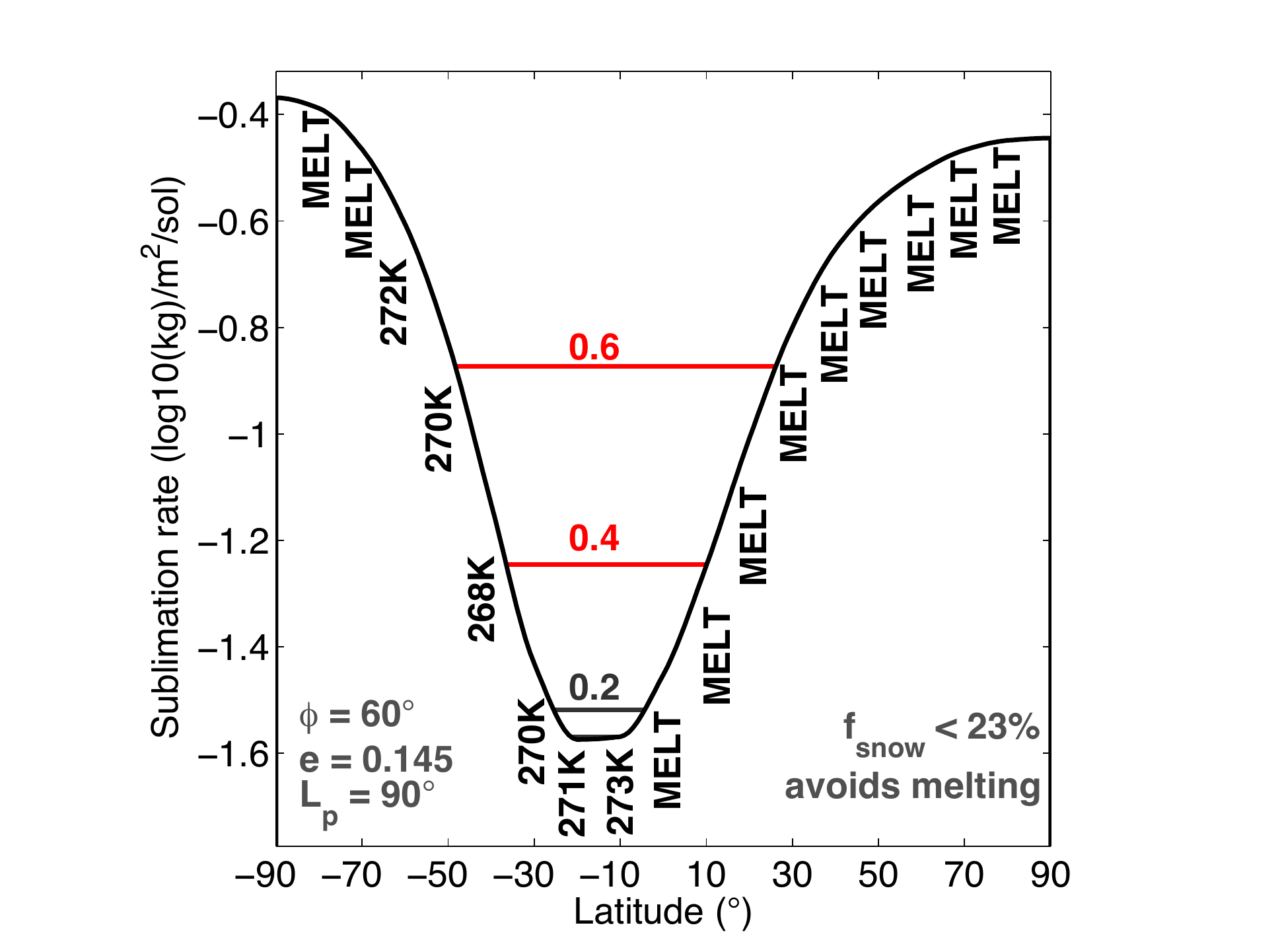}
\end{array}$
\end{center}
\caption{To illustrate the potential-well approximation for finding warm-season snow locations. Nominal parameters (Table 1), 146 mbar atmosphere, flat topography. Curve corresponds to potential sublimation during a year. Temperatures are annual maxima. ``MELT'' denotes melting of snowpack at some point during the year, if it exists at that latitude. Horizontal lines correspond to $f_{snow}$ values, assuming warm-season snow is found at locations that minimize annually-averaged sublimation, and colored red for values of $f_{snow}$ that lead to melting.
(a) Current orbital conditions. 
Massive ice or buried ice may exist in the southern hemisphere,  but snowpack that persists through the warm-season is only likely in the far north, where temperatures are always below freezing.
(b) Optimal orbital conditions. Melting occurs at all latitudes so melting is inevitable. (c) As for optimal orbital conditions, but for perihelion aligned with northern summer solstice. The short, intense northern summer displaces the potential-sublimation minimum to 20S. $f_{snow}$ $>$ 23\% is needed for melting to occur under these circumstances. The latitude of first melting will be near the equator. Note that `273K' is just below the melting point. }
\label{fig:potentialwell}
\end{figure*}

In the case of no topography (cueball planet), the location of the cold traps depends on orbital conditions only. For this case we find the location of cold traps in the following manner: for each \textbf{O$^\prime$}, the output of the thermal model for all seasons ($L_s$) and geographic locations \textbf{x} is used to determine the \textbf{x} where snow is most likely to be present during the melt season. Melt-season snow is assumed to be only found at locations where the annually-averaged sublimation is minimized (Figure \ref{figure:seasonal}).
Converged thermal-model output for $L_s$ is linearly interpolated in time using Kepler's equation (Figure \ref{figure:seasonal}). All \textbf{x} are then assigned an area-weighted rank, $f$, scaled from 0\% (global minimum in annually-averaged sublimation; most favorable for snow accumulation) to 100\% (global maximum in annually-averaged sublimation; least favorable to snow accumulation). Ice lost by melting is assumed to be recovered by refreezing close ($<$100km) to source. Warm-season snow is assumed not to occur above a critical $f$, termed $f_{snow}$ (the percentage of the planet's surface area that has warm-season snow). Using the $f$(\textbf{x}) and $f_{snow}$, warm-season snow is assigned to favored geographic locations (Figure \ref{fig:potentialwell}). In general,  the critical $f$ for warm-season snow will be greater than the critical $f$ for interannually-persistent snow, so melting does not have to be supraglacial. The most favorable circumstances for aqueous alteration may be where melt occurs during a seasonal accumulation-ablation cycle which leaves bare soil during part of the year. 

Melting is almost certain to occur when orbital forcing leads to annual-maximum temperatures above freezing at all latitudes (Figure \ref{fig:potentialwell}). Thermal barriers $>$10 cm thick can insulate snow against diurnal melting, but a sublimation lag covering all ice is logically impossible, and a debris lag covering all ice is unlikely. The albedo of pure, fresh, fine-grained snow is high enough to prevent melting, but contamination with dust is very likely (Appendix \ref{solidstate}). Twice-yearly transfer of the water ice reservoir across the equator to the cold high-obliquity winter pole would require unreasonably high seasonally reversing mean wind speeds.

With MOLA topography, annual-average sublimation rates are calculated as for the cueball case, but now for a range of $P$ that spans mountaintop pressures and canyon pressures. 
The latitude-$P$ grid is then interpolated onto latitude and longitude using MOLA topography (Figure \ref{figure:ralltall}). This assumes that Mars' long-wavelength bedrock topography was in place before the sedimentary rock era, in agreement with geodynamic analysis \citep[e.g.,][]{phillips2001}. This also neglects the effect of the adiabatic lapse rate on surface $T$, which is an acceptable approximation at the $P$ of interest here (\S\ref{discussion}.1).


\subsection{Integrating over all orbital states}

\noindent For each \textbf{O$^\prime$}, $f$ is obtained for each \textbf{x}, together with the potential snowpack temperatures and potential melt rates. 
Then, $f_{snow}$ maps out the snow distribution, and $f_{snow}$ and $\Delta T$ map out the melt distribution. The resulting maps of snow and snowmelt are weighted using the  \textbf{O$^\prime$} probability distribution function \citep{laskar2004}, and the weighted maps are summed. (Note that this is not the same as the median melt column from a large ensemble of solar system integrations; \citet{laskar2004}.) Melt likelihood is then given by

\begin{equation}
\hspace{-0.1in}{M_\textbf{x} \! = \!\! \int \! \left(T_{max,\textbf{x}} > (273.15 - \Delta T) \right) \left( f_{snow} > f_\textbf{x} \right) \mathrm{p}( \textbf{O$^\prime$}) \, \mathrm{d}  \textbf{O$^\prime$}}
\end{equation}

\noindent where the ``greater than'' operator returns 1 if true and 0 if false, and p() is probability.

\newpage
\clearpage

\begin{center}
 \begin{table}
{\scriptsize
\renewcommand{\arraystretch}{0.85}
\begin{tabular}{l l @{\hspace{-1.5cm}} r l l}
\textbf{Symbol} & \textbf{Parameter} & \textbf{Value and units} & \textbf{Source / rationale} \\ [0.5ex]
\hline
\\
Fixed parameters:\\
$A_{vonk}$  & von Karman's constant & 0.4   &    \\
$C_p$  &Specific heat, Mars air & 770  J/kg/K &  \\
$C_s$ & Specific heat capacity of snow & 1751  J/kg/K & \citet{carr2003a}  \\
$D_{air}$  & Mechanical diffusivity of air & 14 $\times$ 10$^{-4}$ m$^2$/s &  \citet{hecht2002}  \\
$g$  & Mars surface gravity & 3.7 m/s$^2$ &    \\
$k_b$  & Boltzmann's constant & 1.38 $\times$ 10$^{-23}$  m$^2$ kg s$^{-2}$ &    \\ [1ex]
$k_{snow}$ & Thermal conductivity of snowpack & 0.125   W/m/K & \citet{carr2003a} \\
$m_c$  & Molar mass of CO$_2$ & 0.044  kg &   \\
$m_w$  & Molar mass of H$_2$O & 0.018  kg &  \\
$M_w$  & Molecular mass of H$_2$O & 2.99 x 10$^{26}$  kg &    \\
$P_{atm,0}$  & Pressure of atmosphere now & 610  Pa & NSSDC \\
$L_{e}$  & Latent heat of water ice sublimation & 2.83 $\times$ 10$^6$  J/kg & \citet{hecht2002} \\
$L_{e}$  & Latent heat of water ice melting & 3.34 $\times$ 10$^5$  J/kg & \citet{hecht2002} \\
$r_{h}$ & Relative humidity & 0.25  &   \\
$R_{gas}$  & Gas constant & 8.3144  J/(mol K) &    \\
$u_{s,ref}$  & Reference near-surface wind speed & 3.37   m/s & \citet{millour2008} ``MY24'' average \\
$z_o$  & Roughness length & 0.1  mm & Polar snow. \citet{brock2006} \\
$z_{anem}$  & Anemometer height & 5.53  m &  \citet{millour2008} \\
$\alpha$  & Albedo & $\approx$0.28 & Albedo of dusty snow. Appendix \ref{material}. \\
$\epsilon$ & Emissivity of ice at thermal wavelengths & 0.98  &  \\
$\nu_{air}$  & Kinematic viscosity of air & 6.93 $\times$ 10$^{-4}$   m$^2$/s & \citet{hecht2002}  \\
$\rho$ & Density of snowpack & 350  kg m$^{-3}$ & \citet{carr2003a}  \\
$\rho_{0}$  & Density of atmosphere (now) & 0.02  kg m$^{-3}$ &NSSDC \\
$\sigma$  & Stefan-Boltzman constant & 5.67 x 10$^{-8}$  W m$^{-2}$ K$^{-4}$ &    \\ 
$\tau$ & Time of interest & 3.5   Gyr ago & \citet{murchie2009a}\\
& Mars semimajor axis & 1.52366 AU & NSSDC \\
 & Duration of 1 Mars sol & 88775  s &   \\
& Dust concentration & $\sim$2\% by volume & Appendix \ref{solidstate}  \\
& Dust radius & 4 $\mu$m & Appendix  \ref{solidstate} \\
& Ice grain radius & 1 mm  & Appendix \ref{solidstate}  \\
 & Solar constant (now) & 1.361 $\times$ 10$^{3}$ W/m$^2$ & \citet{kopp2011}\\
\hline
\\
Selected variables:\\
\textbf{O} & Spin-orbit properties\\
\textbf{O$^\prime$} & Milankovitch parameters \\
\textbf{C} & Climate parameters\\
$\phi$ & Obliquity & 0-80$^\circ$\\
$b_{DB}$ & Dundas-Byrne ``b'' & f($P$) (Appendix \ref{details}) & Extrapolation from GCM runs  \\ 
$e$ & Eccentricity & 0.00 -- 0.16 \\
$L_p$ & (Solar) longitude of perihelion & 0 -- 360$^\circ$ \\
$L_s$ & Solar longitude & 0 -- 360$^\circ$ \\
$M$ & Mean anomaly & 0 -- 360$^\circ$  \\
$P$ & Atmospheric pressure & 24-293 mbar \\
$P_o$ & Atmospheric pressure at zero elevation & 24-293 mbar \\
$\Delta T$ & Non-CO$_2$ greenhouse forcing & 0-15 K \\
$f_{snow}$ & Fraction of planet surface area with warm-season snow & 0-50\% \\
$Q_k$ & Fraction of incident sunlight absorbed at level $k$ & 0-100\% \\
$LW\!\!\downarrow$ & Greenhouse forcing \\
$LW\!\!\uparrow$ & Thermal emission by surface \\
$SW\!\!\downarrow$ & Insolation \\
$s_r$ & Rayleigh scattering correction factor \\
$L_{fo}$ & Latent heat losses by forced convection \\
$L_{fr}$ & Latent heat losses by free convection \\
$S_{fo}$ & Sensible heat lost by forced convection\\
$S_{fr}$ & Sensible heat lost by free convection \\
\hline
\end{tabular}}
\caption{Selected parameters and variables.}
\label{table:nonlin}
\end{table}
\end{center}

%
%



\clearpage


\section{Results and analysis\label{results}}

\subsection{Controls on the occurrence of near-surface liquid water on Early Mars}

%

\begin{figure}
\includegraphics[width=0.8\textwidth, clip=true,trim=20mm 3mm 5mm 10mm]{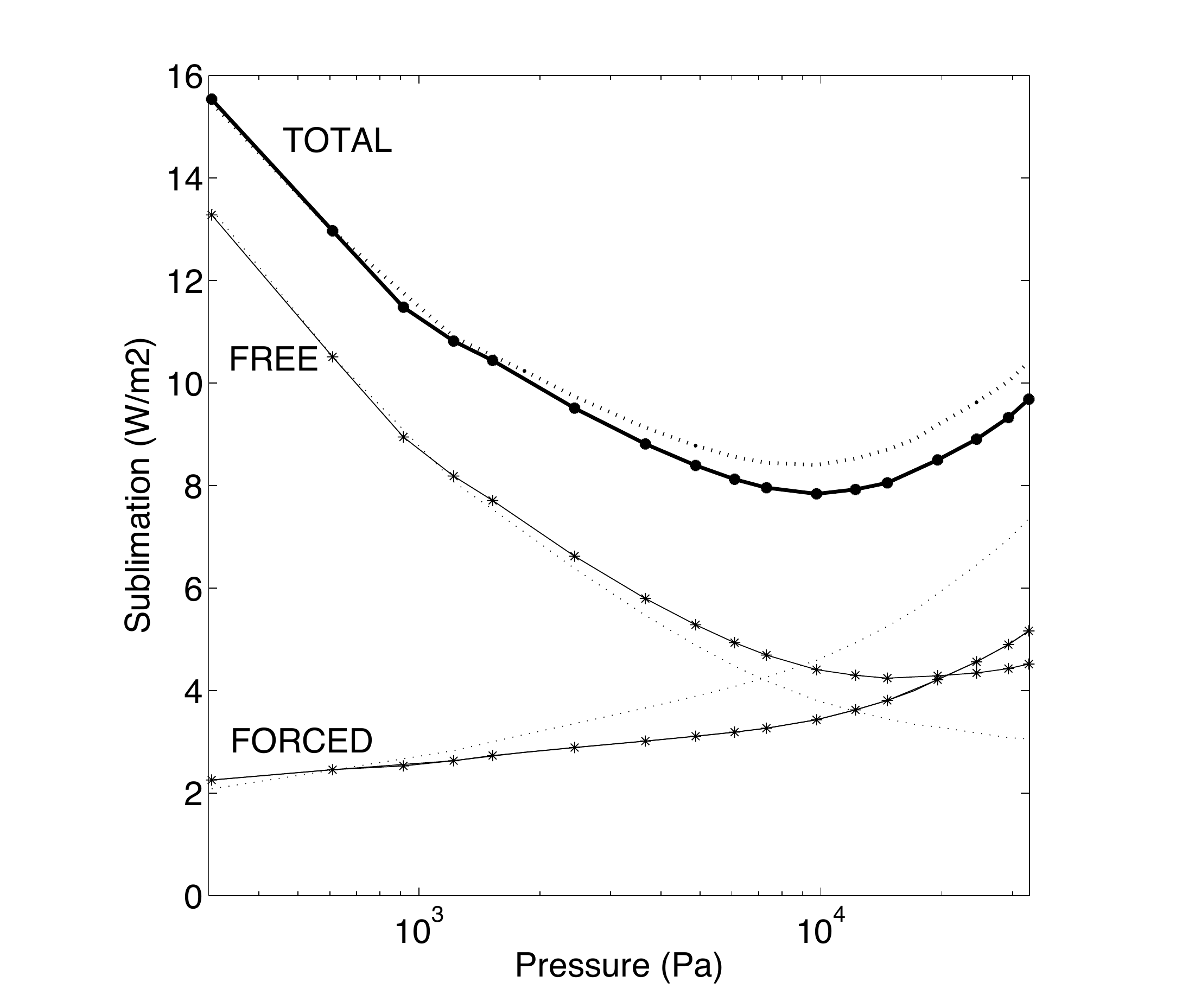}
\caption{Pressure dependence of sublimation rate at an equatorial site. FREE is $L_{fr}$, FORCED is $L_{fo}$. Solid lines with asterisks correspond to a wind speed that declines with increasing $P$, dotted lines correspond to constant near-surface wind speed of 3.37 m/s. $e$=0.11, $\alpha\approx$0.28, $L_p$=0, $L_s$=0, $\phi$=50. \label{figure:pressure}}
\end{figure}

Warm-season snow locations depend on sublimation rates. Diurnal-mean sublimation is shown as a function of $P$ in Figure \ref{figure:pressure}. 
Losses due to free convection decrease with increasing $P$, because the greater atmospheric density dilutes the buoyancy of moist air (Appendix \ref{details}). Losses due to forced convection increase with atmospheric density (Appendix \ref{details}). Surface temperature increases monotonically with increasing greenhouse forcing, leading to an uptick in sublimation rate for $P >$ 100 mbar. Snow is most stable against sublimation when $P$ $\sim$ 100 mbar. Therefore, when the atmospheric pressure at zero elevation ($P_o$) $\ll$ 100 mbar, snow is most stable in topographic lows \citep{fastook2008}. When $P_o$ $\gg$ 100 mbar, snow is most stable on mountaintops. 

\begin{figure}
\includegraphics[width=0.80\textwidth, clip=true,trim=20mm 3mm 5mm 10mm]{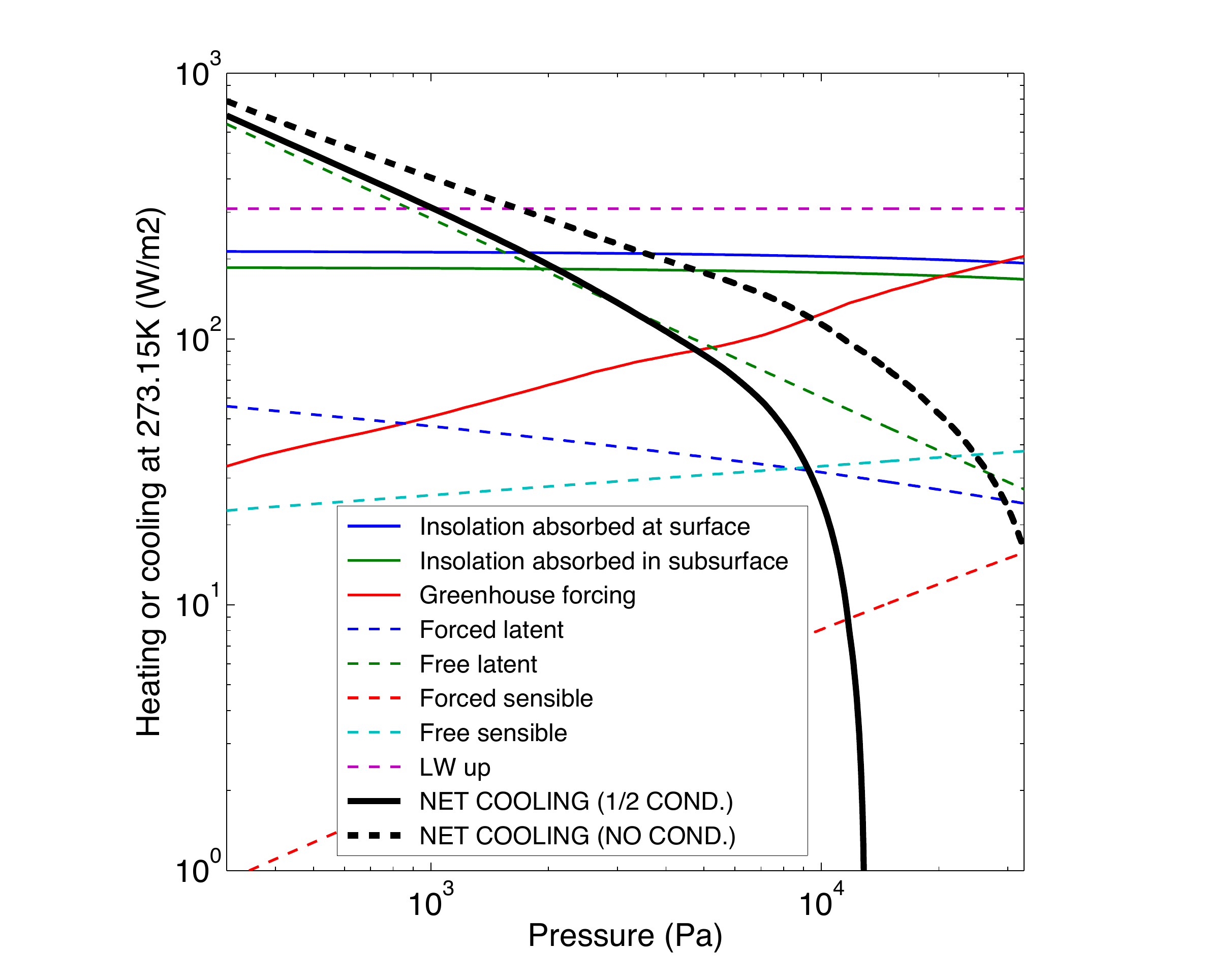}
\caption{Pressure dependence of the surface energy balance for an equatorial site with $T_1$ = 273.15K imposed. Wind speed declines with increasing $P$. Note that subsurface melting can occur for $T_{surf}$ $<$ 273.15K. The absorbed insolation terms decrease slightly at high $P$ due to Rayleigh scattering. Greenhouse forcing is stronger than in the time-dependent case shown in Figure \ref{FigureThermal} because the atmospheric temperature is assumed to be in equilibrium with a surface at the freezing point. If $\nicefrac{1}{2}$ of the subsurface absorbed insolation returns to the surface through conductive heating, then melting will occur (net cooling at the melting point will be zero) for $P$ $\gtrsim$ 130 mbar. However, if none of the subsurface absorbed insolation conductively warms the surface, then surface melting will not occur even for the highest $P$ shown ($\sim$ 330 mbar). $S_{fo}$ and $L_{fo}$ are weak because of the low surface roughness.  $e$=0.15, $\alpha$$\approx$0.28,$L_p$=0$^\circ$,$L_s$=0$^\circ$, $\phi$=50$^\circ$. \label{termbyterm}}
\end{figure}

Melting and runoff depend on energy fluxes around the hottest part of the day. Figure \ref{termbyterm} shows the terms in the energy balance for a snow surface artificially initialized just below the freezing point. At low $P$, $L_{fr}$ exceeds insolation and melting cannot occur.  At high $P$, $L_{fr}$ is much less important. Instead, absorbed insolation and greenhouse warming are balanced principally by radiative losses and melting.  Whether or not surface melting can be sustained will depend on the partitioning of the subsurface absorbed insolation (Figure \ref{termbyterm}). 


Setting $\Delta T$ = 5K raises peak melt rate from 0.44 kg/m$^2$/hr to 1 kg/m$^2$/hr. Melt fraction reaches 1 in the shallow subsurface. Melt is produced for 6 (instead of 4) Mars-hours per sol, and there is some melt in the subsurface for 7.5 (instead of 4) Mars-hours per sol.

Both $\alpha$ and \textbf{O$^\prime$} affect the energy absorbed by the snowpack. This can be shown by considering the energy absorbed by equatorial snow at equinox:

\begin{equation}
\begin{centering}
{ E_{equinox}  \approx  (1 - \alpha) L_{\tau}  \!\!  \underbrace{ \left( \frac{1-e^2}{1 + e \cos\Psi}  \right)}_{distance\,\,from\,\,Sun}} ^{-2} \label{equation:eae} 
 \end{centering}
 \end{equation}

\noindent where $E_{equinox}$ is the sunlight absorbed at noon at equinox, $L_{\tau}$ is the solar luminosity at Mars' semi major axis at geological epoch $\tau$, $\Psi$ is  the minimum angular separation between $L_p$ and either $L_s$ = 0 or $L_s$ = 180 (\citet{murray2000}, their Equation 2.20), and the atmosphere is optically thin. If $e$ is large then peak insolation need not occur at equinox.
Equation \ref{equation:eae} shows that moving from average orbital conditions ($e$ = 0.06, $\Psi$ = 90) to optimal orbital conditions ($e$ = 0.15, $\Psi$ = 0) has the same effect on $E_{equinox}$ as darkening from albedo 0.28 to albedo zero. 


\subsection{Seasonal cycle and snow locations}

Figure \ref{figure:seasonal} shows the seasonal cycle of $T$ and sublimation rate on a flat planet. Annual average sublimation rate controls warm-season snow location, and annual-peak snow temperature determines whether melting will occur. The cold trap latitudes indicated correspond to $f_{snow}$ $\rightarrow$ 0, i.e. a single thick ice-sheet. Suppose instead that warm-season snow covers a wider area -- that the ``potential well'' of Figure \ref{fig:potentialwell} fills up with snow. For modern orbital conditions, the area of snow stability will then extend south from the North Pole. If warm-season snow covers more than 43\% of the planet -- if the cold-trap effect is weak or nonexistent -- then melt is possible even under modern orbital conditions.  For optimal orbital conditions, increasing $f_{snow}$ spreads the melting area to form a broad band equatorward of 30$^\circ$.

\subsection{Flat Mars -- snowmelt on a cueball}

\noindent Next, maps are summed over orbital states to find the latitudinal distribution of surface liquid water at a given $P$. If liquid water supply limits sedimentary rock formation, then the sum over orbital states should correspond to geologic observations of Early Mars deposits. 

Suppose that the planet is flat, with warm-season snow only present in a narrow ring at the latitude that minimizes the annually-averaged sublimation rate. Suppose that we impose climate conditions only just allow melting under the optimal orbital conditions. Then:-- \vspace{0.1in}

\noindent \emph{What is the latitudinal distribution of snow, melt and melt intensity?} Obliquity is the strongest control on Mars snowpack stability. The 1D model predicts that snow is most stable near the equator for $\phi$ $\ge$40$^\circ$, near the poles at $\phi$ = \{0$^\circ$,10$^\circ$,20$^\circ$\}, and at intermediate latitudes ($\pm$55$^\circ$) for $\phi$ = 30$^\circ$, in agreement with other studies \citep{jakosky1985,mischna2003,levrard2004,forget2006,madeleine2009}. Nonzero $e$ drives snow to the hemisphere in which aphelion occurs during summer. Holding $e$ fixed, the width of the latitudinal belt swept out by warm-season snow during a precession cycle decreases with increasing $\phi$, from $\pm$26$^\circ$ at $\phi$=40$^\circ$  to $\pm$6$^\circ$ at $\phi$=80$^\circ$ (for $e$=0.09). Holding $\phi$ fixed, the width of the warm-season snow belt increases with increasing $e$, from $\pm$18$^\circ$ at $e$=0.09 to $\pm$22$^\circ$ at $e$=0.16  (for $\phi$ = 40$^\circ$).

\begin{figure}[h]
\includegraphics[width=0.95\textwidth, clip=true, trim = 0mm 0mm 5mm 5mm]{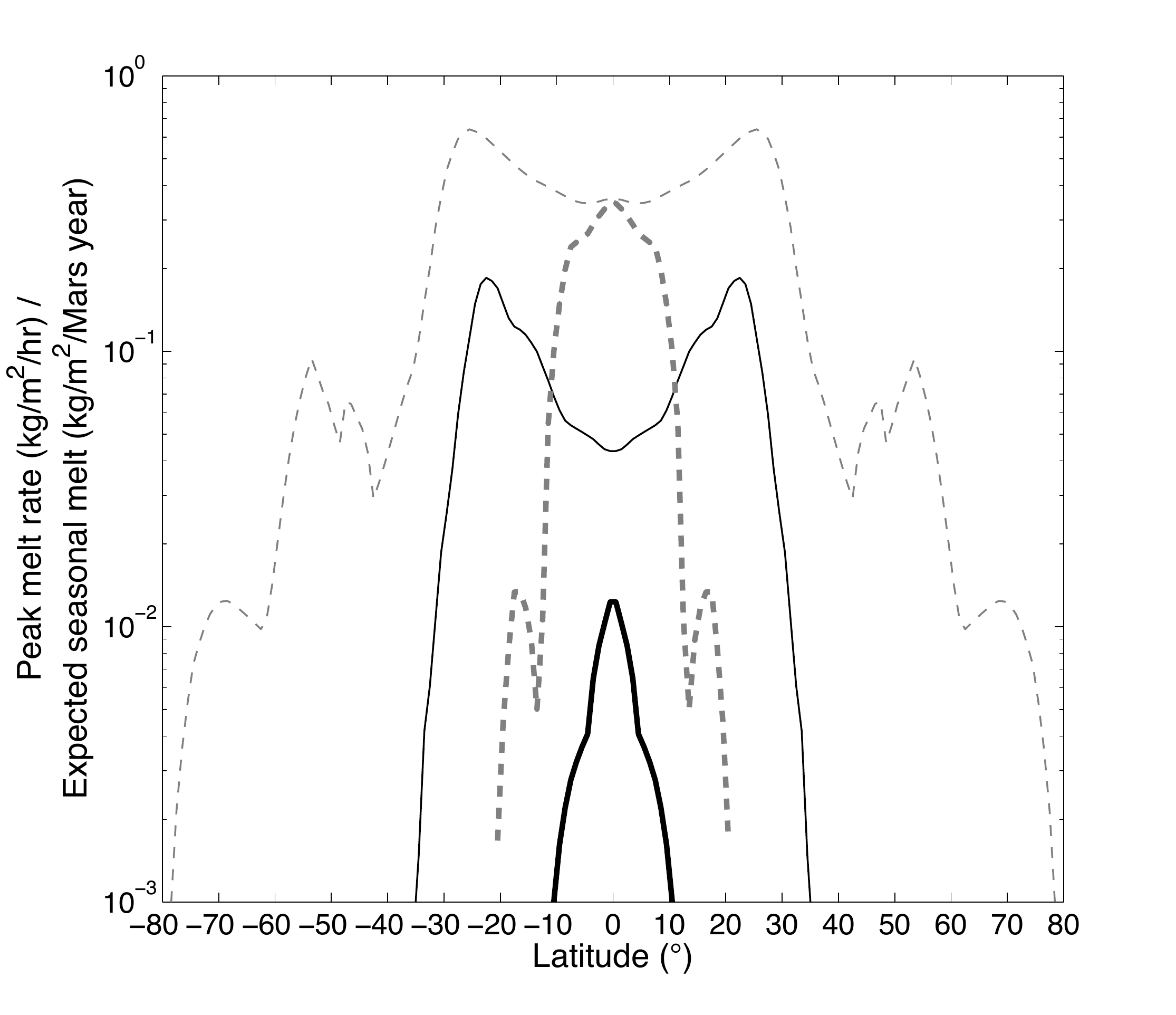}
\caption{Flat-planet results for a climate that only marginally rises above the freezing point on Early Mars. The black solid lines correspond to the orbitally-averaged melt column per Mars year (relevant for aqueous alteration). The gray solid lines correspond to the peak melt rate experienced at any point in the orbits considered (the geomorphically relevant melt rate). The thick lines are for a very small value of $f_{snow}$ (4\%), and the thin lines are for $f_{snow}$ = 50\%. The melt events poleward of 35$^\circ$ for $f_{snow}$ = 50\% correspond to extremely improbable orbital conditions.  \label{figure:cueball}}
\end{figure}

99\% of melting occurs for latitudes $<$10$^\circ$. Annual column snowmelt is sharply concentrated at the equator within the thin melt band (Figure \ref{figure:cueball}). Even though the probability of a melt year is just $\sim$0.05\% , the orbitally integrated expectation for the equatorial snowmelt column is 5 km/Gyr, which is the global spatial maximum on this flat planet. A typical Mars--year of melting produces 9 kg/m$^2$ melt at the latitude of warm-season snow. Peak instantaneous melt rate is of order 0.1 kg/m$^2$/hr, and is less sharply concentrated at the equator than annual column snowmelt (Figure \ref{figure:cueball}). Runoff is unlikely for these low melt rates, so any aqueous alteration must result from infiltration or from alteration of silicates within the snow. Low rates of infiltration may be sufficient to alter aeolian deposits, which volumetrically dominate the Mars sedimentary rock record \citep{grotzinger2012}. 

\begin{figure}
\includegraphics[width=1.00\textwidth, clip=true, trim = 10mm 12mm 0mm 50mm]{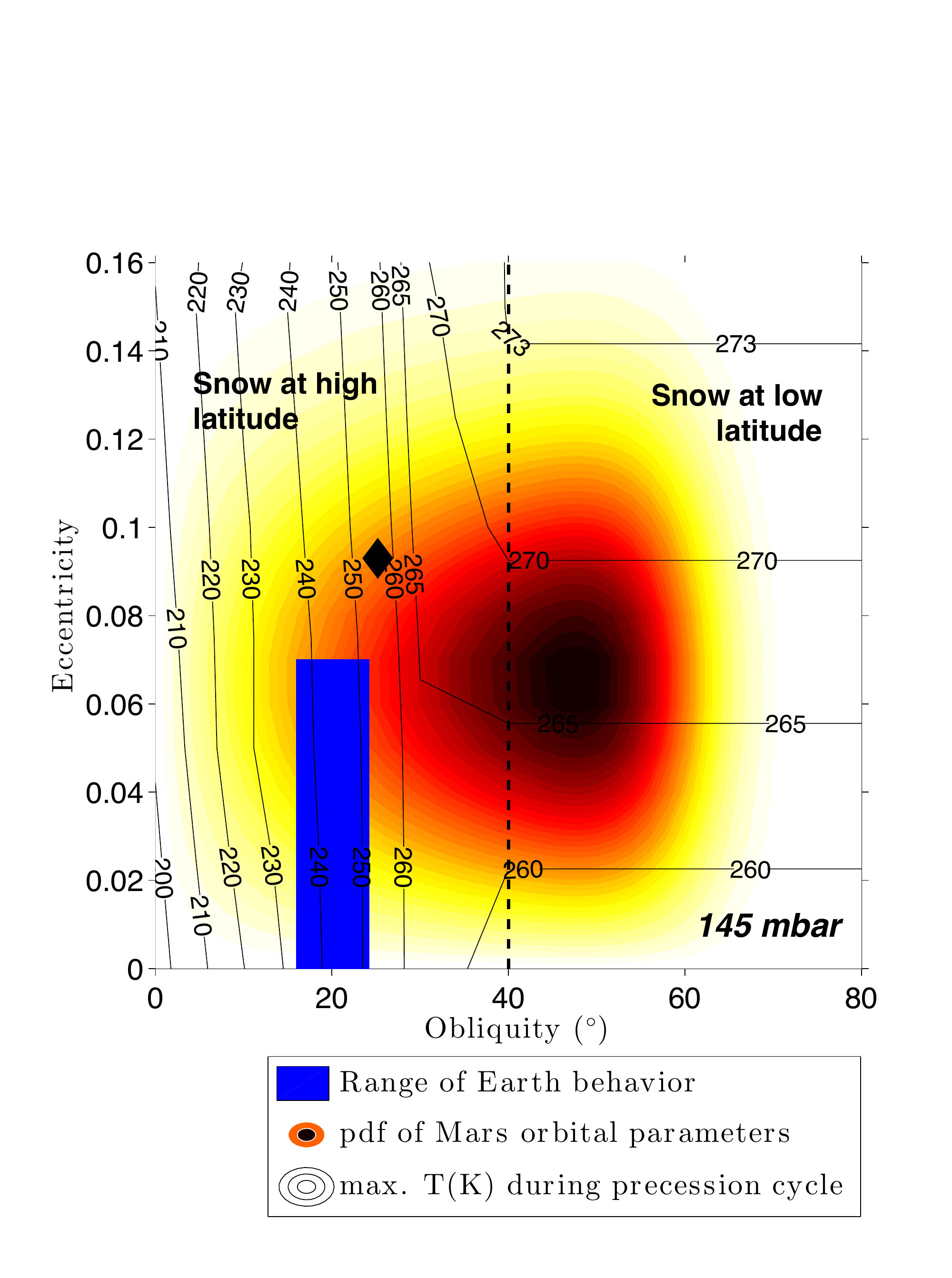}
\caption{The sensitivity of annual-peak snowpack temperature to Milankovitch forcing on an idealized Mars lacking topography. Maximum snowpack temperatures over a precession cycle (black contours) are highest for high obliquity and moderate eccentricity. The probable range of MarsÕ orbital elements (color ramp, with white shading least probable and red shading most probable) is much broader than that of EarthÕs orbital elements (Gyr range shown by blue rectangle). Black diamond corresponds to Mars' present-day orbital elements. Vertical dashed line divides $\phi$ $<$40$^\circ$ (for which warm-season snow is generally found at high latitude), from $\phi$ $\ge$40$^\circ$ (for which warm-season snow is generally found at low latitude). $\Delta T$ = 5K, $P$ = 145 mbar, $\alpha \approx $0.28, Faint Young Sun. \label{figure:orbital}}
\end{figure}

\noindent \emph{What is the distribution of melt and melt intensity with orbital conditions?} As \textbf{C} is moved towards conditions that allow melt, melt first occurs at $\phi$ $\ge$40$^\circ$ and $e$$>$0.13 (Figure \ref{figure:orbital}). \citet{kite2011a,kite2011b} explain that high $\phi$ is needed to drive snow to the equator, and high $e$ is needed to bring perihelion close to the Sun (Figure \ref{figure:orbital}). Melting requires that perihelion occurs when the noontime sun is high in the sky -- for the equator, this requires $Lp$ $\sim$ 0$^\circ$ or $Lp$ $\sim$ 180$^\circ$). Holding $e$ and $\phi$ fixed and moving Mars through a precession cycle, the cueball planet is entirely dry between $Lp$=45$^\circ$ and $Lp$=135$^\circ$ inclusive, and between $Lp$=225$^\circ$ and $Lp$=315$^\circ$ inclusive.


\noindent \emph{What is the seasonal distribution of melt?} All melting occurs near perihelion equinox. The melt season lasts $\le$50 sols (for $e$ $\le$ 0.145).  


\noindent \emph{How does increasing $f_{snow}$ affect the results?} Pinning snow to $\pm$1$^\circ$ of the optimum latitude corresponds to $f_{snow}$ $\sim$ 1\%, similar to the present-day value for warm-season snow.  However, midlatitude pedestal crater heights (44$\pm$22.5m; \citet{kadish2010}) are geologic evidence for high $f_{snow}$ in the Late Amazonian. The modern surface ice reservoir is 2.9$\pm$0.3 x 10$^6$km$^3$ \citep{selvans2010,plaut2007}. Assuming secular loss of ice is slow, and that the midlatitude pedestal craters correspond to the thickness of a single ancient ice layer of uniform thickness, the ice accumulation area is $\sim$46\% of the planet's surface area. This ice accumulation area must be less than $f_{snow}$. 
At $f_{snow}$ = 50\% in the model, the melt belt thickens to $\pm$33$^\circ$, with minor melt activity around $\pm$50$^\circ$ (Figure \ref{figure:cueball}). Maximum annual column melt and melt rates are at $\pm$22$^\circ$, where column melt is 5$\times$ greater (98 km/Gyr) and peak melt rates are 3$\times$ greater than at the equator. Peak melt rates now reach 0.7 mm/hr, so runoff is conceivable (Figure \ref{figure:cueball}). Although there is still a very strong increase of melting with increasing $e$, melting can now occur for any value of $L_p$ and most $e$ (given $\phi$ $\ge$40$^\circ$). Melt is strongest in the Northern Hemisphere at  15$^\circ$ $< L_p <$ 45$^\circ$ and  135$^\circ$ $< L_p <$ 165$^\circ$. It is strongest in the Southern Hemisphere at 195$^\circ$ $< L_p <$ 225$^\circ$ and 315$^\circ$ $< L_p <$ 345$^\circ$. 
The explanation for this behaviour is that the now-broad belt of warm-season snow never entirely shifts out of the hot zone. The edge of this belt closest to the perihelion summer pole sees the sun at zenith shortly before (and shortly after) perihelion solstice equinox. 
This dramatic increase in melting does not require any change in greenhouse forcing or paleopressure, just a change in the way the climate system deposits snow.

\noindent \emph{How does pressure change affect snowmelt on a flat planet?} At the highest $P$ considered, 293 mbar, low values of $f_{snow}$ produce a broad band of melting between $\pm$(15-20)$^\circ$. There is a secondary peak around $\pm$50$^\circ$. The maximum in melt rate snaps away from the equator at $f_{snow}$ $\ge$ 35\%, and above 35\% this maximum moves to gradually higher latitudes.
These patterns are similar at 98 mbar and 149 mbar, although the peak melt rates and expected-seasonal-mean melt rates are both lower because of the reduced greenhouse effect. At 49 mbar, melting only occurs for $f_{snow}$ $>$35\%, and never at equatorial latitudes. Lower $P$ further restricts melting to high $f_{snow}$ and higher latitudes. Physically, these trends correspond to the need for a long day-length (or perpetual sunlight) to warm the snowpack at low $P$. This is not possible at the equator where the day is always $\approx$12 hours long. The melt rate under orbital conditions that are optimal for melting can be thought of as a potential well in latitude, with maxima at high latitudes (for high-$\phi$ polar summer), and a minimum near the equator.  At low $P$ the melt potential is zero at low latitudes, so large values of $f_{snow}$ are needed for melting, which will then occur away from the equator. Increasing $P$ favors melting, so melt rates increase everywhere, and for 98 mbar and above the melt potential is nonzero even at the equator, allowing melting as $f_{snow}$ nears 0.  If Mars had a relatively thick $O$(100 mbar) atmosphere during the sedimentary-rock era, then the progressive inhibition of equatorial melting is a latitudinal tracer of Mars atmospheric escape.


\noindent The ``cueball'' results presented in the above sections have strong echoes in the geologic record of ancient Mars. The distribution of evidence for liquid water shows strong latitudinal banding: sedimentary rocks are concentrated at latitudes $<$15$^\circ$ (\S 3.1) but have ``wings'' at 25-30S and 20-30N, and alluvial fans are most common at 15S-30S \citep{kraal2008,wilson2012}. 


\vspace{-0.0in}

\subsection{With MOLA topography}

\noindent The main meridional trends in model output without topography hold when MOLA topography is used. This is because nominal model parameters produce snow distributions that are more sensitive to latitude than to elevation.

\begin{figure*}
\includegraphics[width=1.0\textwidth, clip=true, trim = 48mm 65mm 45mm 55mm]{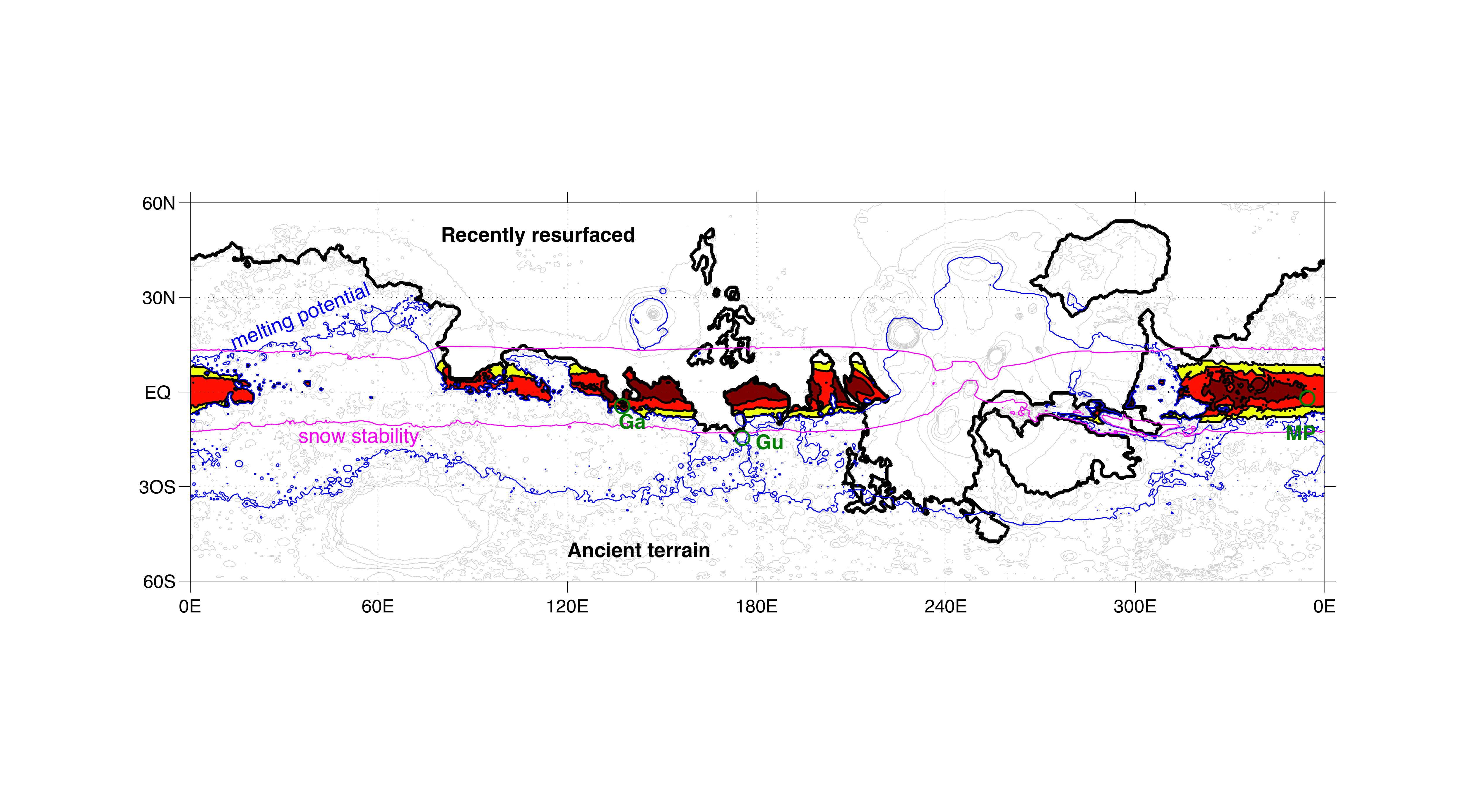}
\vspace{0.0in}
\caption{A snapshot of snowmelt distribution for a single example of orbital forcing, showing role of snow stability and melting potential. $\phi$=50$^\circ$, $e$=0.145, $L_p$ = 0$^\circ$, $P_o$ = 49 mbar, $\Delta T$ = 5K. Areas equatorward of the magenta line correspond to $f_{snow}$ $<$ 20\% -- likely snowpack locations. Areas poleward of the blue line are hot enough to see melting at some point during the year, if snow were present. Where the snow zone intersects the hot zone, some melting will occur. Melt zones for small values of $f_{snow}$ are shaded in warm colors.  $f_{snow}$ $<$ 10\% is shaded yellow,  $f_{snow}$ $<$ 5\% is shaded red, and  $f_{snow}$ $<$ 2\% is shaded maroon. Notice that extensive melting in Valles Marineris requires high $f_{snow}$ or a different phase of the precession cycle. Thick black line corresponds to the boundary of terrain resurfaced since sedimentary rocks formed. This terrain is not included in the warm-shaded areas. Landing sites of long-range rovers are shown by green circles: -- Ga = Gale Crater; Gu = Gusev Crater; MP = Meridiani Planum. Grayscale contours in background are topographic contours at intervals of 1.5 km from -5 km up to +10km. 
\label{figure:ralltall} 
}
\end{figure*}

At $P_o$ = 48 mbar and optimal $\textbf{O$^\prime$}$ ($\phi$=50$^\circ$, $e$=0.16, $L_p$ = 0$^\circ$), snow distribution shows only a weak preference for low points. However, peak temperature is controlled by elevation because $L_{fr}$ is stronger (and $LW\!\! \downarrow$ weaker) at high elevations.  Therefore  melting only occurs in planetary topographic lows close to the equator at low $f_{snow}$ and $\Delta T$=2K (Figure \ref{figure:ralltall}).

For $f_{snow}$ = 50\%, melting at optimal orbital conditions occurs for all low-lying locations equatorward of 30$^\circ$. 
When perihelion is aligned with solstice at $\phi$=50$^\circ$, the snow distribution shifts away from the equator, and no melting occurs below $f_{snow}$ $\sim$ 60\%. 

Low $\phi$ is much less favorable for snow melting than high $\phi$ \citep{jakosky1985}. For $\phi$ $\le$ 30$^\circ$ and perihelion aligned with solstice, snow is most stable poleward of 60$^\circ$, but these most favored locations never reach the freezing point. As $f_{snow}$ is raised, melting will first occur at lower latitudes because these receive more sunlight. The most favored locations are S Hellas and the lowest ground around 40N. These are the midlatitude locations where scalloped depressions are most prominent \citep{soare2007,zanetti2010}, although these features might not require liquid water to form \citep{lefort2009,lefort2010} and the model is not directly applicable to Upper Amazonian features.

Snow and melt distributions on MOLA topography depend on the trade-off between $P$ and sublimation rate, which controls snow stability (Figure \ref{figure:pressure}). 
For example, suppose wind speed on Early Mars was much higher than modelled. Then the relative importance of wind-speed-dependent turbulent losses in the surface energy balance would increase. This would increase the importance of elevation ($\sim$1/$P$) in setting snow location, relative to latitude which sets $SW\!\!\!\downarrow$. The snow and melt distributions for this ``windy early Mars'' (not shown) are broader in latitude and more concentrated in low areas (especially Northern Hellas, but also Northern Argyre and the Uzboi-Ladon-Margaritifer corridor).

\subsection{Effects of climate on snowmelt distribution with MOLA topography}

Summing the orbital--snapshot maps of melt likelihood ($\int \!\! p(\textbf{O}) \mathrm{d}$\textbf{O}) shows the effect of \textbf{C} = \{$P$, $\Delta T$, $f_{snow}$\} on melt likelihood averaged over geological time.

\begin{figure*}
\includegraphics[width=1.0\textwidth, clip=true, trim = 40mm 50mm 35mm 30mm]{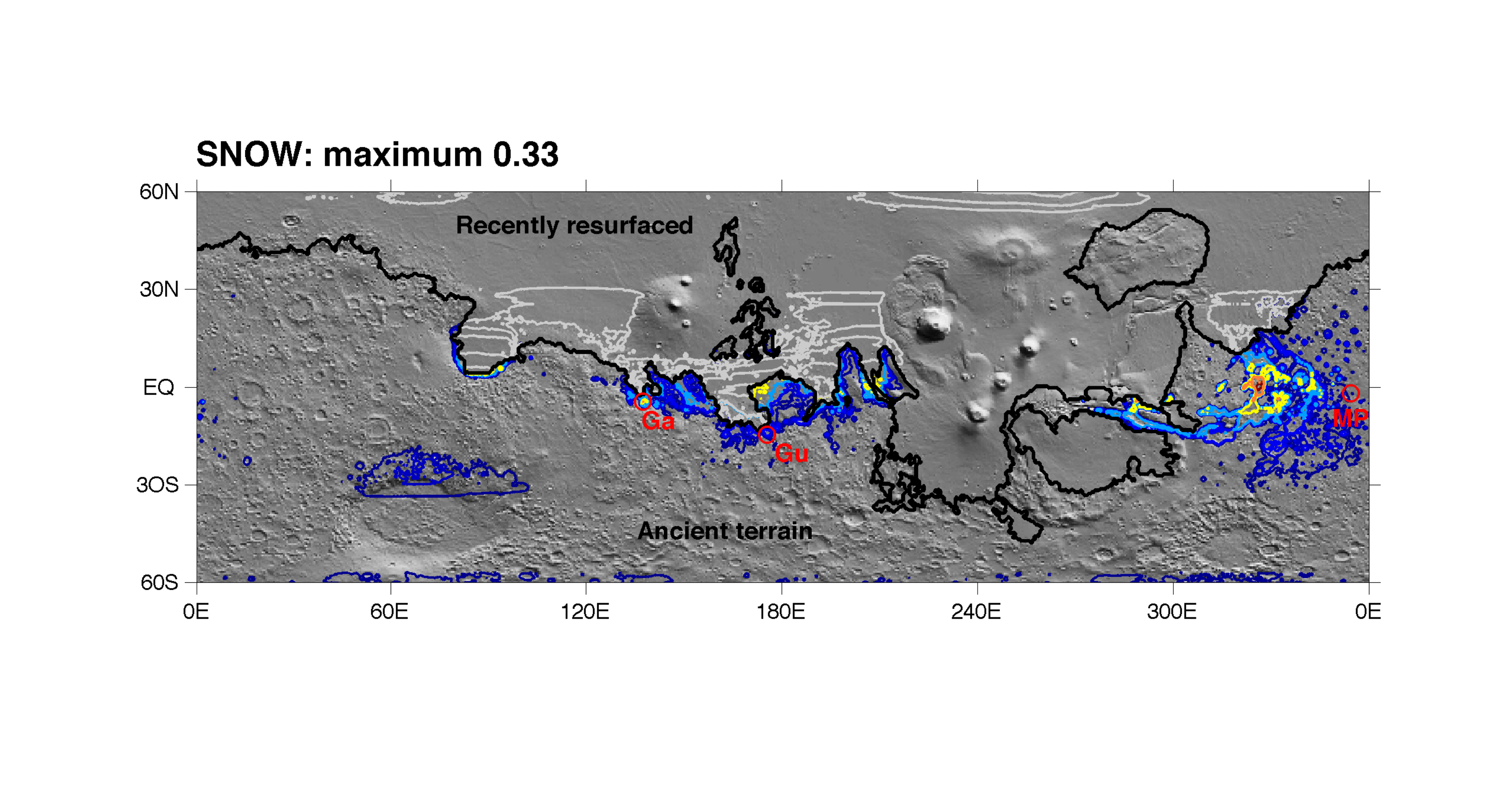}
\vspace{-0.1in}
\includegraphics[width=1.0\textwidth, clip=true, trim = 40mm 50mm 35mm 30mm]{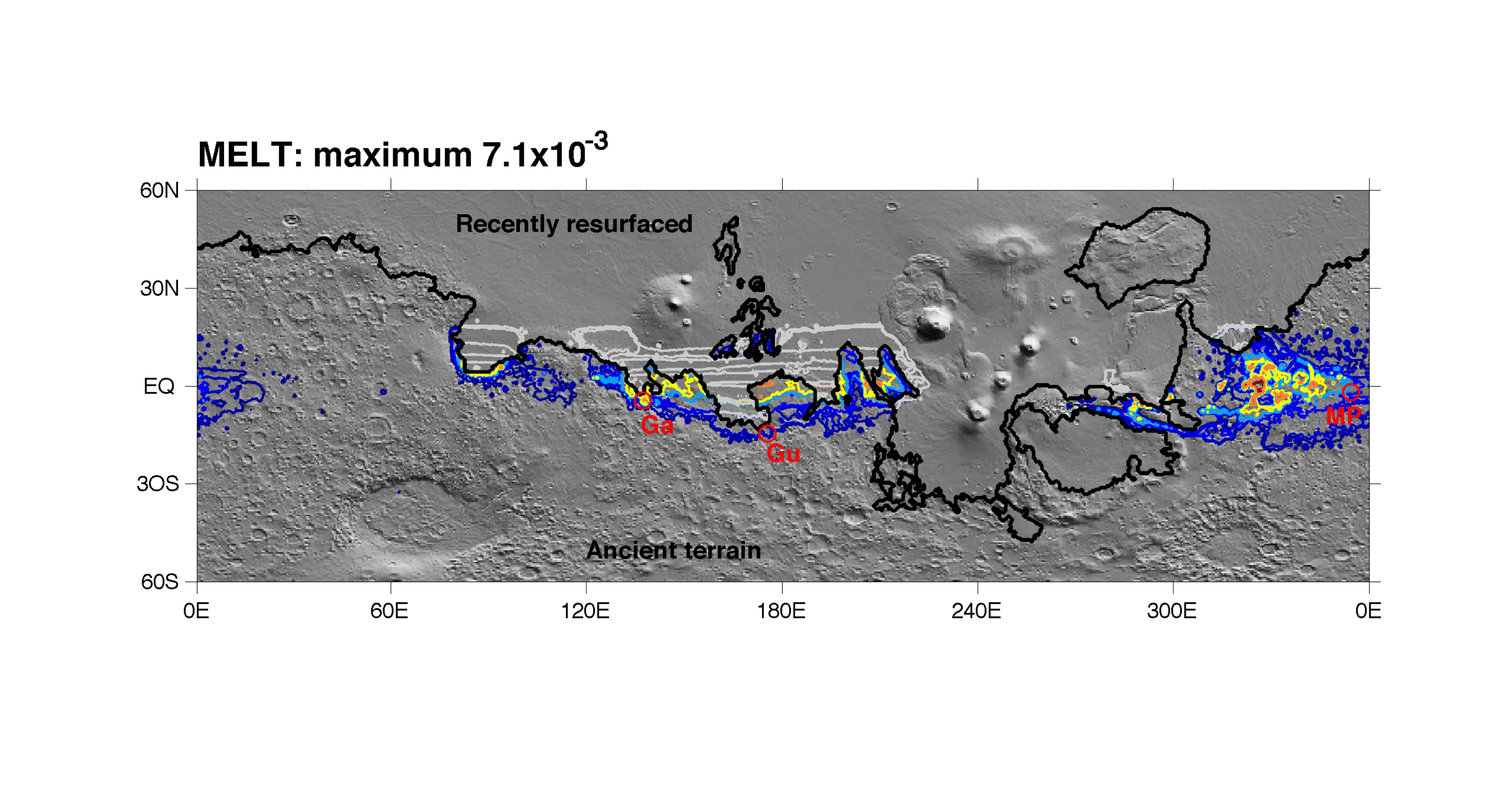}
\vspace{0.0in}
\caption{ Probabilities of (upper panel) warm-season snow and (lower panel) melting for $P$ = 49 mbar, $\Delta T$ = 5K, and $f_{snow}$ = 2\%. Background is shaded relief MOLA topography, illuminated from top left. Maximum probability on the warm-season snow map is 0.33, maximum of the melt map is 7.1 $\times$10$^{-3}$ -- the location is Hydaspis Chaos for both maxima. Contours are at 1\%, 5\%, 10\%, 25\%, 50\%, 75\% and 90\% of the maximum value. Because melting requires unusual orbital conditions, while warm-season low-latitude snow only requires high obliquity, the lowest colored contour in the snow map is greater than the highest colored contour in the melt map. Thick black line corresponds to border of ancient terrain, and grayed-out contours are probabilities on recently resurfaced terrain. Long-range rover landing sites are shown by red circles:-- Ga = Gale Crater; Gu = Gusev Crater; MP = Meridiani Planum.
\label{figure:INTORB1} }
\end{figure*}

For  $P_o$ = 49 mbar and for small values of $\Delta T$ (5K) and $f_{snow}$  (2\%), warm-season snow is found primarily in (Figure \ref{figure:INTORB1}a) Valles Marineris, the circum-Chryse chaos, the Uzboi-Ladon-Margaritifer corridor, craters in W Arabia Terra, the Isidis rim, northern Hellas, Gale, Aeolis-Zephyria Planum, and parts of the Medusae Fossae Formation, as well as at high ($>$50$^\circ$) latitudes. However, warm-season snow only melts very close to the equator (Figure \ref{figure:INTORB1}b) -- in Gale, the circum-Chryse chaos, Meridiani Planum, Aeolis-Zephyria Planum, the Isidis rim, and the floors of the Valles Marineris canyons.  Even in central Valles Marineris, among the wettest parts of the planet under this climate, melting occurs with probability $<$0.5\% (e.g., 5 Myr of melt years during 1 Gyr). As $f_{snow}$ is increased to 5-10\% at $\Delta T$ = 5K, melting in Meridiani Planum and Valles Marineris becomes more frequent. Melting in Northern Hellas does not occur until either $f_{snow}$ or $\Delta T$ is greatly increased. 

As the atmosphere is lost, melting becomes restricted in space as well as time ($\Delta T$ = 5K, $P_o$ = 24 mbar, $f_{snow}$ = 0.1\%, Figure \ref{figure:INTORB2}a). The last holdouts for surface liquid water on Mars are Gale Crater, du Martheray Crater, and Nicholson Crater in the west-of-Tharsis hemisphere, and the floors of the Valles Marineris canyons in the east-of-Tharsis hemisphere (Figure \ref{figure:INTORB2}a). Gale Crater (near 6S, 135E) is usually a hemispheric maximum in snowmelt for marginal-melting climates. Melting can only occur for very improbable orbital combinations under this climate. If they occurred at all, wet periods would be separated by long dry intervals. 

At $P_o$ =  293 mbar and low $f_{snow}$, low-latitude snow is restricted to high ground and so is melt. 
Figure \ref{figure:INTORB2}b shows the melt distribution for $f_{snow}$ = 10\% and $\Delta T$ = 7.5K. For $f_{snow}$ $\ge$ 20\%, snow is still most likely at high ground, but the melt pattern flips: melt occurs at all elevations, but it is most common at low ground as in the low-$P$ case. As $f_{snow} \rightarrow$ 100\%, melt extent is limited only by temperature. This is maximized at low elevations, because of the greater column thickness of greenhouse gas (and because of the adiabatic atmospheric temperature lapse rate, neglected here). 

Hellas is usually the most favored area for snowmelt within the midlatitude ancient terrain. A climate relatively favorable for melting in Northern Hellas is shown in Figure \ref{figure:INTORB2}c ($\Delta T$ = 15K, $P_o$ = 24 mbar, $f_{snow}$ = 40\%). The contour of locally maximal snowmelt extending from deepest Hellas, to the Northern Hellas floor, to crater Terby, is intriguing in view of recent descriptions of thick packages of sedimentary rock in these locations \citep{wilson2007,ansan2011,wilson2010}.
The Uzboi-Ladon-Margaritifer corridor of fluvial activity \citep{grant2002,grant2008,milliken2010pmag,mangold2012} is not as favorable for snowmelt.

For the wettest conditions considered (e.g., $\Delta T$ = 15K, $P_o$ = 49 mbar, $f_{snow}$ = 40\%, Figure \ref{figure:INTORB2}d), melt occurs more than 25\% of the time in most places equatorward of 30$^\circ$. Such wet global climates grossly overpredict both the spatial extent of sedimentary rock formation on Early Mars and the extent of surface aqueous alteration, as discussed in the next section.

\vspace{-0.15in}

\newpage
\pagebreak

\begin{figure*}[h]
\includegraphics[width=1.0\textwidth, clip=true, trim = 40mm 50mm 35mm 30mm]{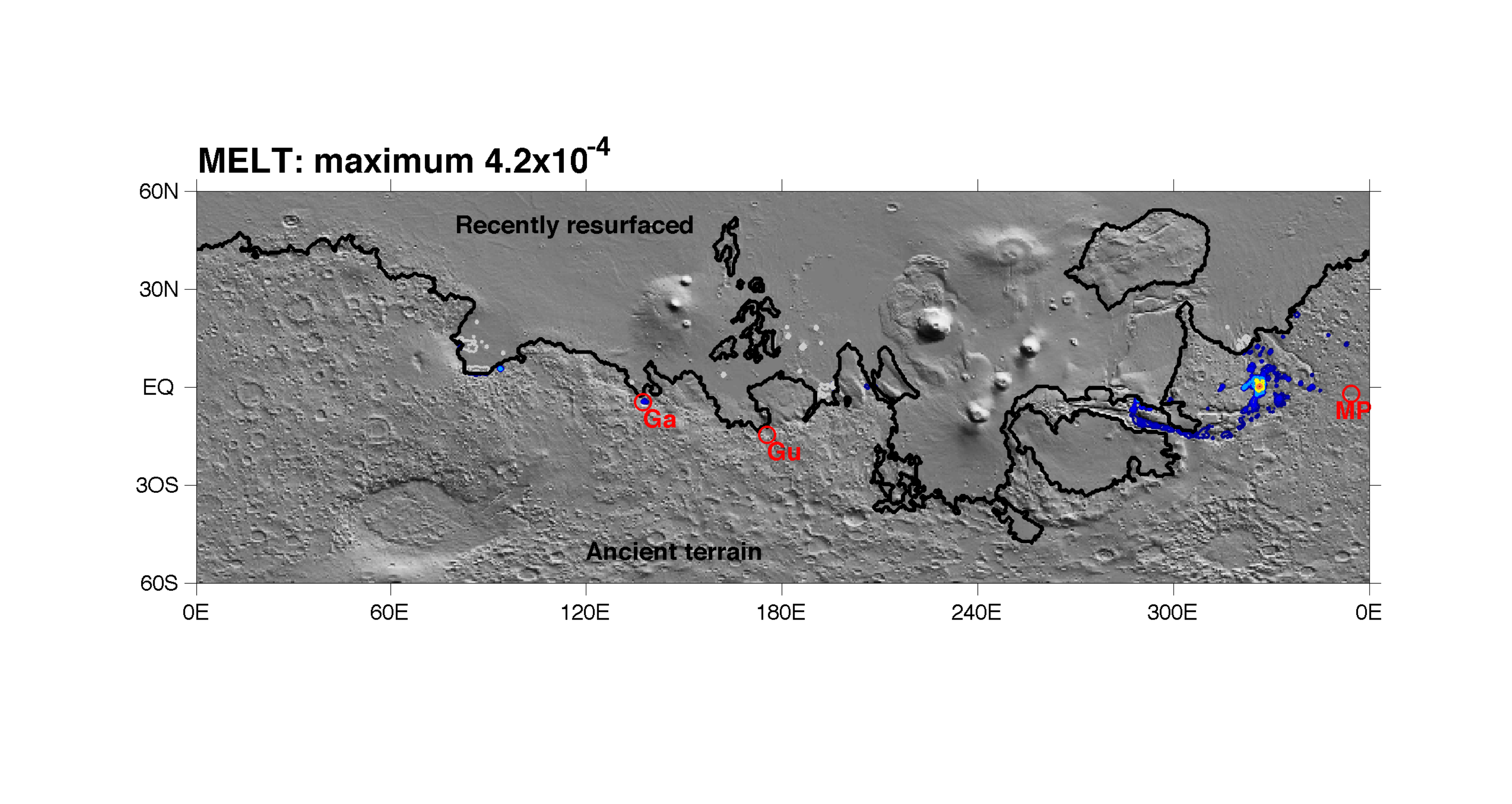}
\vspace{0.0in}
\includegraphics[width=1.0\textwidth, clip=true, trim = 40mm 50mm 35mm 30mm]{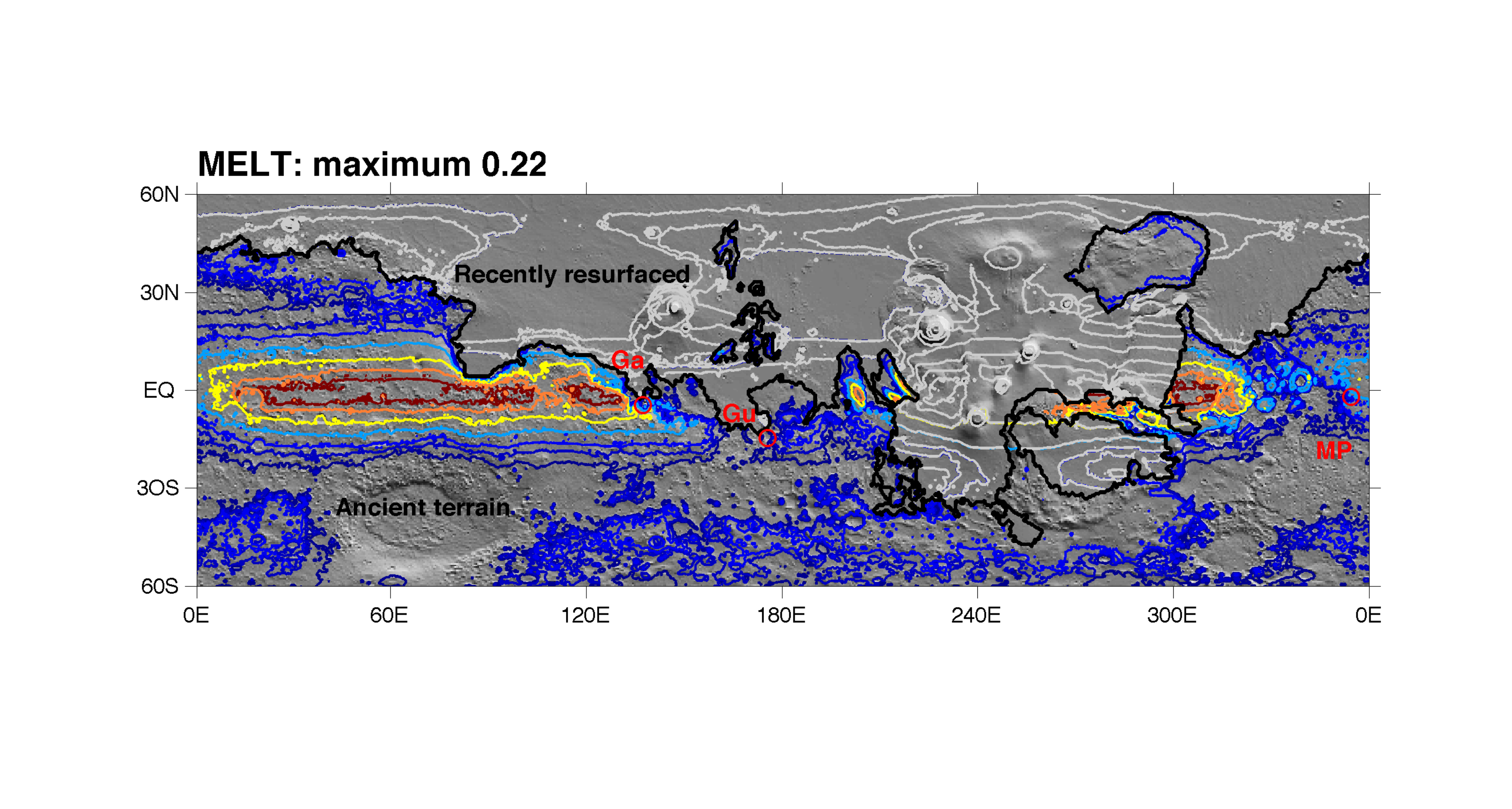}
\vspace{0.0in}
\includegraphics[width=1.0\textwidth, clip=true, trim = 40mm 50mm 35mm 30mm]{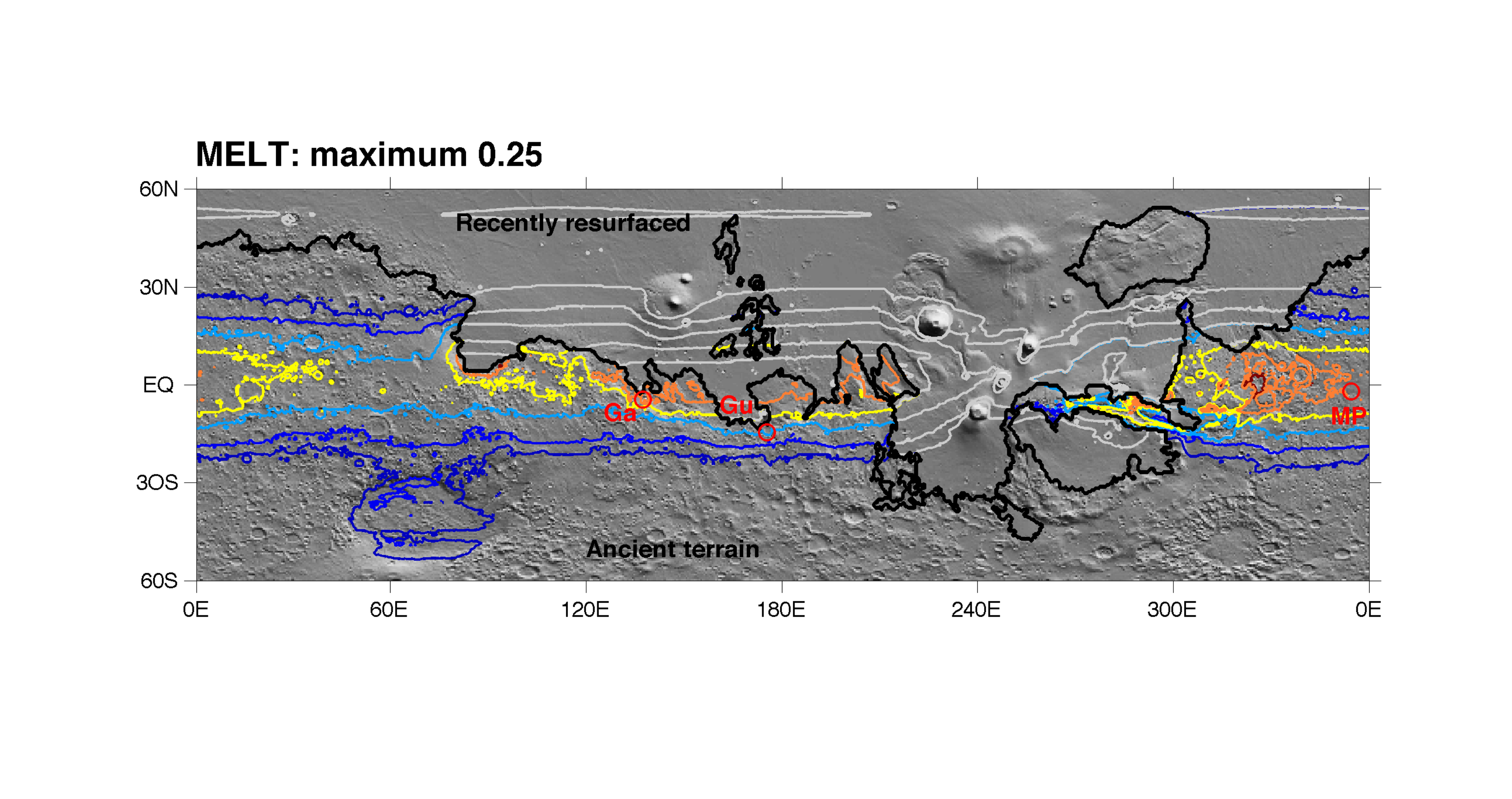}
\vspace{-0.1in}
\end{figure*}

\newpage
\clearpage
\pagebreak

\begin{figure*}
\includegraphics[width=1.0\textwidth, clip=true, trim = 40mm 50mm 35mm 30mm]{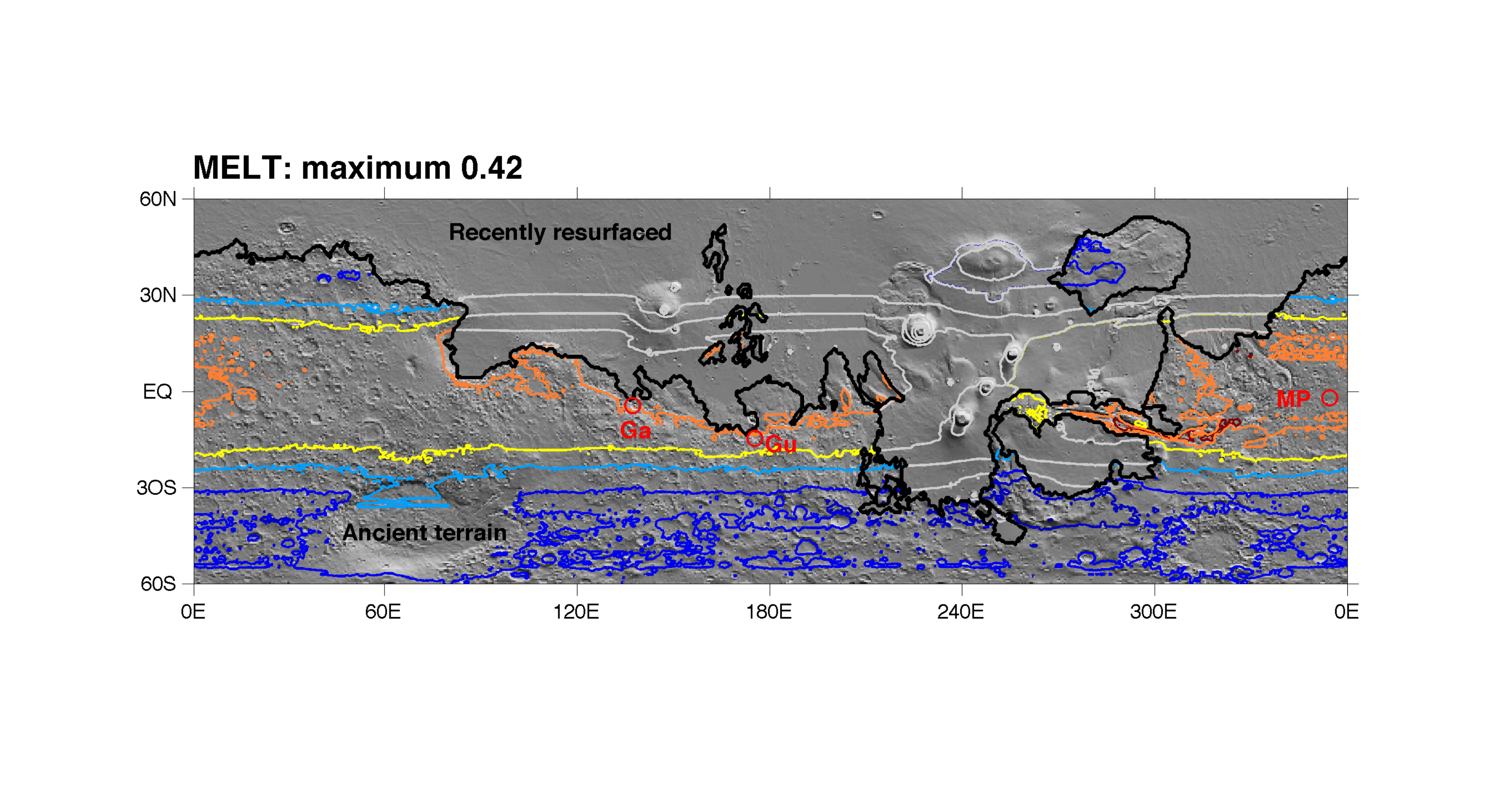}
\vspace{0.0in}
\caption{Sensitivity of snowmelt maps to extreme variations in model parameters. Background is shaded relief MOLA topography, illuminated from top left. Colored contours correspond to snowmelt probabilities on ancient terrain.  Contours are at 1\%, 5\%, 10\%, 25\%, 50\%, 75\% and 90\% of the maximum melt likelihood, which is given to the top left of each panel.
Black line corresponds to border of ancient terrain, and grayed-out contours are snowmelt probabilities on recently-resurfaced terrain. Long-range rover landing sites are shown by red circles:-- Ga = Gale Crater; Gu = Gusev Crater; MP = Meridiani Planum. \textbf{(a)} Parameters that only marginally allow melting even under optimal orbital conditions: $P$ = 24 mbar, $\Delta T$ = 5K, $f_{snow}$ = 0.1\%. \textbf{(b)} High $P$ drives snow (and melt) to high ground: $P$ = 293 mbar, $\Delta T$ = 7.5K, $f_{snow}$ = 10\%. This is inconsistent with the observed concentration of sedimentary rock at low elevations, but may be relevant to the distribution of older valley networks. \textbf{(c)} Parameters that produce snowmelt in Hellas: $P$ = 24 mbar, $\Delta T$ = 15K, $f_{snow}$ = 20\%. For legibility, the 1\% contour is not shown for this subfigure.
\textbf{(d)} Very high $f_{snow}$ and $\Delta T$ predict a latitudinally broader distribution of sedimentary rocks than observed: $P$ = 49 mbar, $\Delta T$ = 15K, $f_{snow}$ = 40\%. 
\label{figure:INTORB2} 
} 
\end{figure*}


\newpage
\clearpage
\pagebreak

\section{Snowmelt in space: understanding the distribution of sedimentary rocks on Mars\label{distribution}}

\subsection{Comparison of global data to model output: implications for Early Mars climate state }

\noindent  The full climate ensemble consists of 343 orbitally integrated melt-likelihood maps similar to those in Figure \ref{figure:INTORB2}. To reduce this to a manageable number for analysis, $k$-means clustering was used \citep{press2007}. 
The spatial variability of the melt-likelihood maps was normalized by the within-map mean and within-map standard deviation, and clustering was carried out on these self-standardized maps. Representative results are shown in Figure \ref{figure:cluster}, together with the mean melt-likelihood maps for each of the climate clusters identified.


\setlength\fboxsep{0pt}
\setlength\fboxrule{2pt}

\begin{figure*}[ht]
\begin{center}
\includegraphics[width=1.0\textwidth, clip=true, trim = 20mm 10mm 10mm 15mm]{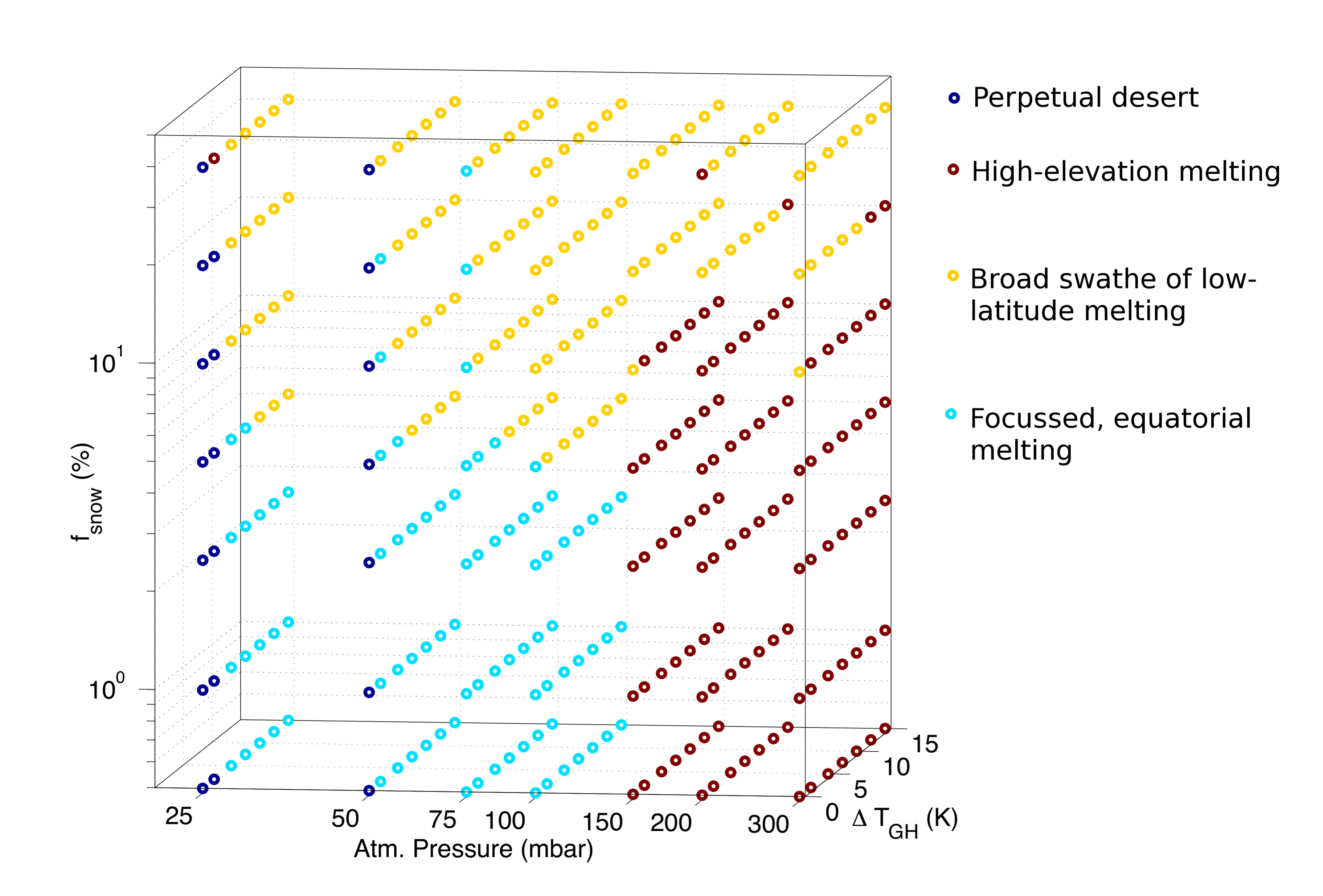}
\end{center}
\vspace{0.2in}
\cfbox{red}{
\includegraphics[width=1.0\textwidth, clip=true, trim = 22mm 20mm 22mm 20mm]{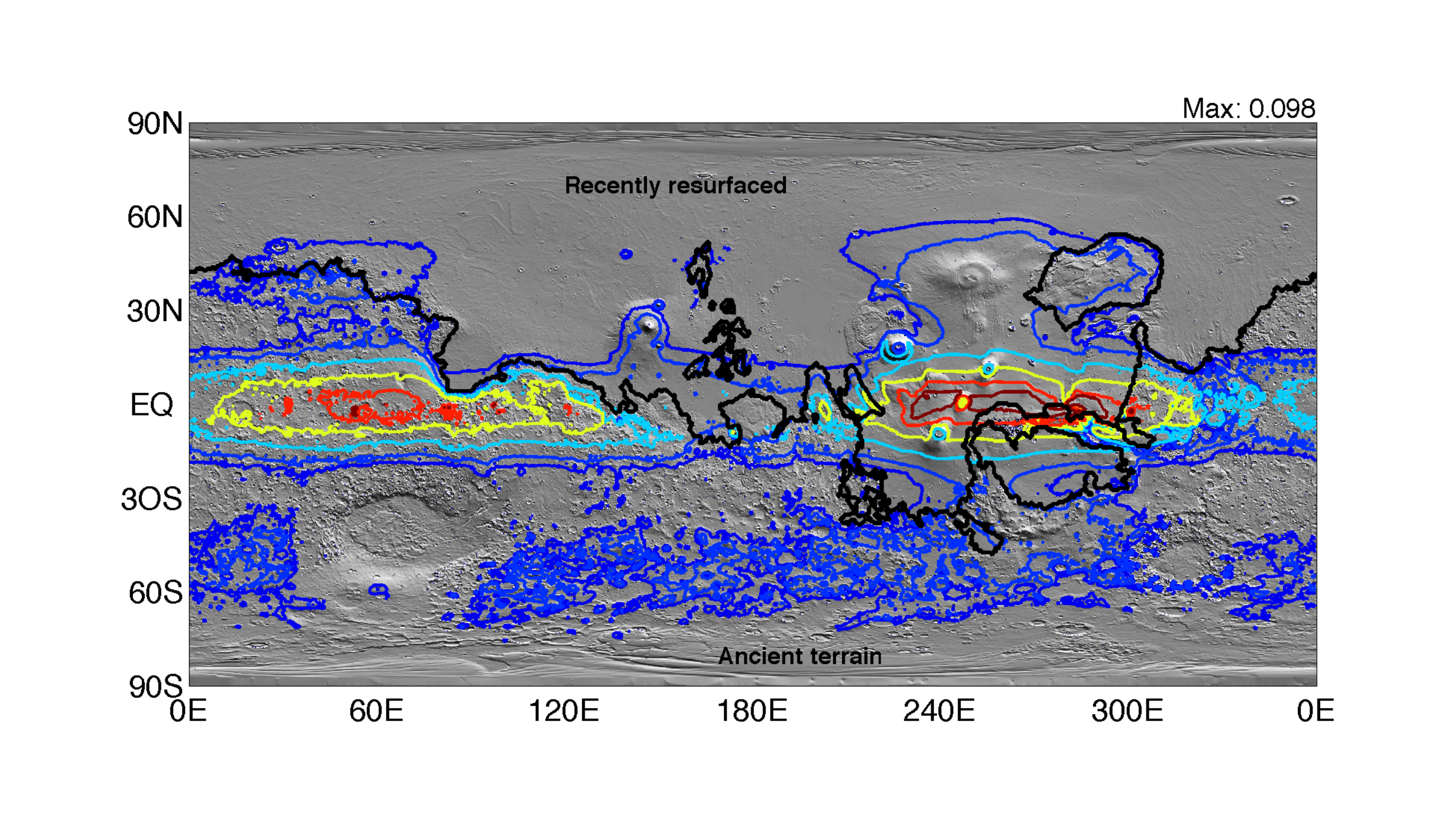}
}
\end{figure*}

\setlength\fboxrule{2pt}

\begin{figure*}

\cfbox{yellow}{
\includegraphics[width=1.0\textwidth, clip=true, trim = 22mm 20mm 22mm 18mm]{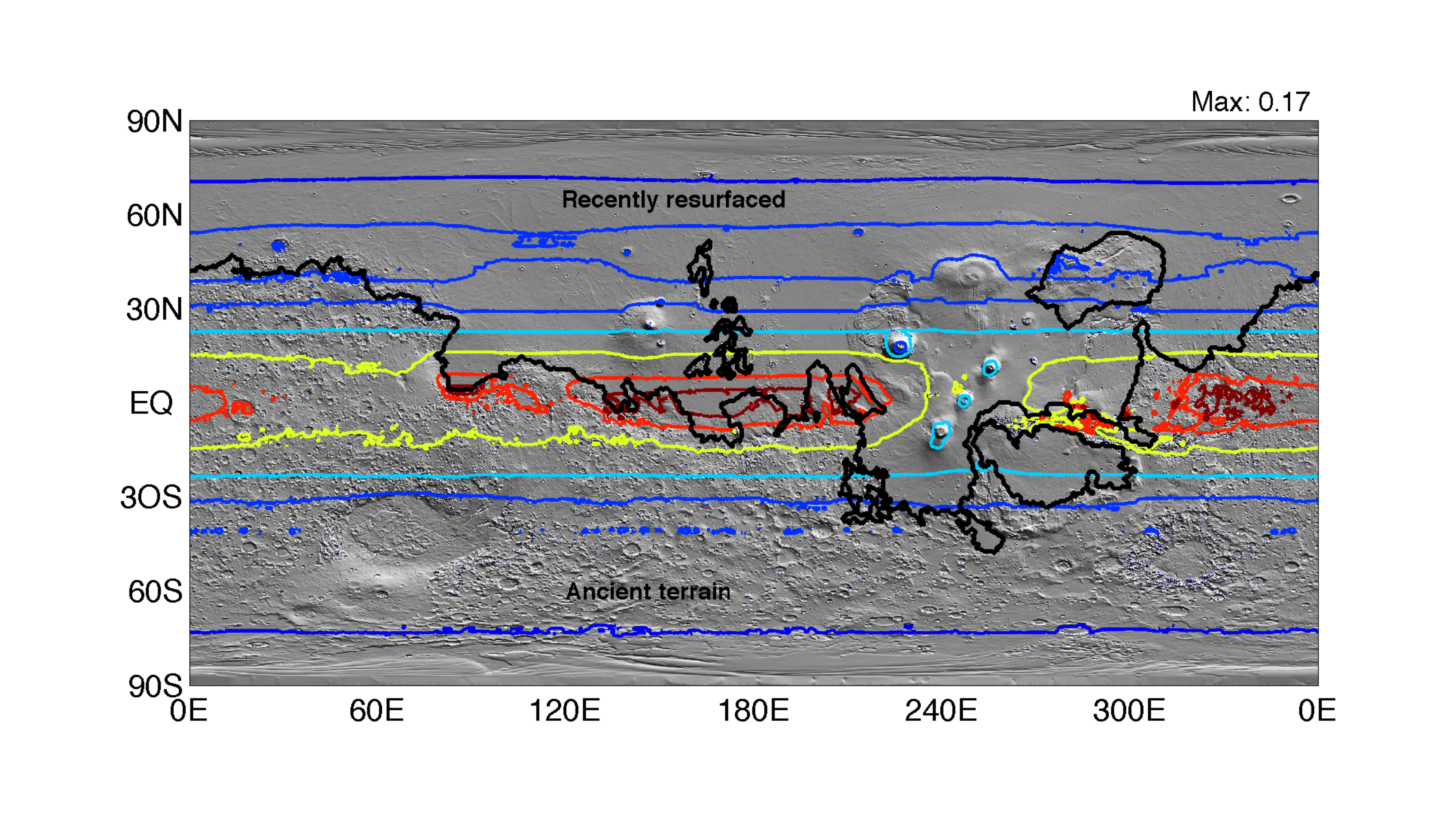}
}

\cfbox{cyan}{
\includegraphics[width=1.0\textwidth, clip=true, trim = 22mm 20mm 22mm 18mm]{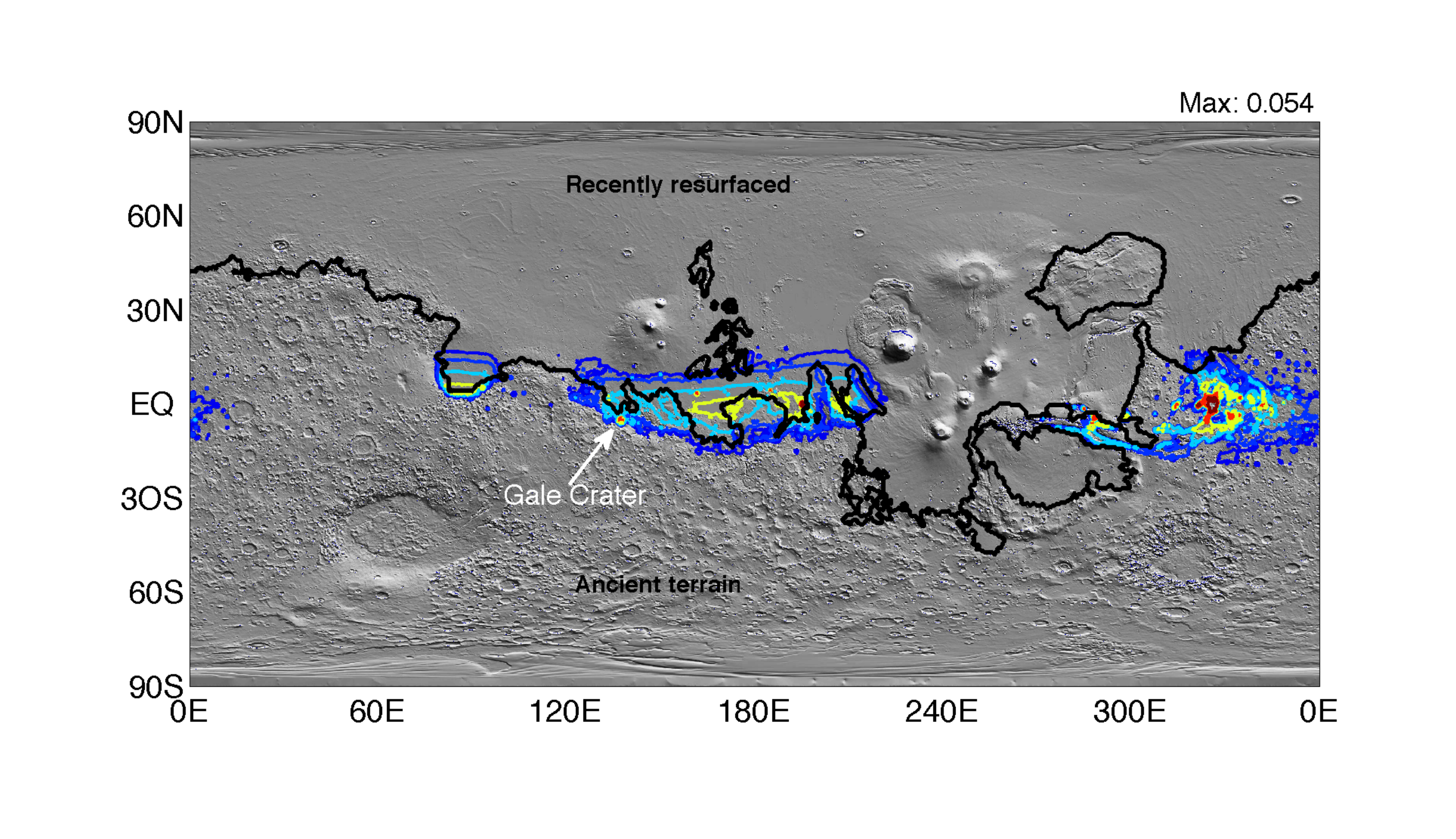}
}

\caption{Effect of climate on Mars sedimentary-rock distribution, assuming a snowmelt water source. Top panel shows clustering of climates into three melt-producing classes, plus perpetual--global--desert climates (dark blue dots). Remaining panels show maps of the mean of each of the melt-producing climate classes. Border colors correspond to the dots in the climate parameter space which contribute to that map. ``Max'' to the top right of each map refers to the spatial maximum in melt likelihood, which is the probability that a given location sees some melting during the year. The colored contours correspond to melt likelihoods of  5\% , 10\%, 25\%, 50\%, 75\%, and 90\% of the spatial maximum for that climate class. Thick black line shows the boundary between ancient terrain and recently--resurfaced terrain.  \label{figure:cluster}} 
\end{figure*}


Dark blue dots correspond to perpetual-global-desert climates. Zero melting is predicted on horizontal surfaces under all orbital conditions. 

At $P_o$ $\ge$ 150 mbar, all $\Delta T$, and low-to-moderate $f_{snow}$, melting occurs at high elevation (red cluster). Some melt also occurs at  mid-southern latitudes. Neglect of the adiabatic lapse rate will lead to growing inaccuracy at high $P_o$, but will not alter the conclusion that warm-season snow will be driven to high ground at high $P_o$, far from the places where sedimentary rocks are observed (Figure \ref{figure:cluster}b). 

For a wide range of $P_o$, all $\Delta T$, and moderate-to-high $f_{snow}$, the framework predicts a broad swath of low latitude melting (amber cluster). Figure \ref{figure:cluster}c is effectively a map of maximum snowpack temperature -- as $f_{snow}$ becomes large, warm-season snow is no longer restricted by elevation. Melting is most intense at low elevation because of the increased CO$_2$ column, but the overall pattern is diffuse in both elevation and latitude. This contrasts with the strongly-focussed observed sedimentary rock distribution (Figure \ref{figure:sedrocksmap}). Melt probabilities are large over a broad part of the planet, which does not sit easily with thermal infrared data indicating that most soil on Mars did not experience volumetrically important aqueous alteration \citep{bandfield2011}.


For $P_o$ $<$ 150 mbar and at least one of low $\Delta T$ or low $f_{snow}$, the model predicts focused, equatorial melting (cyan cluster), in excellent agreement with observations (Figure \ref{figure:cluster}d). The agreement is especially good given the simplicity of the model physics (\S4) and the fact that we are considering one of three objectively-defined \emph{classes} of paleoclimates rather than the optimum \textbf{C}. Independently of the snowmelt model, this result suggests that latitude and elevation are the main controls on sedimentary rock distribution on Mars, because our model physics does not include 3D effects. 
We highlight seven points of data/model agreement:--

\begin{enumerate}

\item The thickest sedimentary rock exposures on Mars are in Valles Marineris (up to 8km), Gale Crater (5km), and Terby Crater (3km) \citep{murchie2009b,anderson2010,wilson2007,ansan2011}. Sedimentary layered deposit thicknesses in the chaos source regions are up to $\sim$1km (Aram; \citet{glotch2005}). The Medusae Fossae Formation is a sedimentary accumulation up to 3km thick \citep{bradley2002} which may also be aqueously cemented sedimentary rock in its lower part \citep{burr2009,burr2010}. With the exception of Terby, this is the same set of locations where the focussed, equatorial melting paleoclimate class predicts maxima in orbitally-integrated snowmelt. The Northern Valles Marineris canyons contain thicker sedimentary-rock mounds than the southern Valles Marineris canyons, and are correspondingly more favored for snowmelt in the model. 

\item Gale is a hemispheric maximum in ancient-terrain sedimentary rock thickness, and is a hemispheric maximum in ancient-terrain snowmelt in the model.

\item Snowmelt is strongly focussed in the Valles Marineris, the chaos source regions, and Gale. Predicted deposit thickness dies away quickly from these regions.

\item Meridiani Planum is correctly predicted to be a local maximum within a broader wedge-shaped Sinus Meridiani outcrop narrowing and thinning to the East \citep{edgett2005,hynek2008epsl,andrewshanna2010,zabrusky2012,wiseman2011}. The concentration of Western Arabia sedimentary rock in mound-filled craters (e.g. Crommelin, Firsoff, Danielson, Trouvelot, and Becquerel) is reproduced by the model. 
Alignment of Meridiani Planum with snowmelt maximum implies net True Polar Wander $<$10$^\circ$ since sediment deposition \citep{matsuyama2006,perron2007,kite2009,matsuyama2010}. 

\item The southern Isidis rim is identified as a regional maximum for post-Noachian surface liquid water, consistent with data (eg., \citet{jaumann2010}).

\item In the Northern Plains, deep equatorial craters are commonly modified by sedimentary infill (e.g., Nicholson, Reuyl). This correlation is reproduced by the model. 

\item The focussed equatorial melting climate cluster predicts strong enhancement of melting in Northern Hellas relative to other locations in the same latitude band (similar to Figure \ref{figure:INTORB2}). However, $\Delta T$  $\ge$10K is needed for non-negligible melting away from the equator, so this longitudinal enhancement is diluted in the class-average map and is not visible. A secondary enhancement within this southern latitude belt is the Uzboi-Ladon-Margaritifer corridor. These longitudinal enhancements match data on the distribution of sedimentary rocks and alluvial fans \citep{kraal2008}. However, the model underpredicts the thickness of Terby fill, relative to the equatorial belt of sedimentary rocks. 



\end{enumerate}


\noindent The model predicts snowmelt in the circum-Chryse region out of proportion to the observed sedimentary rock deposits. However, chaos and outflow channels continued to form around Chryse through the Early Amazonian  \citep{warner2009,carr2010}, and would have destroyed sedimentary rocks deposited earlier. If supraglacial snowmelt crevassed to the base of the ice mass and inflated subglacial lakes, seasonal melting could have contributed to chasm flooding and overflow. Sedimentary rocks overly chaos in Aram and Iani \citep{glotch2005,warner2011}. The model does not predict snowmelt at Mawrth, consistent with Mawrth's interpretation as a Noachian deposit formed under a earlier climate \citep{mckeown2009}, nor does it predict snowmelt at Terra Sirenum, consistent with the nominally Late Noachian age of the inferred paleolake deposits there  \citep{wray2011}. Finally, the model predicts snowmelt mounds within deep Northern Lowlands craters that appear little-modified in CTX, such as the crater at 94.5E, 10N. If the snowmelt hypothesis is correct, then these craters must postdate the sedimentary-rock era in order to avoid infilling by sedimentary rock. This prediction can be tested with crater counts on ejecta blankets.



The colors assigned to the climate classes correspond to a hot--to--cold sequence in Early Mars climate parameter space. The high elevation (red) and broad-swath (amber) classes have melt likelihoods as high as 0.17. The focussed equatorial (cyan) class shows much lower melting probabilities ($\leq$ 0.054) and is wrapped around the perpetual-global-desert climates (dark blue). To obtain the distribution of snowmelt that is in best agreement with sedimentary rock data, at least one of $P_o$, $\Delta T$ or $f_{snow}$ must be small. This assumes that sedimentary rock accumulation is proportional to the number of years with snowmelt. Sedimentary rock formation involves nonlinear and rectifying processes, which could allow the snowmelt production predicted by the broad-swath-of-low-latitude melting (amber) class to yield a sedimentary rock distribution consistent with observations. Alternatively, a broad initial sedimentary-rock distribution could be focussed by aeolian erosion, which would preferentially efface thin deposits. Therefore, data-model comparison supports the focussed-equatorial-melting climate class (cyan), and rules out the high-elevation melting paleoclimates (red class) and the perpetual global desert (dark blue). It disfavors, but does not rule out, the broad-swath-of-melting (amber) climate class.

In summary, if the seasonal-melting hypothesis is correct, Mars paleoclimate has left a fingerprint in the sedimentary rock distribution. Sedimentary rocks are distributed as expected if Mars only marginally permitted snowmelt, even under near-optimal orbital conditions. The climates that give the best fit to data predict planets on which the wettest geographic location would have been dry for $\gtrsim$90\% of the time. During the sedimentary-rock era, Mars was a dry place.

\subsection{Possible implications for other geologic data: valley networks, chlorides, and alluvial fans}


\noindent Regionally-integrated valley networks record overland flow \emph{prior} to the sedimentary rock era. We find that Mars valley-network elevation distribution is biased high by 600m relative to ancient terrain, although this may reflect the generally higher elevation of mid-Noachian (as opposed to Early Hesperian) outcrop  \citep{hynek2010}. 
High elevation is the fingerprint of high $P_o$ (Figure \ref{figure:INTORB2}b; Figure \ref{figure:cluster}b). This suggests a geologic record of progressive atmospheric loss:-- $P_o$ is $>$100 mbar at valley-network time (to drive snow to high ground as suggested by valley network elevations), falls to $\sim$100 mbar by sedimentary-rock time (high enough to suppress evaporative cooling, low enough to allow sedimentary rock formation at low elevation), and falls further to the current situation (6 mbar: $L_{fr}$ prevents runoff on horizontal surfaces). 
The melt rates predicted by our model with nominal parameters are $\lesssim$ mm/hr, probably insufficient to form the classical valley networks. Processes that could link runoff from snowmelt to the formation of classical highland valley networks include:-- (1) a stronger greenhouse effect than considered here, with or without the orbital variability considered in this paper; (2) increasing $e$ to $\sim$0.22, as can occur transiently during the restructuring of solar system orbital architecture predicted by the Nice model \citep{agnor2012}; 
(3) transient darkenings from impact ejecta and ash and transient heating from impact ejecta.

Chloride deposits ($n$ = 634) are generally older than the sedimentary rocks, extremely soluble, rare in the equatorial sedimentary rock bracelet, and regionally anticorrelated with sedimentary rock \citep{osterloo2010}. This excludes an erosional mechanism for the latitudinal distribution of sedimentary rocks. One possibility is that dust obscures chlorides at low latitudes. Another possibility is that chlorides were dissolved in the equatorial band during the melt events that lithified the sedimentary rocks. This would imply that melt rarely occurred far from the equator.

Peak runoff production \emph{during} the sedimentary-rock era is constrained to $\sim0.3\pm0.2$ mm/hr 
\citep{irwin2005,jaumann2010}. Melt production at these rates is possible in the climate-ensemble shown. However, runoff production will be some fraction of the melt rate, because of refreezing and infiltration. 
Similar to the case of the classical highland valley networks, additional energy could be supplied by transient darkenings (ash-on-snow, ejecta-on-snow) or transient volcanogenic warming. 
An alternative way to maximize runoff at low $P_o$ is a phase lag between and the position of cold traps (which is set by orbital forcing, e.g. \citet{montmessin2007}) and the position of ice deposits. For example, an ice deposit built up at 20S at high $\phi$ while $L_p$ $\sim$ 90$^\circ$ may melt if it is not removed by sublimation before $L_p$ swings back to 270$^\circ$. This phase lag contrasts to the snow considered in this paper, which is always in equilibrium with orbital forcing.


Evidence that the alluvial fans are \emph{younger} than most sedimentary rocks \citep{grant2011} is consistent with loss of CO$_2$ over time, because low $P_o$ suppresses equatorial melting (\S5.3). Figure 10 shows that at high $f_{snow}$ on a cueball planet, melt rates peak at $\pm$22$^\circ$, being negligible at higher latitudes and several-fold lower at the equator. The latitude of the wing ``peak'' increases with $f_{snow}$, but wings exist for a broad range of moderate-to-high $f_{snow}$ and $P_o$. Wings are observed in the latitudinal distribution of alluvial fans on Mars \citep{kraal2008,wilson2012}. 

\section{Snowmelt in time: predictions for MSL at Gale Crater \label{formation}}


\begin{figure*}
\includegraphics[width=1.0\textwidth, clip=true, trim = 0mm 0mm 0mm 0mm]{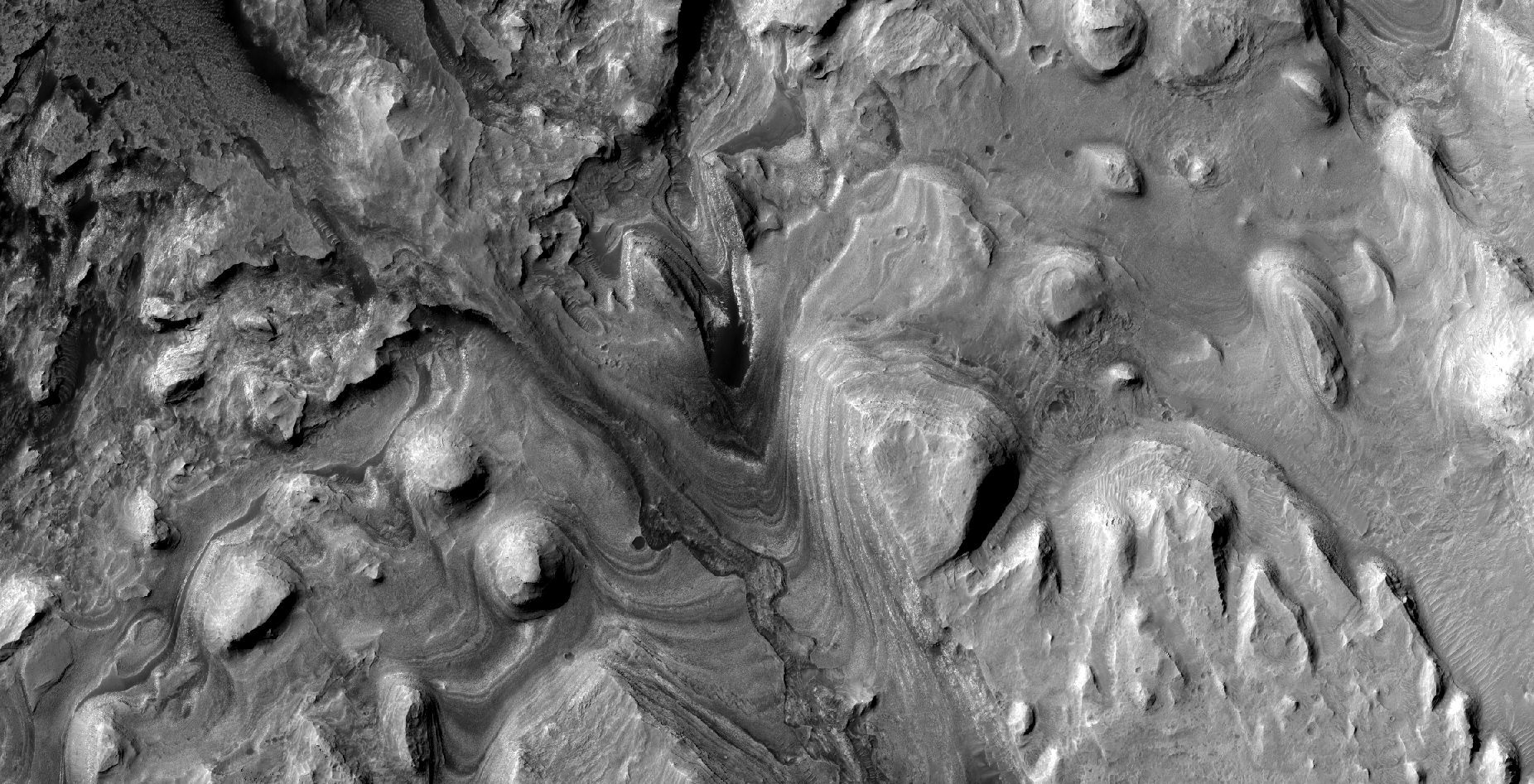}
\caption{Possible Mars Science Laboratory primary mission exploration targets in the foothills of the Gale Crater mound. Finely layered sulfate--bearing and phyllosilicate--bearing sedimentary rocks, locally reworked by a channel. See \citet{anderson2010} for geologic context. Part of HiRISE PSP\_009294\_1750. Image is $\sim$4500 m across, illumination is from top left.  \label{figure:galehirise}} 
\end{figure*}

\subsection{Testing snowmelt at Gale Crater}

\noindent The base of the Gale mound is a good place to test the snowmelt hypothesis, because snowmelt is predicted at Gale for most of the paleoclimates that permit surface liquid water anywhere on Mars.\footnote{Of the subset of climate states considered that predict snowmelt anywhere on the planet, 66\% predict snowmelt at the base of the Gale Crater mound. If we say that Gale Crater has a ``robustness'' of 66\%, then $>$99\% of ancient surface area scores lower for robustness. In addition, for 55\% of climates modeled, the base of the Gale Crater mound is in the top 1\% of the planet for melt likelihood. If we say that the base of the Gale Crater mound has a ``maximality'' of 55\%, then $>$99.9\% of ancient surface area scores lower for maximality.} Model predictions are for the lower unit of the Gale Crater mound, which is known to contain aqueous minerals \citep{milliken2010}, not the spectroscopically bland upper unit \citep{thomson2011} for which much less snowmelt is predicted. We think that it is not a coincidence that Gale is a good place to test the snowmelt hypothesis. MSL was sent to Gale because it hosts one of the thickest sedimentary rock packages on Mars, with mineralogic and stratigraphic hints of climate change \citep{milliken2010}. The snowmelt model predicts relatively abundant snowmelt at Gale, even in a changing Early Mars climate. If snowmelt is the limiting factor in sedimentary--rock production, then naturally Gale would be a place that would sustain sedimentary--rock formation for the widest range of climate conditions.

\noindent \textbf{Hypothesis:} We hypothesize that the Gale Crater mound is an accumulation of atmospherically-transported sediments pinned in place and subsequently reworked by seasonal-meltwater-limited processes. (This hypothesis is adumbrated in a unpublished M.S. thesis by \citet{cadieux2011thesis}, and in conference abstracts by \citet{cadieux2011} and by \citet{niles2012}.)


\noindent\textbf{Tests:} The snowmelt model predicts:--

\begin{itemize}

\item{\emph{Wet-dry cycles on orbital timescales,}} with dry conditions most of the time. 

\item{\emph{Mound-scale geochemistry records a succession of closed systems, not a flow-through geochemical reactor.}}
If the fluids responsible for alteration were in contact with the atmosphere, as is true of all the ancient waters yet sampled by meteorites and rovers \citep{halevy2011}, then $\bar{T}$ $<$ 273K implies restriction of diagenesis to perched aquifers within meters of the surface (more beneath lakes). The Gale Crater mound is 5km high, so this predicts that the Gale Crater mound is a succession of tens-to-thousands of closed systems. If on the other hand the layers near the top of the mountain were precipitated from groundwater that had flowed from the bottom of the mountain, then the mountain is a flow-through geochemical reactor. Basal layers would then be vulnerable to alteration by subsequent upwelling fluids. If smectite layers are found between Mg-sulfate layers, this would place a tight upper limit on flow-through aqueous chemistry \citep{vaniman2011}. 


\item{\emph{Clay/sulfate transitions correspond to a change in silicate input, not a change in global environmental chemistry.}} At Gale and many other sites on Mars, sedimentary rocks transition upsection from irregular to rhythmic bedding \citep{grotzinger2012}. This suggests a change over time in the relative importance of transient darkenings from volcanism and impacts, versus orbital forcing. Early on, large explosive eruptions and large impacts were more frequent -- so many melt events were assisted by regional-to-global albedo reduction. As volcanism and impacts declined, darkening events became less frequent, so eccentricity change (Figure \ref{figure:orbital}) emerged as the key regulator of melt events. Therefore, we predict that the phyllosilicate layers in the base of the Gale Crater mound were altered in-situ, and are stratigraphically associated with impact ejecta  \citep{barnhart2011} or volcanic ash layers.

\item{\emph{Generally homogenous chemistry and mineralogy on ascending the mound}.} With the exception of these events, the protolith is globally-averaged atmospherically-transported sediment, and most alteration is local. This leaves little scope for unmixing of major-element chemistry.

\item{\emph{No Gale-spanning lakes (except immediately after the Gale-forming impact)}}. Local perennial lakes are possible, as in the Antarctic Dry Valleys \citep{doran1998}.


\item{\emph{Isotopic gradients}}. Within a unit representing a single identifiable melt event, isotopic trends will depend on the water loss mechanism. If the water evaporated, earlier deposits will be isotopicallsedry lighter (in H and O isotopes) and later deposits heavier. This is due to the preferential evaporation of light water and will give an O isotope trend similar to that seen within ALH84001. If, on the other hand, the water froze rather than evaporated, later deposits will be lighter or no time dependent trend in isotopic composition will be observed. By contrast, in a groundwater model, if the supply of groundwater is $\sim$constant during mineralization, then the isotopic composition of the evaporating fluid will be some steady-state value, which would depend on the isotopic composition of the upwelling fluid and the evaporation rate. Lesser variability is expected within a single deposition event.

\item{\emph{No organic carbon}}. Slow, orbitally-paced sedimentation and oscillation between reducing and oxidizing conditions would disfavor preservation of organic carbon.

\end{itemize}


\subsection{From snowmelt time series to the Gale Crater stratigraphic logs}



\noindent\emph{Seasonal cycles and runoff.} Early in the melt season, melt will percolate vertically and refreeze \citep{marsh1984}. Vertical infiltration of snowmelt can indurate and cement aeolian dust and sand. Draining and channelization of melt will lengthen the lifetime of subsurface melt, especially late in the melt season. The impermeable ice table constructed by refreezing of early-season melt favors late-season runoff. Runoff and ponding of snowmelt in ice-covered lakes requires that water reaches channels before it refreezes. Once water reaches channels, ice cover protects against further freezing. Because the daily average temperature is below freezing (in general this is not a requirement for seasonal-melting models, but it is a feature of the model output considered here), this requires that drainage times through firn are $<$1 sol, in turn requiring high drainage density. Channel deposits with high drainage density are sometimes seen within the sedimentary rocks of Mars (e.g., HiRISE ESP\_020602\_1755 and PSP\_007474\_1745), and feed into much larger (and much more frequently preserved) inverted channels. A possible terrestrial analog for these processes is the Coastal Thaw Zone of the Antarctic Dry Valleys.
%



\begin{figure*}[h]
\includegraphics[width=1.0\textwidth, clip=true, trim = 80mm 0mm 0mm 0mm]{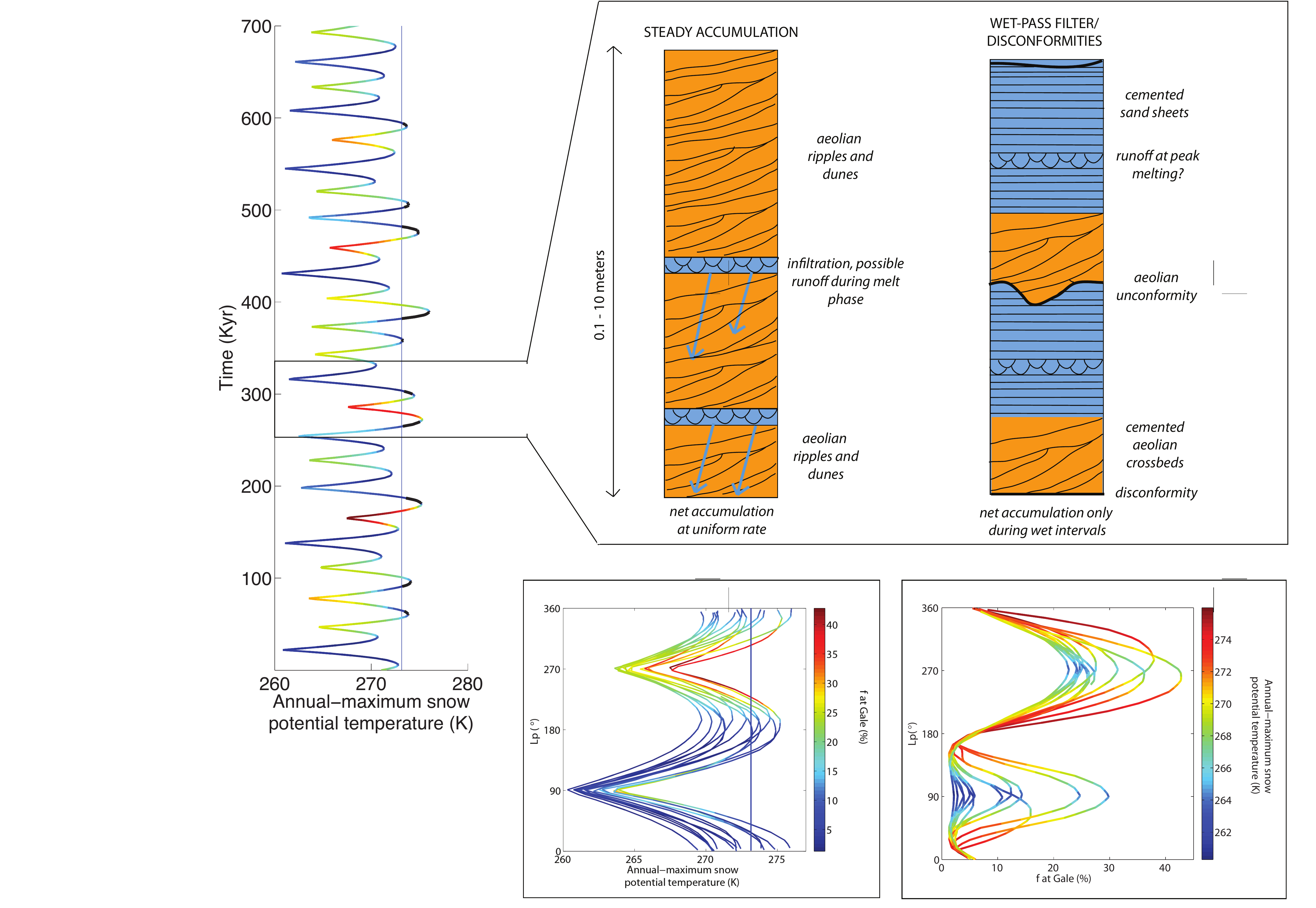}
\caption{Snowmelt model predictions for seven hundred thousand years at Gale Crater. \emph{Left time series:} Potential temperature of snowpack at Gale. Because the Solar System cannot be deterministically reverse integrated to 3.5 Gya, the orbital forcing is necessarily fictitious, but it is realistic (Laskar solution \texttt{301003BIN\_A.N006} for 73.05--73.75 Mya, but with 0.02 added to the eccentricity). The color scale corresponds to $f$ at Gale. Red is unfavorable for warm-season snow at Gale, and blue is most favorable for warm-season snow at Gale.  The vertical blue line corresponds to the melting threshold. Black highlights intervals of melting at Gale for  $f_{snow}$ = 10\%. $\Delta T$ = 6K, $P_0$ = 49 mbar. \emph{Stratigraphic logs:} Two end-member stratigraphic responses to orbitally-paced wet-dry cycles over $\sim$ 50 Kyr interval. Orange corresponds to sediment accumulated during dry intervals, and  blue corresponds to sediment accumulated during wet intervals. In the left column, the Gale Crater mound accumulates steadily with time, and layers are cemented by infiltration during wet intervals. In the right column, both accumulation and diagenesis are restricted to wet intervals.  \emph{Lower panels:} Precession cycles of temperature and $f$. Imperfect cyclicity results from varying eccentricity. Perihelion during northern-hemisphere summer is especially favorable for snow accumulation at Gale. Gale is dry when perihelion occurs during southern-hemisphere summer: snow accumulation is unlikely, and any snow that does accumulate fails to reach the melting point.
 \label{figure:galelog}} 
\end{figure*}

\noindent\emph{Milankovitch cycles.} Snowmelt predictions are mapped onto sedimentology and stratigraphy in Figure \ref{figure:galelog} (compare Figure \ref{wetpassfilter}). Wet--dry cycles with period $\sim$ 20 Kyr are inevitable unless $\Delta T$ $\sim$ 15K. Early in the wet phase of a wet-dry cycle, infiltration can provide water for diagenesis of layers that were deposited under dry conditions (Figure \ref{figure:galelog}). As cementation reduces permeability, infiltration will decline and runoff will be increasingly favored. The primary control on temperature cycles is precession, with secondary control by $\sim$100 Kyr eccentricity cycles. The ``steady accumulation" column in Figure \ref{figure:galelog} shows sedimentological predictions for the case where atmospherically-transported sediment is lithified by infiltration of snowmelt. The ``wet-pass filter/disconformities" column shows the case where rock formation only occurs during wet intervals. This produces major disconformities. Quasi-periodic liquid water availability at Gale will not necessarily produce quasi-periodic sedimentology. On Earth, orbitally-paced climate signals are recorded with high fidelity by abyssal sediments, but are shredded by fluvial processes and so are barely detectable in fluviodeltaic sediments \citep{palike2006,jerolmack2010}. 

Sequence stratigraphy on Earth divides stratigraphic packages into periods of relative sea-level rise and fall. Snowmelt model output suggests that the equivalent of sequence stratigraphy for Mars will involve mound accumulation during rare wet periods, and mound degradation (and apron accumulation?) during more common dry periods.

%
%
%
%
%

\section{Discussion\label{discussion}}

\subsection{Validity of model assumptions}
\noindent The snowmelt model assumes that the \emph{gross} accumulation rate of atmospherically transported sediment integrated over 4 Ga is enough to build up thick sedimentary rocks everywhere on Mars, but that the supply of liquid water is the limiting step for \emph{net} accumulation. Alternative limiting factors include sediment availability and preservation/exposure. 



\citet{malin2000a,malin2001} and \citet{edgett2002} have suggested sedimentary rocks were once much more widespread. In this case, restricted exposure today would correspond to blanketing or erosion since sedimentary--rock time. 

Sediment availability is unlikely to have limited sedimentary rock formation on Mars, because gross deposition rates for atmospherically-transported sediment on today's Mars (10$^{1-2}$ $\mu$m/yr; \citet{arvidson1979,geissler2005,geissler2010,johnson2003,kinch2007,drube2010}) are not much less than
past accumulation rates of sedimentary rocks: 20-50 $\mu$m/yr at the best-measured site \citep{lewis2008}. 
Planet-wide sand motion occurs on Mars \citep{bridges2012}.
Sand transport and dust lifting on Mars is sensitive to small increases in $P$ \citep{newman2005}. Because Mars has lost atmospheric CO$_2$ over time \citep{barabash2007}, gross accumulation rates for atmospherically-transported sediment were probably $\ge$ 10$^{1-2}$ $\mu$m/yr at the \emph{O}(10$^2$) mbar level required for snowmelt. Present-day reservoirs of airfall sediment are large. For example, dust deposits at Tharsis and E Arabia Terra are \emph{O}(10)m thick \citep{bridges2010,mangold2009}. On Early Mars, background dust supply would be supplemented by sediment produced during impacts and volcanic eruptions. We do not have good constraints on the current surface dust (and sand) budget and how important finite dust reservoirs are for the current dust cycle, let alone on Early Mars. Therefore, applying current deposition rates to make the argument that sediment availability is not a limiting factor is fairly speculative. Neverthless, if fluxes and reservoirs were as large in the past as today, then sedimentary rock formation would not have been limited by the availability of atmospherically-transported sediment. The difficulty then is to pin the sediment in place for $>$3.3 Gyr, and cementation by snowmelt is one mechanism that can resolve this difficulty.


Spatially varying precipitation is ignored. On Earth, ``[a]dvances and retreats of glaciers are broadly synchronous''  \citep{cuffey2010}, because small changes in Earth \textbf{x}, \textbf{O$^\prime$}, and $\Delta T$ overwhelm regional variations in precipitation. This ablation sensitivity is what makes glaciers good dipsticks for Earth's paleoclimate. On Mars, recent glaciations have laid down geomorphic strips near-parallel to lines of latitude, suggesting that longitudinally variable precipitation is less important than insolation in controlling precipitation and snowmelt \citep{kreslavsky2000,kreslavsky2003,heldmann2004,neumann2003,kadish2009,hauber2008,fassett2010}. 
The most recent equatorial ice deposits formed at intermediate elevations on Tharsis and Terra Sabaea \citep{forget2006,shean2010}, associated with 3D effects such as orographic precipitation. Models disagree about where ice should precipitate under different orbital conditions \citep{mischna2003,levrard2004,forget2006,mischna2006,madeleine2009}. 
This motivates follow-up GCM work.



Atmospheric collapse to form permanent CO$_2$ ice caps is more likely for Faint Young Sun insolation and for $\sim$100 mbar initial $P$ \citep{kahre2011agu} (Forget et al., 2012, submitted manuscript). However, snowmelt requires high $\phi$, which is less favorable for atmospheric collapse. Will a CO$_2$ atmosphere that has collapsed at low $\phi$ reinflate on return to high $\phi$? A straightforward calculation suggests that atmospheres do not stay collapsed. Dividing a 100 mbar atmosphere by the current seasonal CO$_2$ exchange rate of $\sim$3 mbar/yr gives a reinflation time of 30 yr, much shorter than orbital change timescales of 10$^4$ yr. Therefore the atmosphere is relatively unlikely to be collapsed for orbital conditions that optimize snowmelt. 

Neglecting the lapse rate in surface temperature is a good approximation for current Mars, where surface temperature is set by radiative fluxes \citep{zalucha2010}. Results from the LMD GCM (\citet{wordsworth2012}; Forget et al., submitted manuscript) show that the adiabatic lapse rate is not large at 250 mbar but is important for $P_o$ $\sim$ 500 mbar. To cross-check, the Ames Mars GCM was run at 80 mbar for modern orbital conditions, topography, and luminosity. Only a weak increase in surface-temperature coupling to the adiabatic lapse rate was found relative to the 6 mbar case. 
Therefore, neglect of the adiabatic lapse rate coupling to surface temperature appears to be adequate for $P_o$ $\sim$ 100 mbar. 

The model assumes that instantaneous values of $e$, $\phi$ and $L_p$ are independent. 
Reverse integrations of the Solar System (obtained from \texttt{http://www.imcce.\\ fr/Equipes/ASD/insola/mars/DATA/index.html}) show statistically significant correlation (p $<$ 0.0003) between $e$ and $\phi$, but with a very small correlation coefficient  ($|$R$|$ $<$0.08) and a sign that varies between integrations. The weakness of these correlations justifies  treating each orbital parameter independently. Mean probabilities exceed median probabilities for high $e$, but the exceedance probability for $e$ = 0.15 is $\sim$0.8 over 4 Gya \citep{laskar2008}.


We assume \textbf{C} changes more slowly than \textbf{O}, because post-Noachian rates of volcanic degassing, weathering, and loss to space are small compared to the atmospheric reservoir of CO$_2$. This assumption does not consider volcanic-- or impact--driven transients in $\Delta T$.

We assume the freezing point depression for melting is not very large, which is appropriate for sulfates (e.g., $\Delta T$ $\lesssim$ 4K for the magnesium sulfate - H$_2$O eutectic brine). Chloride brines allow liquid at much lower temperatures  \citep{pollard1999,fairen2009}. 

\subsection{Comparison with other proposed mechanisms for sedimentary rock formation}
\noindent Mechanisms for sedimentary rock formation on Mars must define sources of water, sediment, sulfur, and heat. In the ice-weathering model of \citet{niles2009}, the water source is an ice sheet. Sediment and sulfur is sourced from dust and gas trapped within the ice sheet. The heat source for weathering is the solid state greenhouse effect at shallow depths, and geothermal heating as the ice is buried. In the global-groundwater model \citep{andrewshanna2007,andrewshanna2010,andrewshanna2011}, strong greenhouse forcing warms the low latitudes to $>$ 273K (long term average). The water source is a deep, regional--to--global groundwater reservoir, which is recharged by precipitation or basal melting. Sulfur can be either from pyrite or from the atmosphere. The seasonal melting model implies conditions that are warmer and wetter than the ice-weathering model, but much colder and drier than the global-groundwater model. Snowmelt under a moderately thicker atmosphere is the water source, and insolation under unusual orbital conditions supplies heat. Sediment is atmospherically transported -- ice nuclei, dust--storm deposits, saltating sand, ash, and fine-grained impact ejecta -- and it is trapped in the snowmelt area by aqueous cementation. The sulfur source is the atmosphere.


The main strength of the ice-weathering model is that it is (near-)uniformitarian. Ice-sheet sulfate weathering is ongoing on Earth, and there is evidence for recent sulfate formation on Mars \citep{mangold2010,masse2012}.
Current gaps in the ice-weathering model include the difficulty of explaining interbedded runoff features \citep{grotzinger2006}, except as post-sulfate reworking, and a lack of a physical model for the proposed weathering mechanism.

Global groundwater models can explain the location of sedimentary rocks and the diagenetic stratigraphy at Meridiani \citep{andrewshanna2007,andrewshanna2011,hurowitz2010}. The global groundwater model is internally self-consistent and complete. Upwelling rates are consistent with inferred sediment accumulation rates. The discovery of gypsum veins in material eroded from the Shoemaker Formation ejecta in Endeavour Crater has been interpreted as evidence for bottom-up groundwater flow \citep{squyres2012}. Chaos terrain strongly suggests Mars had cooled enough to form a cryosphere that could modulate groundwater release. Therefore, even in the groundwater model, post-chaos interior layered deposits must have formed via a mechanism consistent with $\bar{T}$ $<$273, such as spring flow \citep{pollard1999,grasby2003}.

The advantages of the snowmelt model over previous 
models for the sedimentary-rock water source are as follows. The snowmelt model arises from a self-consistent climate 
solution (\S4 -- \S5), liquid water production can ``start and stop'' rapidly relative to Milankovitch cycles, and the equatorial concentration of sedimentary rocks 
emerges naturally (\S 5). The snowmelt model can account for the global distribution of sedimentary rocks (\S 6). 
In the snowmelt model, the sedimentary rocks form more or less in their current locations, with their current layer orientations, and in their current shapes. Most sedimentary rocks are now in moat-bounded mounds, filling craters and canyons. Groundwater models imply
removal of $\gg$10$^6$ km$^3$ of siliciclastic rock to an unknown sink \citep{zabrusky2012,andrewshanna2012vm}. 
This removal is mediated by a major phase of aeolian erosion which produces the moats. Structural deformation is also required to tilt the near-horizontal primary dips expected for playa-like deposition to the observed present-day draping dips. There is no need to appeal to large-scale postdepositional modification in either the snowmelt model or the ice-weathering model. 
Notwithstanding these advantages, the snowmelt model assumes that precipitation is uniform, but in reality it must have been spatially variable. The snowmelt model also does not include a physical model for any of the steps linking melt generation to bedrock formation. 

\subsection{Atmospheric evolution and the decline of sedimentary rock formation}

Few sedimentary rocks form on Mars now, and there is minimal surface liquid water. The only evidence for surface liquid water at the \emph{Opportunity} landing site since the current deflation surface was established is minor Na/Cl-enriched veneers and rinds \citep{knoll2008}. The simplest explanation for these changes is CO$_2$ escape to space. The 2013 MAVEN mission will constrain the present-day rate of escape to space. Supposing a 50-150 mbar atmosphere at sedimentary--rock time (Figure \ref{figure:cluster}, marginally consistent with \citet{manga2012}), a modern reservoir of 12 mbar \citep{phillips2011}, and that soil carbonate formation has been unimportant, a loss to space of $\sim$40-140 mbar over 3.5 Gya is predicted. Total loss of $\sim$40-140 mbar is higher than previous estimates of 0.8-43 mbar over 3.5 Gya from extrapolation of ASPERA-3 measurements \citep{barabash2007}, and 2.6-21.5 mbar from fits to MHD models by \citet{manning2011}. An alternative loss mechanism for CO$_2$ is uptake by carbonate weathering \citep{kahn1985,manning2006,boynton2009,kite2011c}. 
However, many sedimentary rocks contain sulfates, and small amounts of SO$_2$ prevent carbonate precipitation \citep{bullock2007,halevyschrag2009}. Another alternative is that orbital conditions needed to drive melting were sampled early in Mars history, but not subsequently.

\subsection{The Early Mars climate trade space}

\begin{figure}[h]
\includegraphics[width=1.00\textwidth, clip=true, trim = 2mm 10mm 50mm 5mm]{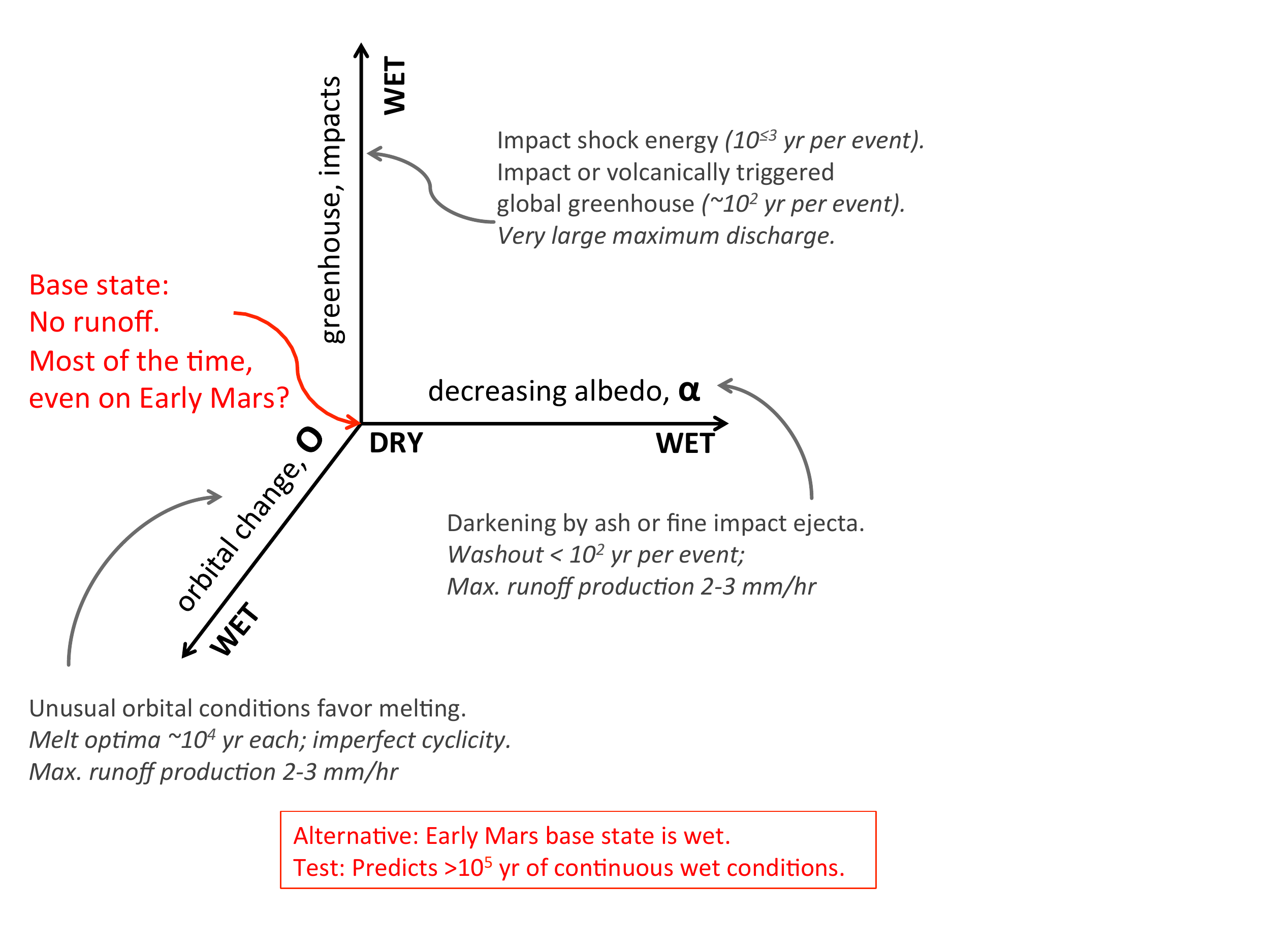}
\caption{The Early Mars climate trade space. Assuming runoff did not occur during most years on Early Mars, nonzero runoff can be produced by perturbing orbital conditions, reducing albedo, heating from greenhouse forcing or impact shock energy, or some combination (black axes). All mechanisms can produce runoff $\sim$1 mm/hr, but are distinguishable (gray arrows) by their limiting runoff and by their timescale. 
\label{earlymarsclimatetradespace}}
\end{figure}

\noindent This paper has emphasized unusual orbital conditions, but more than one mechanism could lead to melting of snow or ice on Mars (Figure \ref{earlymarsclimatetradespace}). 

Deposition of ash or fine-grained impact ejecta can lower albedo, driving transient runoff events on Early Mars (Equation \ref{equation:eae}). Unweathered silicates provide the trigger for their own alteration by darkening the snow. In addition to Gale Crater (\S 7.1), this $\alpha$--hypothesis is relevant for phyllosilicate formation at Mawrth: a regionally extensive, layered deposit which may be consistent with top-down alteration of ash \citep{noedobrea2010,mckeown2009,bishop2008,wray2008,michalski2007,loizeau2007}. Layered clays generally predate the sulfate rocks, consistent with decline of volcanism and impacts in the Early Hesperian \citep{ehlmann2011}.  
Albedo effects are the primary regulator of spatial and temporal \citep{hall2010} melt production in the Antarctic Dry Valleys. 
Experimentally dusted (13-100 g/m$^2$ fine sand) patches of an Antarctic Dry Valleys glacier surface showed sharp increases in melt rate \citep{lewis2001}. 
Antarctic Dry Valleys snowpack melts faster when it is buried beneath sand \citep{heldmann2012}, provided the sand cover is mm-dm thick. 

Heating from longwave forcing or conduction ($\Delta T$ or impact ejecta heating) could also drive melting. Increased $LW\!\! \downarrow$ could result from clouds (but see \citet{colaprete2003}) or short-lived pulses of volcanogenic gases \citep{halevy2007,johnson2008,tian2010,halevy2012}). Melting by impact ejecta is discussed by \citet{mangold2012} and \citet{noedobrea2010}.

Solar luminosity is the final dimension of the Early Mars climate trade space. This paper uses a standard solar model \citep{bahcall2001} that is consistent with solar neutrinos and helioseismology, but not elemental abundances in the photosphere \citep{asplund2005}. Enhanced mass loss from the young Sun would help resolve this discrepancy, and would make the young Sun more luminous \citep{guzik2010,turckchieze2011}. To change the conclusion that the Sun was faint at the time the sedimentary rocks formed, the Sun's subsequent mass loss rate must have been 2 orders of magnitude higher than inferred from nearby solar-analog stars \citep{wood2005, minton2007}. 


Future work could determine which mechanism is responsible using geologic observations that constrain discharge and timescale. Albedo reduction events are short-lived, with runoff production that cannot exceed 2-3 mm/hr. Optimal orbital conditions are relatively long-lived, but again runoff production is limited by sunlight   energy and cannot exceed 2-3 mm/hr. Volcanic- or impact-driven events are short-lived, but with potentially very large discharge. 

The climates considered in this paper are extremely cold, comparable in melt production to the coast of Antarctica \citep{liston2005}. These climates can produce enough water for aqueous alteration, but struggle to match peak-runoff constraints. Therefore, there is scope for exploring warmer snowmelt-producing climates, comparable to the coast of Greenland. One possibility is that cementation of the sedimentary rocks is the result of optimal orbital conditions, but that river deposits interbedded with those rocks \citep{williams2007} record additional transient (non-orbital) warming.

\section{Summary and conclusions\label{conclusions}}





\noindent The work presented here has two parts. First, a seasonal melting framework for Mars has been developed that relates candidate paleoclimate parameters to the production of seasonal meltwater. This framework has potentially broad applications. Second, the model has been applied to a specific problem: the origin and distribution of sedimentary rocks on Mars.  

Seasonal melting on Mars is the product of tides of light and tides of ice, which move around the planet on Milankovitch frequencies. The peaks of these tides rarely intersect. When they do, melting occurs. This water source may contribute to sedimentary rock formation.
The main conclusions from this work are as follows:- 

\begin{itemize}

\item Order-of-magnitude calculations indicate that snowmelt is a sufficient water source for sedimentary rock formation on a cold Early Mars.

\item The distribution of sedimentary rocks on Mars is narrowly concentrated at equatorial latitudes and at low elevations. 

\item The optimal spin-orbital conditions for snowmelt in cold traps on Mars are high obliquity, longitude of perihelion aligned with equinox, and eccentricity as high as possible. Melting then occurs in the early afternoon, at the equator, during perihelion equinox season. 

\item A model of snowmelt on Early Mars has been presented, which uses a potential-well approximation to track cold traps for all orbital conditions. Integrated over all orbital conditions on an idealized flat planet, and assuming snowpack with the albedo of dust and a $\sim$100 mbar pure CO$_2$ atmosphere, the model predicts a narrow equatorial concentration of snowmelt is predicted if warm-season snow is tightly confined to cold traps. A broad low-latitude belt of snowmelt is predicted if warm-season snow is more broadly dispersed.




\item With MOLA topography, atmospheric pressure $>$ 100 mbar drives snow to high ground. 
High $f_{snow}$ allows snowmelt on low ground even at high $P$. Sedimentary rocks are not on high ground, so either $f_{snow}$ was high, snowmelt was not the water source for the sedimentary rocks,
or $P$ was $\lesssim$100 mbar at time of sedimentary rocks. 

\item With MOLA topography, a large swathe of parameter space produces a snowmelt distribution that is a good match to sedimentary rock locations. Enough water is produced to satisfy mass balance for aqueous alteration of sedimentary rock. 


\item Early Mars climate states that produce the best fit to the spatial distribution of sedimentary rocks on Mars are cold. Much warmer climates would lead to snowmelt over a large swath of the planet, inconsistent with observations.

\item Climates that permit surface liquid water on Mars usually predict snowmelt at Gale Crater. Therefore, if MSL does not find evidence for a snowmelt contribution to sedimentary rock formation at Gale Crater, this would be a decisive failure of the model presented here. 

\item Specific predictions for MSL at Gale Crater include generally homogenous aqueous chemical processing on ascending the mound, with clay layers corresponding to a change in siliciclastic input, rather than a change in global environmental chemistry. The Gale Crater mound should have experienced wet/dry orbital cycles, with wet events only during optimal conditions. Evidence for vertical fluid flow over distances comparable to the height of the Gale Crater mound would be a major failure of the model presented here.


\item This is the first physical model to identify Gale Crater as a hemispheric maximum for sedimentary rock formation on Mars. The model therefore has the potential to relate observations at Gale Crater to global habitability.

\end{itemize}

\appendix

\section{Data analysis  \label{masking}}

Locations in the MOC NA sedimentary rock database are likely to be strongly correlated with the true distribution of sedimentary rocks on Mars, even though MOC NA did not sample the planet uniformly. 
MOC NA took 97,000 images of Mars;  4\% showed sedimentary rocks \citep{malin2010} (\texttt{http:// \\ marsjournal.org/contents/2010/ 0001/files/figure16.txt}) . Although MOC NA imaged only 5.5\% of Mars' surface \citep{malin2010}, the Mars Reconnaissance Orbiter Context Camera (CTX) has surveyed $>$75\% of the planet at comparable resolution to MOC NA (April 2012 Malin Space Science Systems press release, \texttt{http://www.msss.com/news/index.php?id=43}) and has not found large areas of sedimentary rock missed by MOC NA. MOC NA targets were selected on a 1-month rolling cycle on the basis of Viking imagery, previous MOC images, and the demands of other Mars missions \citep{malin2010}. Sedimentary rocks were among the highest scientific priorities of the MOC NA investigation \citep{malin2000a,malin2010}. In the same way that oil wells are drilled more frequently in productive basins, there is a high density of MOC NA images in areas of sedimentary rocks identified early in the mission. Maps of the relative abundance of sedimentary rocks show only minor changes when defined using the fraction of MOC NA images showing sedimentary rocks within a given spatial bin instead of the absolute number of sedimentary rock observations in a given spatial bin.

The definition of sedimentary rock used by the MOC NA team excludes at least two areas that are sedimentary in origin, the Terra Sirenum drape deposit and a large part of the Medusae Fossae Formation \citep{grant2010,bradley2002}. However, the Terra Sirenum drape deposit has a distinct phyllosilicate-rich mineralogy from the sulphate-bearing sedimentary rocks that are the focus of this paper \citep{ehlmann2011}, and the entire area of the Medusae Fossae Formation is consistently predicted to be a global near-maximum in sedimentary rock accumulation by our orbitally-integrated model output. Therefore, neither of these omissions from the database is important to the data-model comparison.  

Because formation of sulfate-bearing sedimentary rocks peaked in the Hesperian, Terminal Hesperian and Amazonian terrain may conceal underlying sedimentary rocks and should be excluded from the analysis. The currently available global geological map of Mars \citep{skinner2006} is a digital renovation of Viking-era hardcopy maps \citep{scott1986,greeley1987,tanaka1987}. Instead of using the old map, the edge of young materials was traced using the USGS Mars Global GIS 
as a base. The resulting (``K12'')mask covers 45\% of the planet, but only 3.5\% of the images of sedimentary rocks ($n$ = 105). These 3.5\% are mostly from the Medusae Fossae Formation and the plateaux surrounding Valles Marineris. These rocks appear to represent a late tail in sedimentary rock formation, and so are retained in the plots of elevation and latitude dependence (Figure 2). 
 Omitting them does not change our conclusions. To check that our results do not depend on the idiosyncrasies of the tracing, Figure 2 from \citet{nimmo2005} was georeferenced onto MOLA topography and areas that they mapped as Late Hesperian or Amazonian materials (34\% of the planet) were traced (the ``NT05 map''). K12 draws the dichotomy boundary close to the highland break-in slope, whereas NT05 draws the boundary near the lowland edge of the fretted terrain. Unlike NT05, K12 includes the Medusae Fossae Formation as potentially dating from the sedimentary rock era, in virtue of recent results proving a Hesperian age for large parts of the formation (\citep{kerber2010,zimbelman2012}). The elevation and latitude results shown in Figure \ref{figuredata} show little change between NT05 and K12 masks. 
 Earlier versions of the analysis presented in this paper \citep{kite2011a,kite2011b} used NT05, and reached unchanged conclusions. NT05 reflects the modern understanding of Mars geology better than \citet{skinner2006}, but it was intended as a low resolution overview. We believe that K12 is better suited than either the Viking-era maps or NT05 for the purpose of masking out post-sedimentary-rock terrain.

Image--center coordinates are assumed to be close to the locations of sedimentary rocks. Image footprints can be large for orbits early in the MOC NA mission, so these phases are excluded. 

The Valles Marineris are a unique tectonic feature containing many sedimentary rocks. To make sure that conclusions are insensitive to this unique tectonic feature, all data was excluded within a large ``Valles Marineris box'' (260E - 330E, 20S - 20N). This did not significantly change the latitude or elevation dependence. 

The equatorial concentration of sedimentary rocks was previously noted in uncorrected data by \citet{lewis2009} (unpublished PhD thesis).

\section{Details of thermal model \label{details}}

\noindent
\emph{Radiative terms:} A line-by-line radiative transfer model of the atmosphere \citep{halevy2009} is used to populate two look-up tables:-- $LW\!\!\downarrow$ as a function of $T_1$ and $P$; and $SW\!\!\downarrow$ as a function of $P$ and solar zenith angle. The radiative transfer model, which for simplicity assumes a clear-sky, pure CO$_2$ atmosphere with no clouds or dust, is not run to radiative-convective equilibrium. Instead, for each combination of surface $P$, $T$, $\alpha$, and solar zenith angle, an atmospheric $P$-$T$ structure is prescribed and the resulting radiative fluxes are calculated. Following the approach of \citet{kasting1991}, the tropospheric lapse rate is dry adiabatic and the stratosphere is approximated as isothermal with a temperature of 167 K.  A two-stream approximation to the equations of diffuse radiative transfer (which accounts for multiple scattering) is solved over a wavelength grid with a spectral resolution of 1 cm$^{-1}$ at frequencies lower than 10,000 cm$^{-1}$ and a spectral resolution of 10 cm$^{-1}$ at higher frequencies. The error induced by this spectral resolution relative to high resolution calculations is small compared to the uncertainties in the other model parameters \citep{halevy2009}. The parameterisation of collision-induced absorption is the same  as in \citet{wordsworth2010a}, and is based on measurements by \citet{baranov2004} and calculations by \citet{gruszka1997,gruszka1998}.

The atmospheric temperature profile corresponding to $LW\!\!\downarrow$ is pinned to the diurnal average $T_1$. Mars' bulk atmospheric radiative relaxation time is $\sim$2 days at 6 mbar surface pressure \citep{goody1967,eckermann2011}, 
and increases in proportion to atmospheric density. It is assumed to be large for the $P$ relevant to melting ($>$50 mbar). 


\noindent
\emph{Free convective terms:} The turbulent flux parameterizations closely follow \citet{dundas2010}. Sensible heat loss by free convection is:

\begin{equation}
{S_{fr} = 0.14 (T - T_{a}) k_{a} \left(  \left( \frac{C_p \nu_{a} \rho_{a} }{ k_{a} } \right) \left( \frac{g}{\nu^2_{a}} \right) \left( \frac{\Delta \rho}{\rho_a} \right) \right)^{1/3} }
\end{equation}

\noindent where 
$T_{a}$ is the atmospheric temperature, $k_{a}$ is the atmospheric thermal conductivity, $C_p$ is specific heat capacity of air, $\nu_a$ is viscosity of air, $\rho_{a}$ is density of air, $g$ is Mars gravity, 
and $\Delta \rho / \rho_{a}$ is the difference in density between air in equilibrium with the ground and air overlying the surface layer. $\Delta \rho / \rho_{a}$ is given by

\begin{equation}
\begin{centering}
{ \frac{\Delta \rho}{\rho} = \frac{(m_c - m_w) e_{sat}  (1 - r_{h} )}{m_c P} }
\end{centering}
\end{equation}

\noindent Here, $m_c$ is the molar mass of CO$_2$, $m_w$ is the molar mass of H$_2$O, $r_{h}$ is the relative humidity of the overlying atmosphere, and $e_{sat}$ 
is the saturation vapor pressure over water ice. 
The expression for $\Delta$ $\rho$ assumes that water vapor is a minor atmospheric constituent.

$T_{a}$ is parameterized as \citep{dundas2010}

\begin{equation}
\begin{centering}
{T_{a} = T_{min}^{b_{DB}} T^{1 - b_{DB}}}
\end{centering}
\end{equation}

\noindent where $T_{min}$ is the coldest (nighttime) surface temperature experienced by the model, and $b_{DB}$ is the Dundas-Byrne `b', a fitting parameter. This is an empirical model motivated by Viking 2 measurements \citep{dundas2010}. $b_{DB}$ decreases as $P$ increases, because atmosphere-surface turbulent coupling strengthens. 
$b_{DB}(P)$ is obtained by fitting to the output of GCM runs at 7, 50, and 80 mbar which employed a version of the NASA Ames Mars GCM described in \citet{haberle1993} and \citet{kahre2006}. Specifically, $b_{DB}(P)$ is fit to the global and annual average of the temperature difference between the surface and the near-surface atmosphere for local times from 11:00-13:00.

We let

 \begin{equation}
{L_{fr} = L_{e} 0.14 \Delta \eta \rho_{a} D_{a}
\left(
\left(
 \frac{ \nu_{a}}{D_{a}}
 \right)
 \left(
 \frac{g}{\nu^2_{a}}
 \right)
 \left( \frac{\Delta \rho}{\rho}
 \right)
 \right)^{1/3}
 }
\end{equation}

\noindent where $L_{e}$ is the latent heat of evaporation, $\Delta \eta$ is the difference between atmosphere and surface water mass fractions, and $D_a$ is the diffusion coefficient of H$_2$O in CO$_2$.

\noindent
\emph{Forced convective terms:}  Sensible heat lost by forced convection is given by:

\begin{equation}
{S_{fo} = \rho_a C_p u_{s} A  (T_a - T) }
\end{equation}

\noindent where $u_{s}$ is the near-surface wind speed. Near-surface winds are controlled by planetary boundary layer turbulence which serves to mix the atmosphere vertically, so $S_{fo} \neq 0$ is consistent with the assumption of no meridional heat transport. The drag coefficient $A$ is given by


\begin{equation}
{A = \left( \frac{A_{vonk}^2}{\ln(z_{anem}/z_o)^2} \right)}
\end{equation}


\noindent where $A_{vonk}$ is von Karman's constant, $z_{anem}$ is anemometer height, and $z_{o}$ is surface roughness.

Near-surface wind speed $u_{s}$ in the NASA Ames Mars GCM decreases with increasing $P$ and decreasing solar luminosity. The four-season average of European Mars Climate Database (``MY24" simulation) globally-averaged near-surface wind speeds at the present epoch is 3.37 m/s \citep{millour2008}. This is extrapolated for $P\le$290 mbar using a logarithmic dependence of $u_{s}$ on $P$ fitted to the global and annual average of Ames Mars GCM model surface wind speed for initial pressures of 7, 50 and 80 mbar.  $u_{s}$ is lowered by a factor of 1.08 for the Faint Young Sun using the ratio of wind speeds for two 50 mbar Ames Mars GCM simulations that differ only in solar luminosity. Simulations suggest $u_{s}$ increases with $\phi$ \citep{haberle2003}, but this is ignored. Figure 9 shows the sensitivity of results to $u_{s}$ = f(P) and $u_{s} \neq$ f(P).

Latent heat losses by forced convection are given by:

\begin{equation}
{L_{fo} = L_{e} \frac{M_w}{k T_{bl}} u_{s} (e_{sat} (1 - r_{h}) )}
\end{equation}

\noindent where $M_w$ is the molecular mass of water, and $k$ is Boltzmann's constant. Latent heat fluxes for dirty snow are calculated assuming that the entire exposed surface area is water ice. Dirt concentrations are at the percent level by volume, or less, for all results presented here, so this is acceptable.

The free and forced fluxes are summed together, rather than considering only the dominant term. This matches the functional form of Mars-chamber data \citep{chittenden2008} and is the standard approach in Mars research \citep{dundas2010,williams2008snowpack,toon1980}. However, summing the terms is an idealization that may overestimate cooling.

\noindent \emph{Melt handling:} Melt occurs when $T_K$ $>$ (273.15K - $\Delta T$). $\Delta T$  is a freezing--point depression. It can also be interpreted as any non-CO$_2$ warming due to water vapor, ice clouds, or SO$_2$, stochastic fluctuations in material properties around those assumed in Table 1, or a higher solar luminosity. Additional greenhouse warming (freezing point at 273.15K) implies greater turbulent and $LW \!\!\uparrow$ losses at melting than freezing-point depression (freezing point at 273.15K - $\Delta T$), but $\Delta T$ is small so this difference is ignored here.  

Total melt present and total melt produced are tracked during the sol. Melt is not permitted to drain, and the melt fraction is not allowed to affect snowpack material properties except to buffer temperature during refreezing \citep{liston2005}.   

 

Ablation of the snowpack surface by sublimation is not directly tracked. The effect on sublimation on snowpack survival is treated indirectly, through the potential-well approximation (\S\ref{model}.3). However, ablation also affects snowpack temperature. Implied sublimation rates are $\sim$0.5 mm/sol for conditions favorable to melting. Movement of the snow surface down into the cold snowpack corresponds to advection of cold snow upwards (relative to the surface). Snowpack thermal diffusivity is $\sim$2 x 10$^{-7}$ m$^2$/s. Melting at depths greater than $\sim \kappa / u_{subl}$ $\sim$ 4 cm may be suppressed by this advective effect. 

\noindent \emph{Run conditions}.
\noindent Conductive cooling is found by matrix inversion. Vertical resolution is $\approx$2.5mm for nominal parameters, which is 0.033$\times$ the analytic diurnal skin depth. Time resolution is 12s, and the lower boundary condition is insulating. 

The initial condition at the surface is slightly cooler than radiative equilibrium, decaying to the energy-weighted diurnal average temperature at depth with an e-folding depth equal to the diurnal skin depth. The model is integrated forwards in time for several sols using constant seasonal forcing until the maximum $T_1$ on successive sols has converged (to $<$0.01K) and the diurnal--maximum melt column (if any) has converged to $<$0.018 kg/m$^2$. For polar summers, convergence can take an extremely long time as the melt zone spreads to cover the entire snowpack, so the integration stops after $\sim$8 sols even if the convergence criteria are not met. 
 
There is no meridional heat transport, seasonal thermal inertia, or CO$_2$ cycle. Temperatures are not allowed to fall below the CO$_2$ condensation point. For each spatial location, the model is run for many seasons ($L_s$). The converged output is then interpolated on a grid equally spaced in time to recover annual means.

\noindent \emph{Details about melt-likelihood map construction}.
\noindent Results in this paper are based on grids of runs at $\phi$ = \{0$^\circ$, 10$^\circ$, 20$^\circ$, ..., 80$^\circ$\}, $e$ =  \{0, 0.03, 0.06, 0.09, 0.115, 0.13, 0.145, 0.16\}, $L_p$ = \{0$^\circ$, 15$^\circ$, 30$^\circ$, ..., 90$^\circ$\} (with mirroring to build up a full precession cycle), $L_s$ = \{0$^\circ$, 22.5$^\circ$, 45$^\circ$, ..., 337.5$^\circ$\}, and latitude \{-90$^\circ$, -80$^\circ$, ..., 90$^\circ$\}, giving 1.5 $\times$ 10$^6$ snowpack thermal model runs for each \textbf{C}. Quoted results at intermediate values result from interpolation. Statements about \textbf{C} are based on interpolation in a grid of runs at $P$ = \{4, 8, 16, 24, 48\} $\times$ 610 Pa $\equiv \{24, 49, 98, 146, 293 \}$ mbar, with $\Delta T$ = 0K. $\Delta T$ and $f_{snow}$ were varied in postprocessing. 

To remove longitudinal stripes of high snow probability in the Northern Plains that are artifacts of finite model resolution in \textbf{O$^\prime$} and latitude, the step function in ($f_{snow} - f$) is replaced by a linear ramp in ($f_{snow} - f$). This is a minor adjustment. 

\section{Snowpack radiative transfer \label{solidstate}}
\noindent Crystalline water ice is opaque in the thermal infrared, but
almost transparent to visible light. The resulting solid-state
greenhouse effect enhances snowmelt \citep{clow1987,brandt1993}.
The purpose of the solid-state greenhouse parameterization in this
paper is to self-consistently model the tradeoff between snowpack
broadband albedo ($\alpha$) and subsurface absorption of
sunlight. This does not require precisely calculating $\alpha$ as a
function of dust content, so the model uses simple linear
approximations to the radiative transfer equations developed for
widely--seperated atmospheric aerosols
(e.g. \citet{kieffer1990,calvin2009}).  Although more sophisticated models
can be employed to take account of aspherical particles, near--field
effects, and heterogeneous compositions
\citep[{e.g.}][]{cull2010,yang2002}, the lack of consensus on their
importance leads us to not include them in our algorithm.

The solid-state greenhouse parameterization uses the snow radiative
transfer model of \citet{brandt1993}. Ice refractive indices are from
\citet{warren2008}, and are converted to Henyey-Greenstein parameters
using a standard Mie code following \citet{bohren1983}. Mars dust
optical parameters are calcuted using the refractive indices of
\citet{wolff2006,wolff2009}.  An illustration of these parameters for
the canonical atmospheric dust sizes is shown in
Figure 1 of \citet{madeleine2011}, but we also employ larger
sizes as well.   The 2000 ASTM Standard Extraterrestrial Spectrum Reference
E--490--00 is used to describe the wavelength dependence of the direct
flux component;  diffuse flux is neglected as a being a minor perturbation. The young Sun was $\sim$100K
cooler in the standard solar model. Solar reddening increases $\alpha$
by $<$0.01, so the spectral shift is ignored here.  The effect of
small amounts of meltwater on $\alpha$ is minor \citep{warren1982} and
is also ignored. The effects on wavelength-dependent direct-beam
semi-infinite albedo (not shown) are broadly similar to the idealized
``red dust'' in \citet{warren1980}. Once optical properties are
prescribed, the most important variables are dust content, effective dust grain
radius, and effective ice grain radius. A given $\alpha$ can usually be obtained
by several different combinations of these properties. The
\citet{brandt1993} model is used to build a look-up table of
fractional subsurface absorption as a function of these variables,
plus direct-beam path length. This length is mapped to depth within
soil by multiplying by the cosine of the zenith angle.

The radiative transfer model reproduces the trends found by
\citet{clow1987}. The larger values of the Martian dust single
scattering albedo in the optical
\citep{wolff2003,wolff2006,wolff2009} reduce the amount of melting for
a given dust concentration. Ice grain size growth is slow in Mars'
present day polar caps \citep{kieffer1990} but much faster under the
near-melting conditions that are important for the model presented
here. We adopt an effective size of 1 mm, corresponding to observed ice-grain radii in hoar
layers in Earth snowpacks. Not surprisingly, there are no direct measurements of dust content in
snow on Mars. Dust content in ice has been reported as ``a few percent
(up to at most around 30\%)" by volume in the Northern Plains
\citep{dundas2010}, and $\sim$15\% by mass in the SPLD
\citep{zuber2007}. We assume $\sim$2\% dust mass fraction by volume and a
dust grain radius of 4$\mu$m.

\section{Selection of snowpack material properties \label{material}}
\noindent Snow stability and peak temperatures are affected by material properties such as $\alpha$ and TI. Low (snowlike) TI \citep{carr2003a} is used here because snow precipitation is also predicted by all General Climate Models (GCMs) at high $\phi$ (e.g., \citet{fastook2008,mischna2003,madeleine2009}). In addition, water ice precipitation was observed on Mars by the Phoenix lander (\citet{whiteway2011}, their Figure 1). High (icelike) TI suppresses the diurnal thermal wave and makes melting at the equator much more difficult. 


The impact of parts-per-thousand levels of dust on snowpack albedo and runoff is severe \citep{warren1980,warren1984}. Present day observed and calculated Mars seasonal H$_2$O snow $\alpha$ is 0.25-0.4 \citep{vincendon2010,kereszturi2011}. $\alpha$ on the South Polar water ice cap is 0.30 \citep{titus2003}. To allow melting to create gullies, snowpack surface-layer albedo must have been as low as 0.12 \citep{williams2009}. Dust storms and dust devils occur every year, and caused major changes in regional and global albedo between 1978 and 2000 \citep{geissler2005} and between 2003 and 2007 \citep{putzig2007}. Globe-encircling dust storms, which now occur every few years, are likely to occur twice every year at high $\phi$ \citep{haberle2003}. Dust is required to supply ice nuclei for heterogenous nucleation. Therefore, it is reasonable to expect snowpack at high $\phi$ to be contaminated with dust. Given the likelihood of dust contamination, this paper assumes $\alpha$ = 0.28, the same as Mars' light-toned dust continents. This requires a dust concentration of \emph{O}(1\%) (Appendix \ref{solidstate}). Other likely sources of darkening contaminants on Early Mars are volcanic ash and fine-grained impact ejecta. In the words of  \citet{warren1984} ``When snow melts, the impurities often tend to collect at the surface rather than washing away with the meltwater.'' Using a low $\alpha$ favors melting, which is conservative because the reconstructed paleoclimate will involve the smallest change from the current Mars climate that is consistent with the geological evidence. 

\begin{figure}[H]
\includegraphics[width=1.00\textwidth, clip=true, trim = 12mm 0mm 0mm 5mm]{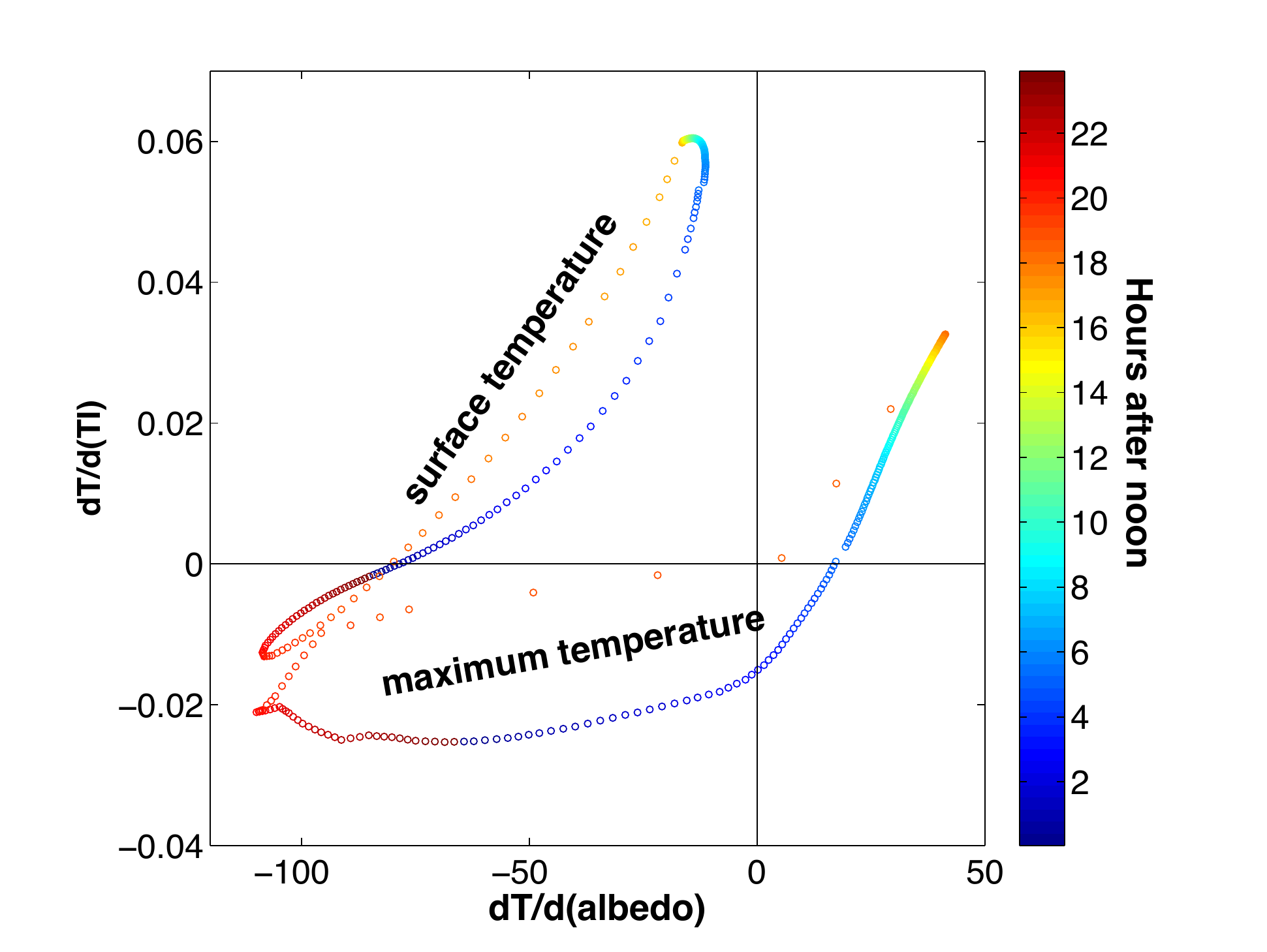}
\caption{Response of snowpack temperatures to adjusting material properties. Color corresponds to local hour angle (time after solar noon). The small jumps in the temperature loops are interpolation artifacts. $e$=0.11, $\phi$=50$^\circ$, and $L_p$=0$^\circ$. \label{figure:jacobiananalemma}}
\end{figure}

Figure \ref{figure:jacobiananalemma} shows sensitivity to material properties. Increased TI damps the diurnal surface temperature cycle. Increasing albedo lowers surface temperature for all times of day, especially near noon. The response of the maximum temperature within the snowpack (lower loop) is complicated by subsurface absorption of sunlight. This occurs at greater depths when dust concentration is decreased, even though the snowpack as a whole is more reflective. Because buried solar energy cannot easily escape, nighttime subsurface temperatures are increased by increasing the albedo. The location of maximum temperature moves to steadily greater depths during the night. The energy-burial effect is abruptly reversed shortly after dawn (18 hours after noon), when the location of maximum temperature returns to the near-surface. 

%
%

\ack
\noindent It is a pleasure to thank the following people for their generosity with time, ideas and data. We are grateful to Richard Brandt and Steve Warren for sharing the radiation code underlying \citet{brandt1993}. Aaron Wolf provided help with statistics. Joannah Metz supplied discharge calculations for the SW Melas Chasma fans. Mikki Osterloo supplied chloride coordinates. This work was triggered by discussions with Oded Aharonson, Jeff Andrews-Hanna, and Devon Burr. We thank the anonymous reviewers of an earlier version of the manuscript, whose comments improved the manuscript. Discussions with Konstantin Batygin, Bill Cassata, Bill Dietrich, Bethany Ehlmann, John Grotzinger, Alex Hayes, Ross Irwin, Vedran Lekic, Alejandro Soto, Ken Tanaka and Robin Wordsworth improved the manuscript.  E.S.K. is grateful to Francois Forget and Robin Wordsworth for sharing their Early Mars manuscript. We are grateful to the HiRISE team for maintaining a responsive public target request program, HiWish, which was useful in this work. 

E.S.K. and M.M. were supported by the U.S. taxpayer through NASA Science Mission Directorate grants NNX08AN13G, NNX09AN18G, and NNX09AL20G, as well as a startup allocation on the NSF Teragrid (TG-EAR100023). Additional computing costs were defrayed by a  NASA Mars Fundamental Research Program grant to Oded Aharonson.


\label{lastpage}

\bibliography{edwin_kite_references5}
\bibliographystyle{plainnat}

\end{document}